\documentclass[fleqn,usenatbib]{mnras}

\usepackage[percent]{overpic}
\usepackage{times}
\usepackage{graphicx}
\usepackage{amsmath}
\usepackage{mathabx}
\usepackage{subfigure}
\usepackage{natbib}
\usepackage{txfonts}
\usepackage{journals}
\usepackage{xcolor}
\usepackage[utf8]{inputenc}
\usepackage[overload]{empheq}
\usepackage{verbatim}
\usepackage{CJKutf8}
\usepackage{CJK}
\newcommand{\teff}{$T_{\rm eff}$}
\newcommand{\feh}{$\rm [Fe/H]$}
\newcommand{\logg}{${\rm log}\,g$}
\def\be{\begin{equation}}
\def\ee{\end{equation}}

\title[3D dust maps by V-net]
{Constructing the three-dimensional extinction density maps using V-net} 

\author[B.-Q. Chen et al.]{
Bing-Qiu Chen,$^{1}$\thanks{E-mail: bchen@ynu.edu.cn.}
Fei Qin$^{2,3}$\thanks{E-mail: feiqin@kias.re.kr.}
and Guang-Xing Li$^{1}$
\\
$^{1}$South-Western Institute for Astronomy Research, Yunnan University, Kunming, 650500, P.\,R.\,China\\
$^{2}$School of Physics, Korea Institute for Advanced Study, Dongdaemun-gu,  Hoegiro 85, Seoul 02455, Republic of Korea\\
$^{3}$Korea Astronomy and Space Science Institute, Yuseong-gu, Daedeok-daero 776, Daejeon 34055, Republic of Korea
}
\begin{document}

\date{Accepted ???. Received ???; in original form ???}

\pagerange{\pageref{firstpage}--\pageref{lastpage}} \pubyear{2022}
\maketitle
\label{firstpage}

\begin{abstract}
One of the major challenges we face is how to quickly and accurately create the three-dimensional (3D) density distributions of interstellar dust in the Milky Way using extinction and distance measurements of large samples of stars. In this study, we introduce a novel machine-learning approach that utilizes a convolution neural network, specifically a V-net, to infer the 3D distribution of dust density. Experiments are performed within two regions located towards the Galactic anti-center. The neural network is trained and tested using 10,000 simulations of dust density and line-of-sight extinction maps. Evaluation of the test sample confirms the successful generation of dust density maps from extinction maps by our model. Additionally, the performance of the trained network is evaluated using data from the literature. Our results demonstrate that our model is capable of capturing detailed dust density variations and can recover dust density maps while reducing the ``fingers of god" effect.  Moving forward, we plan to apply this model to real observational data to obtain the fine distribution of dust at large and small scales in the Milky Way.
\end{abstract}


\section{Introduction} \label{sec:intro}

\begin{figure*}
\centering
\includegraphics[width=0.45\textwidth]{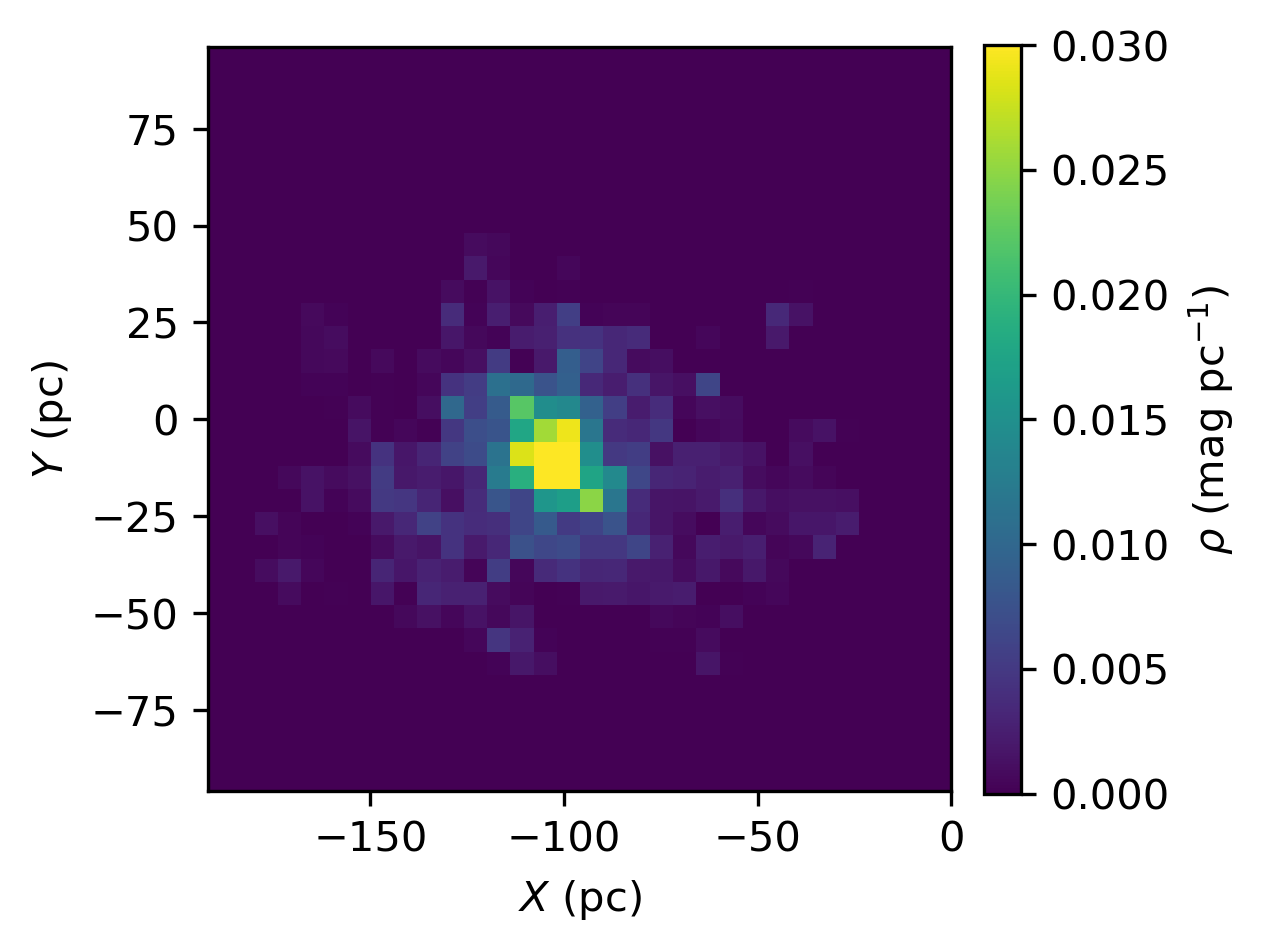}
\includegraphics[width=0.45\textwidth]{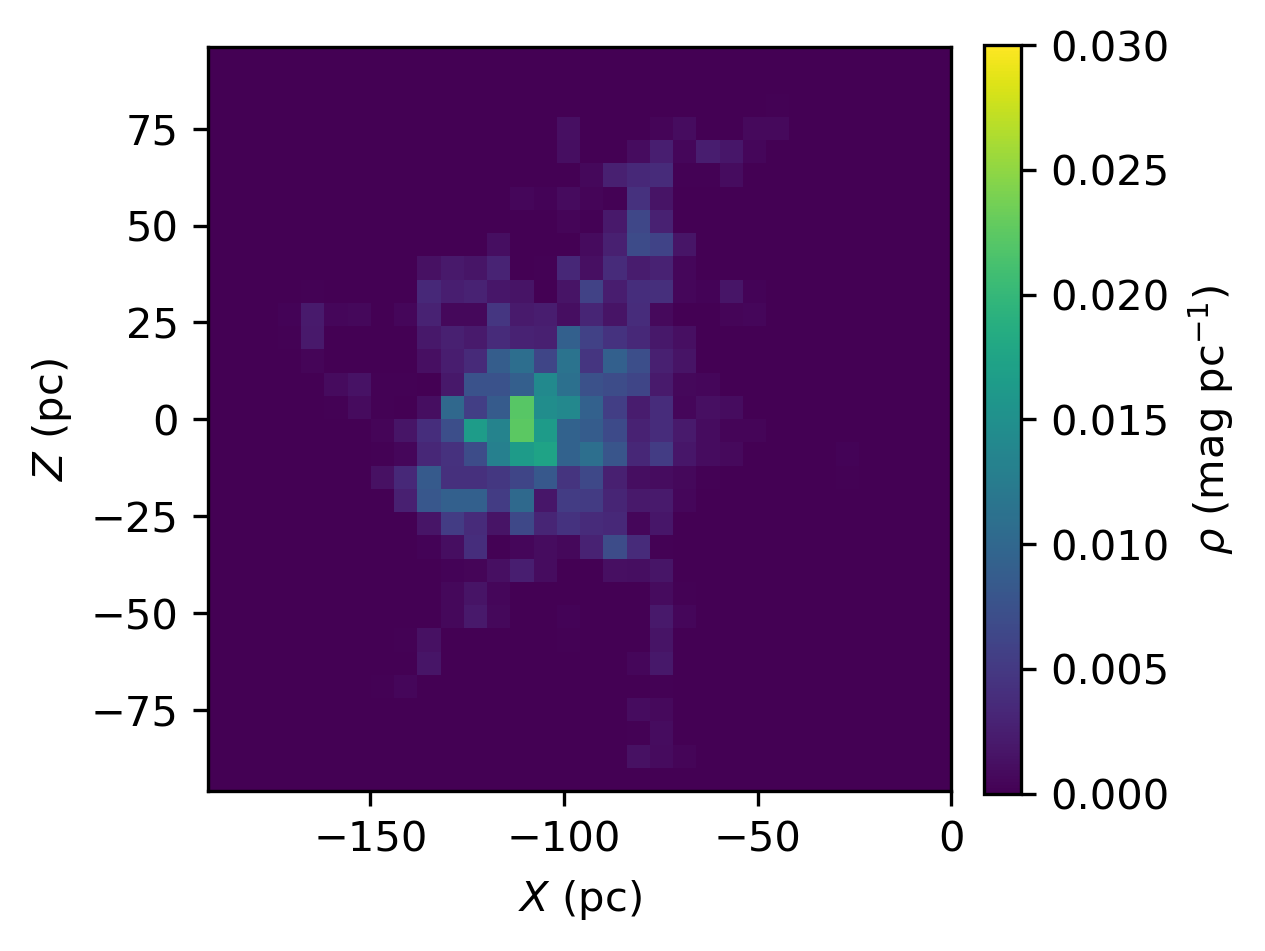}
\includegraphics[width=0.45\textwidth]{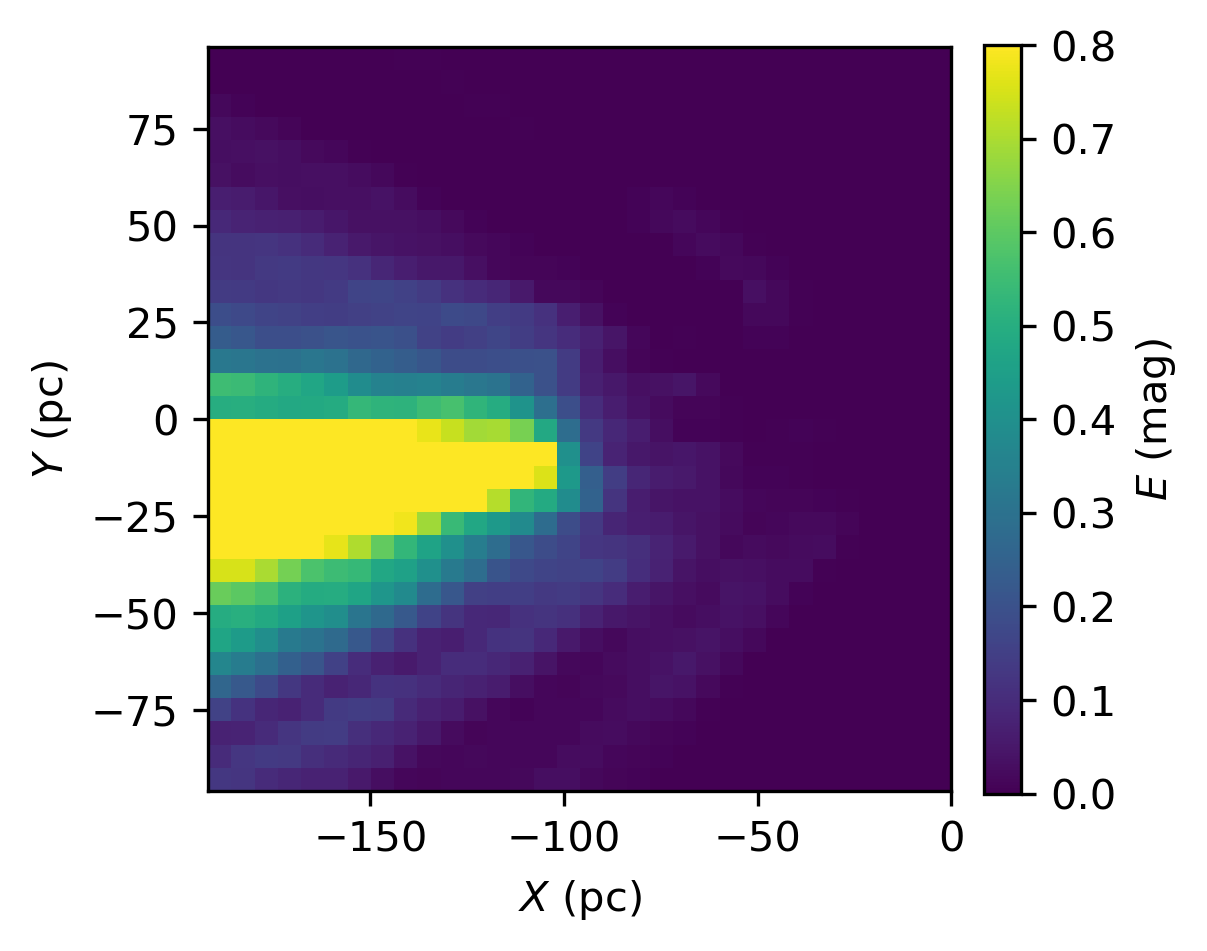}
\includegraphics[width=0.45\textwidth]{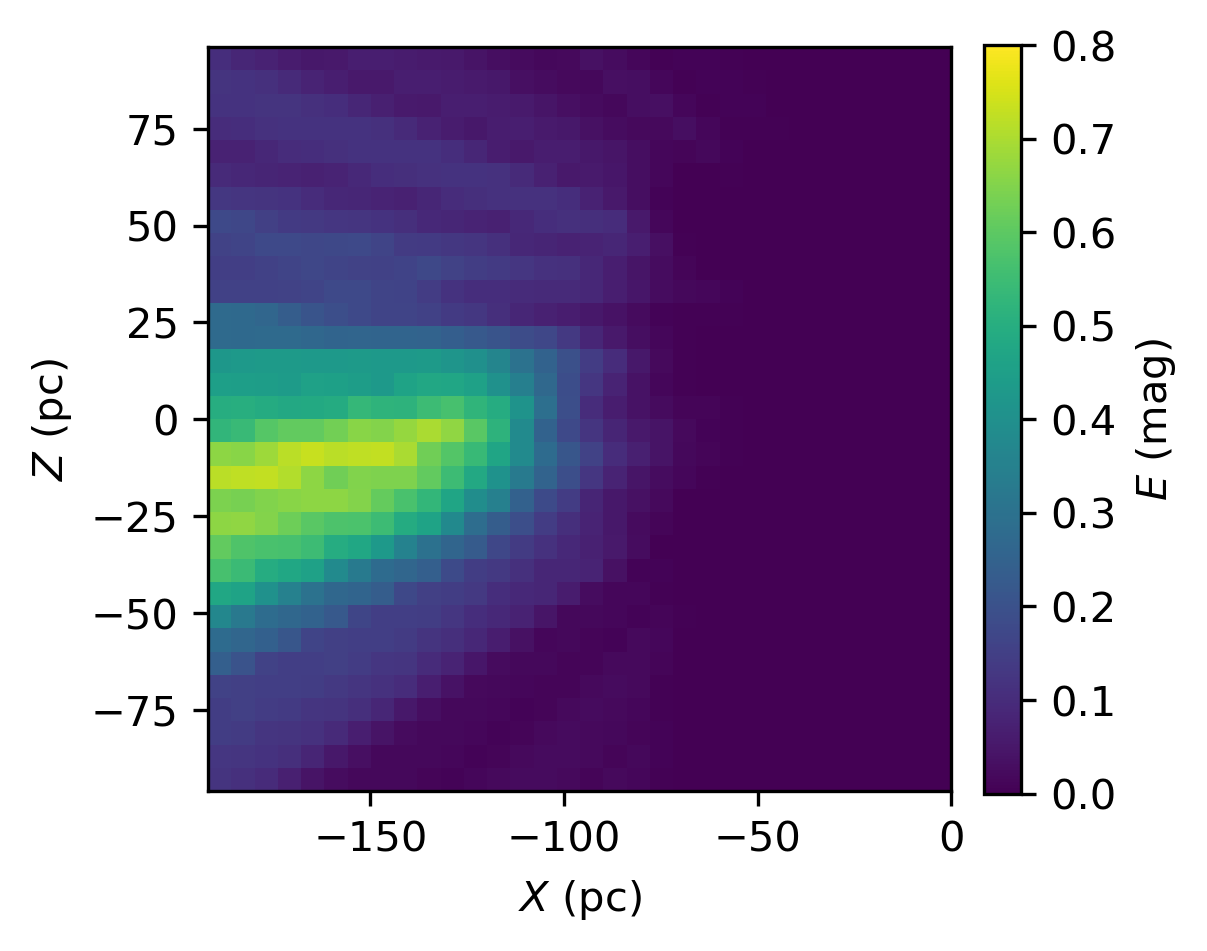}
\caption{The extinction density (upper panels) and the LOS extinction (bottom panels) distributions for an example simulation of the SR region. The left panels illustrate the dust density and extinction distributions in the $XY$-plane, with $Z$ values ranging from 0 to 6\,pc. Conversely, the right panels depict the same quantities in the $XZ$-plane, with $Y$ values ranging from 0 to 6\,pc.}  
\label{rhosim}
\end{figure*}

\begin{figure}
\centering
\includegraphics[width=90mm,height=52mm]{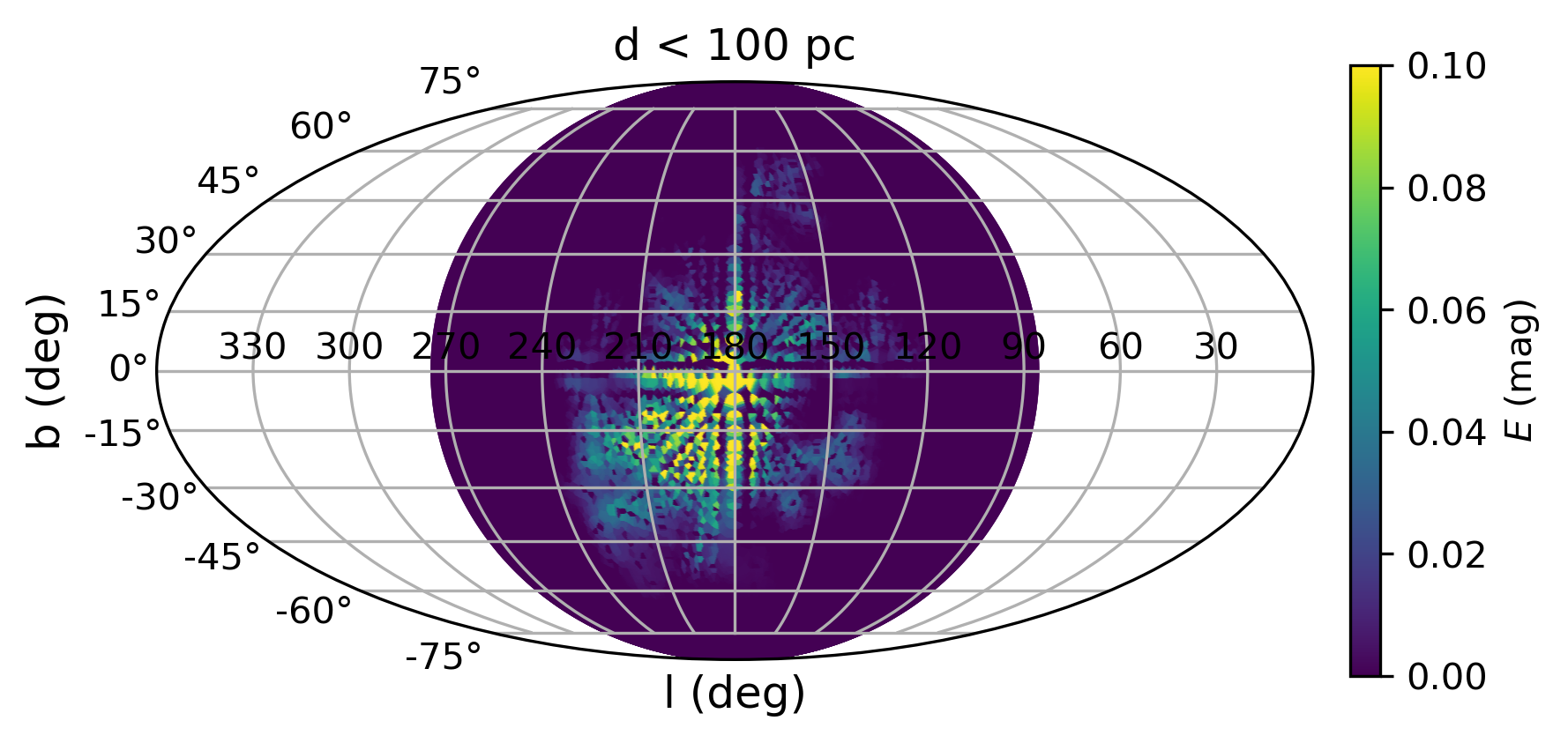}
\includegraphics[width=90mm,height=52mm]{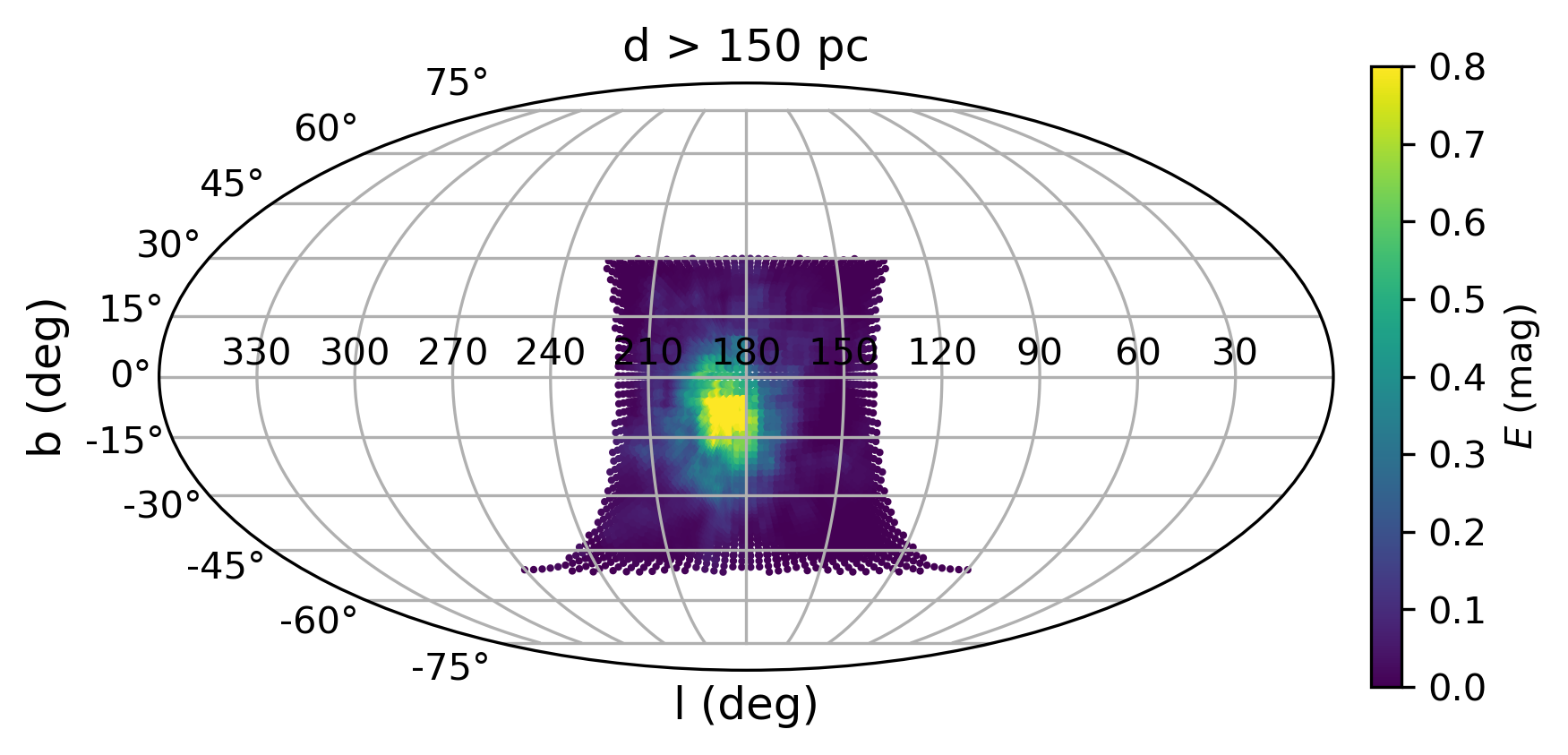}
\caption{Extinction distributions in the Galactic coordinates for the example simulation of the SR region. The upper and bottom panels show grids with distances smaller than 100\,pc and those with distances larger than 150\,pc, respectively.}  
\label{extlb}
\end{figure}

Interstellar dust is a crucial component in our Galaxy that can absorb and scatter light passing through it, leading to the extinction and reddening of background sources. Thus, when conducting optical or near-infrared observations, it is essential to account for the effects of dust. To correct the extinction effects of observed objects, three-dimensional (3D) extinction maps have been created (e.g. \citealt{Marshall2006,Chen2013,Chen2014,Chen2019,Green2015,Green2019,Guo2021}). These maps are also valuable tools in studying the nature of dust in the Milky Way (e.g. \citealt{Chen2015,Chen2020,Lallement2019,Zucker2022}). The distribution of dust in the Galaxy is complex and inhomogeneous, influenced by various processes such as gas dynamics, star formation and feedback, and the Galactic magnetic field. Therefore, understanding the distribution of Galactic dust is crucial to comprehend the processes that shape our Galaxy.

Over the years, numerous studies have been conducted to create 3D dust maps of the Milky Way. The conventional approach involves dividing the sky into different lines of sight (LOS) and obtaining the extinction and distance profiles for each subfield to construct the 3D extinction maps (e.g. \citealt{Chen2014, Sale2014, Green2015, Hanson2016, Guo2021}).  While this method is straightforward and effective, it has a limitation: it treats each LOS independently and does not propagate information between neighbouring LOS. This means that the dust that causes extinction is to be correlated, but this correlation is not taken into account, resulting in discontinuities in many published extinction maps that resemble the ``fingers of god'' artefact.

In an effort to reduce or eliminate this effect, astronomers have been exploring various methods in recent years. \citet{Lallement2014} used an inversion method based on a regularized Bayesian approach to construct a map of the interstellar medium (ISM) within 2.5\,kpc of the Sun. \citet{SaleM2014} presented a mapping method that models extinction as a Gaussian random field. \citet{Kh2017} used a non-parametric model that employs a Gaussian process to infer the 3D distribution of dust density in the Milky Way by connecting all LOS. \citet{Green2019} incorporated a spatial prior that correlates dust density across nearby sightlines. \citet{Leike2019} modelled dust as a log-normal process using a hierarchical Bayesian model and inferred the kernel of the log-normal process non-parametrically. In a more recent study, \citet{Edenhofer2023} employed a Gaussian Process within a spherical coordinate system to model the logarithmic dust extinction distributions. %However, these methods all essentially rely on strong prior assumptions and are computationally intensive, requiring significant computational resources.

%% the V-net method
In recent years, the field of astronomy has witnessed the integration of artificial intelligence (AI) technology to construct 3D density maps (e.g. \citealt{He2019, Hong2021, Veena2023, Qin2023}). Among the various deep learning algorithms, convolution neural networks (CNNs) have gained popularity due to their remarkable success in computer vision tasks. CNNs are designed to automatically learn hierarchical representations of visual data by processing it through multiple convolution layers that perform local operations on small regions of the input. One of the salient features of CNNs is their ability to capture spatial relationships between different parts of an image. Recently, \citet{Cornu2022} utilized a CNN that trained on synthetic colour-magnitude diagrams to derive extinction and distance profiles for individual LOS.

In this study, we introduce a novel CNN architecture V-net to infer the 3D distribution of dust density. To train and evaluate the V-net, we employ simulations of dust density and LOS extinction maps, which are introduced in Sect.~2. We provide a detailed description of the V-net architecture and its performance in Sect.~3. In Sect.~4, we apply our trained model to the measured data from the literature. Finally, we summarize in Sect.~5.  

%\begin{figure}
%\centering
%\includegraphics[width=\columnwidth]{levyflights.png}
%\caption{An illustration of the generated L\'{e}vy flights in $XYZ$-space for an example %simulation.}  
%\label{levy}
%\end{figure}

\begin{figure*}
\centering
\includegraphics[width=0.45\textwidth]{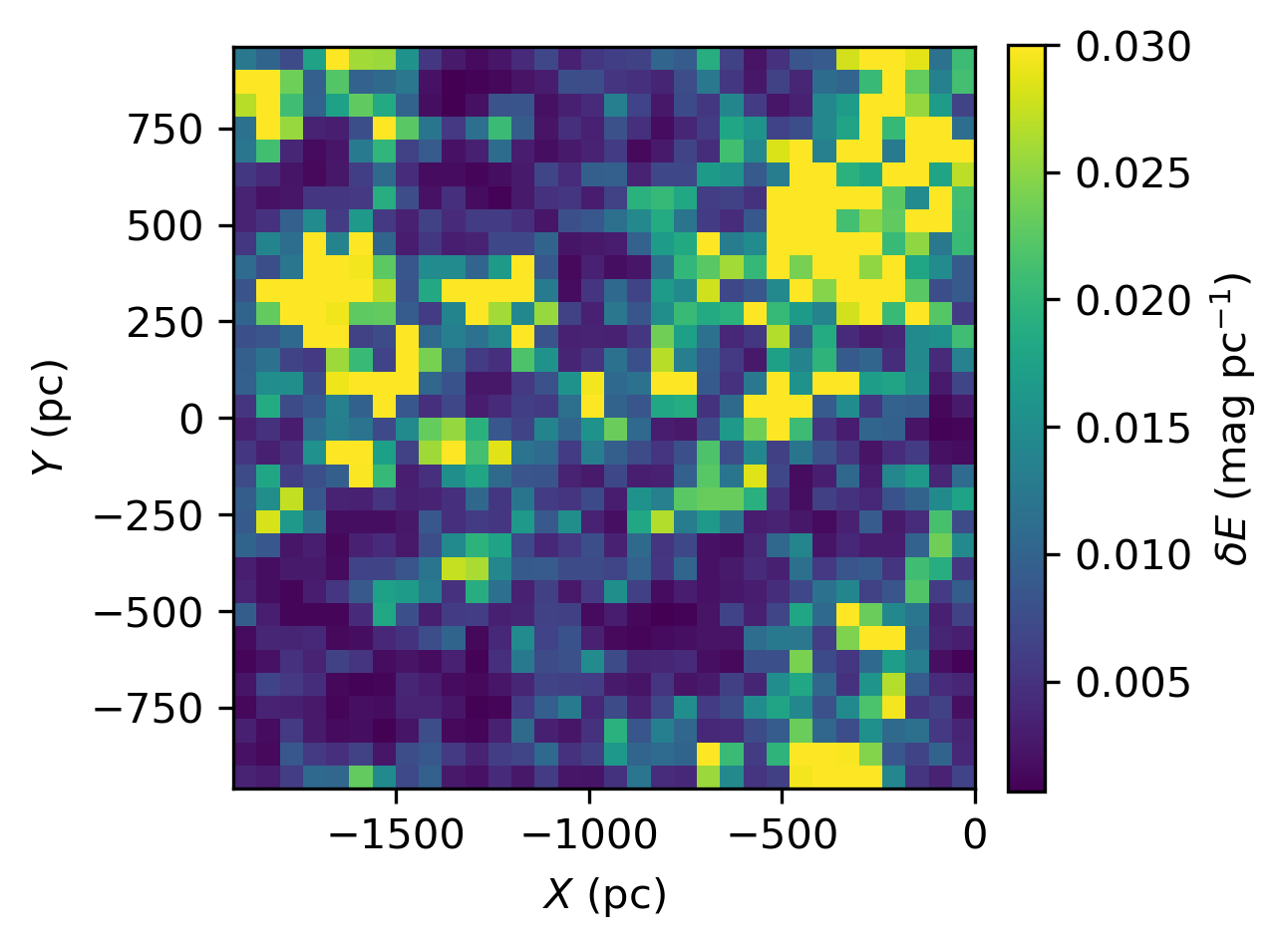}
\includegraphics[width=0.45\textwidth]{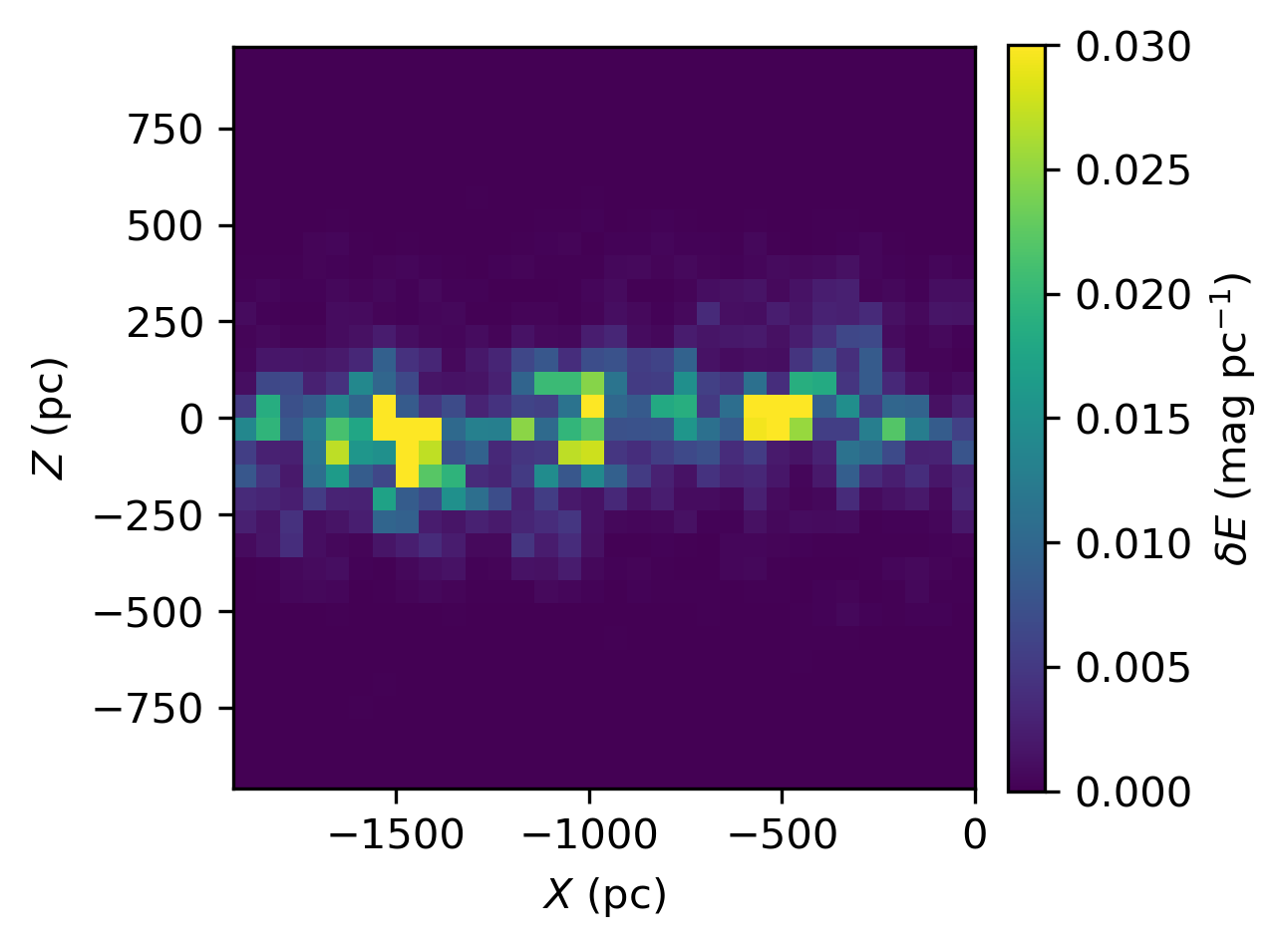}
\includegraphics[width=0.45\textwidth]{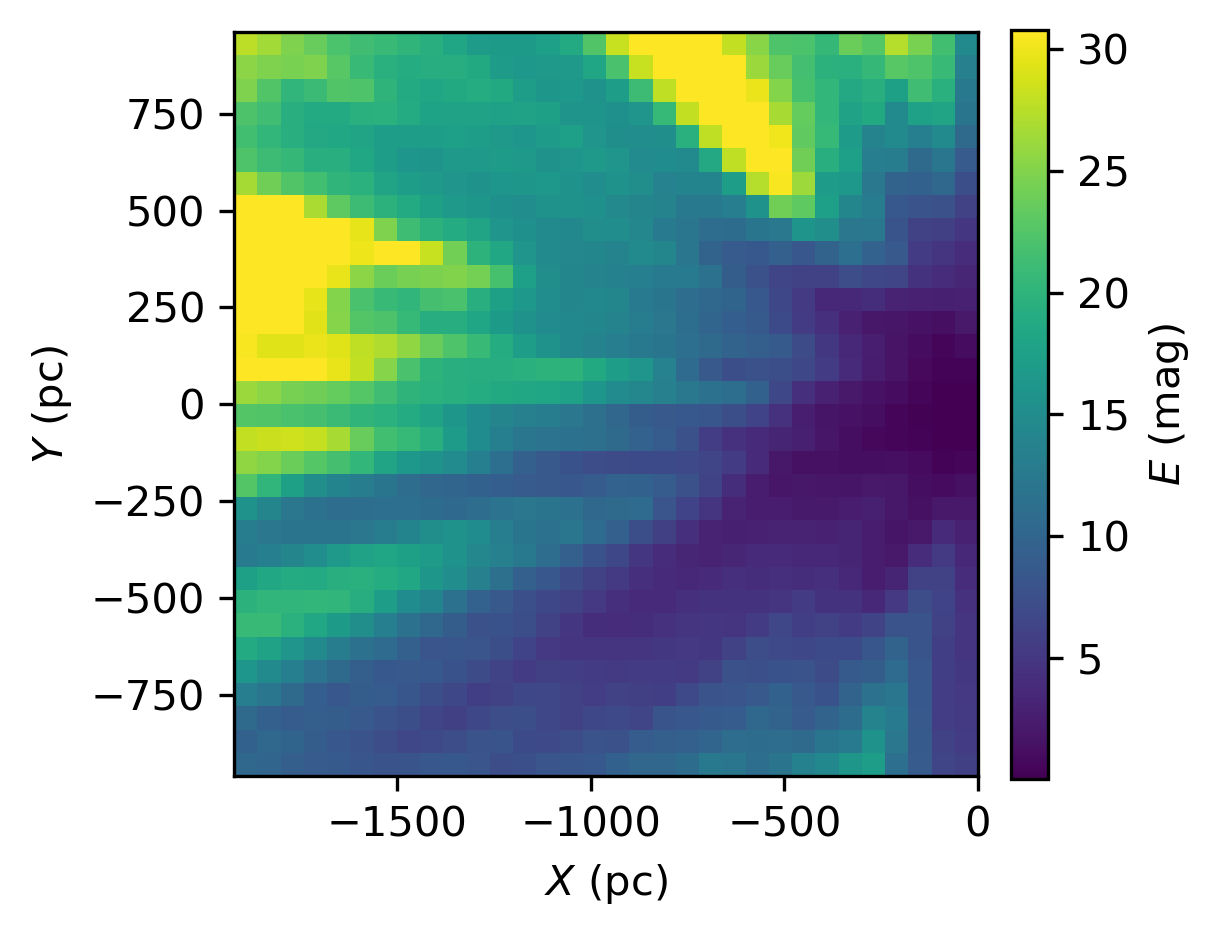}
\includegraphics[width=0.45\textwidth]{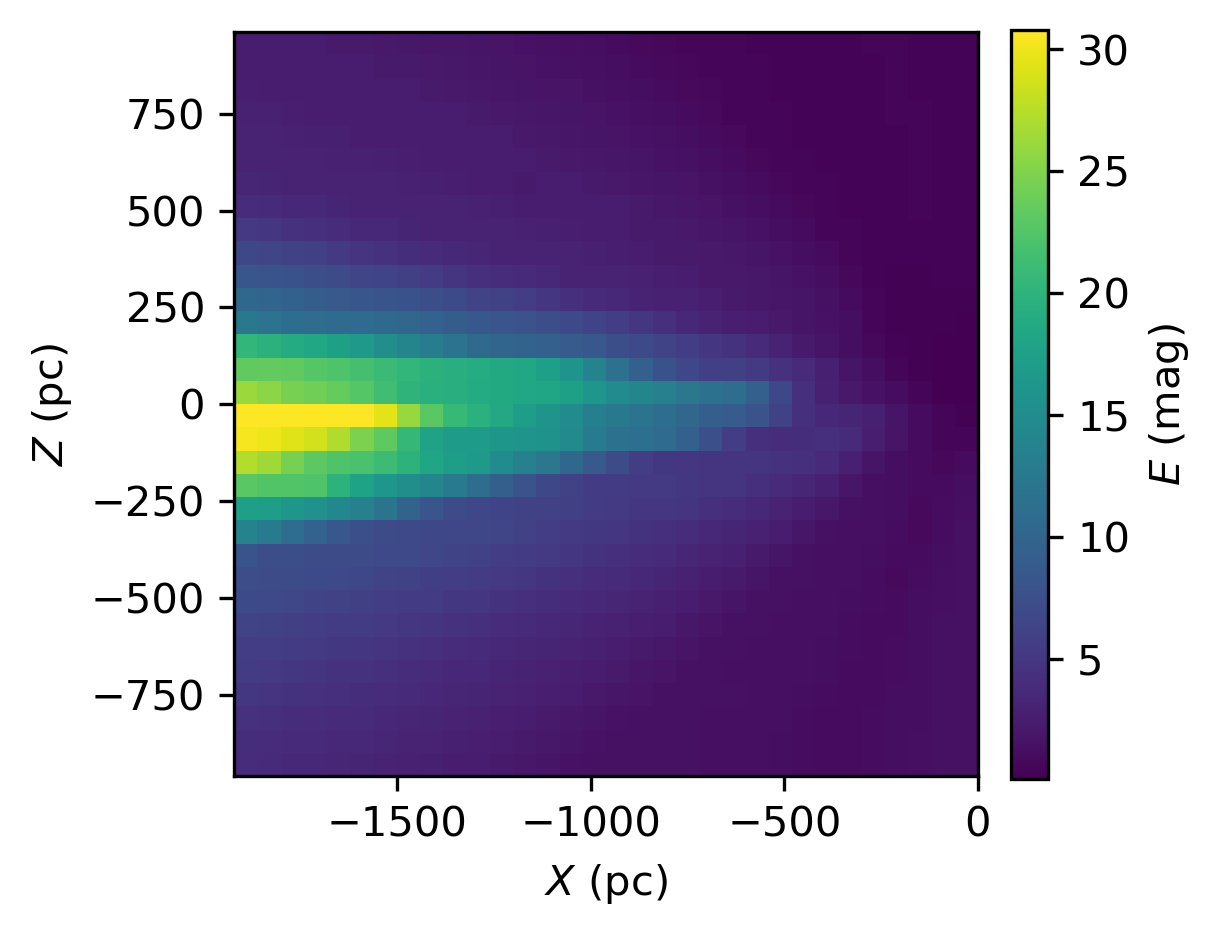}
\caption{The extinction density (upper panels) and the LOS extinction (bottom panels) distributions for an example simulation of the LR region. The left panels illustrate the dust density and extinction distributions in the $XY$-plane, with $Z$ values ranging from 0 to 60\,pc. Conversely, the right panels depict the same quantities in the $XZ$-plane, with $Y$ values ranging from 0 to 60\,pc. } 
\label{rhoLR}
\end{figure*}

\begin{figure}
\centering
\includegraphics[width=90mm,height=52mm]{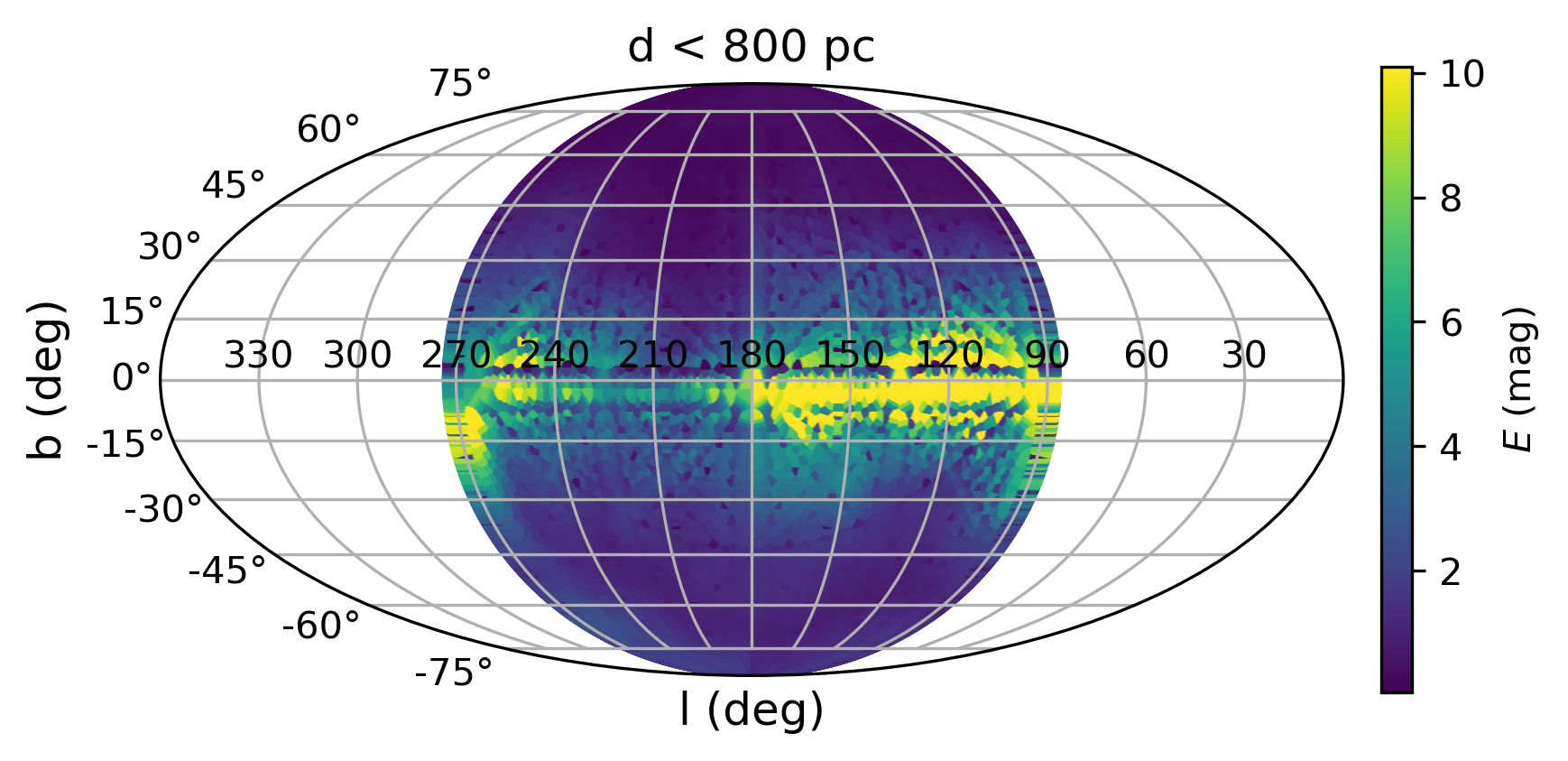}
\includegraphics[width=90mm,height=52mm]{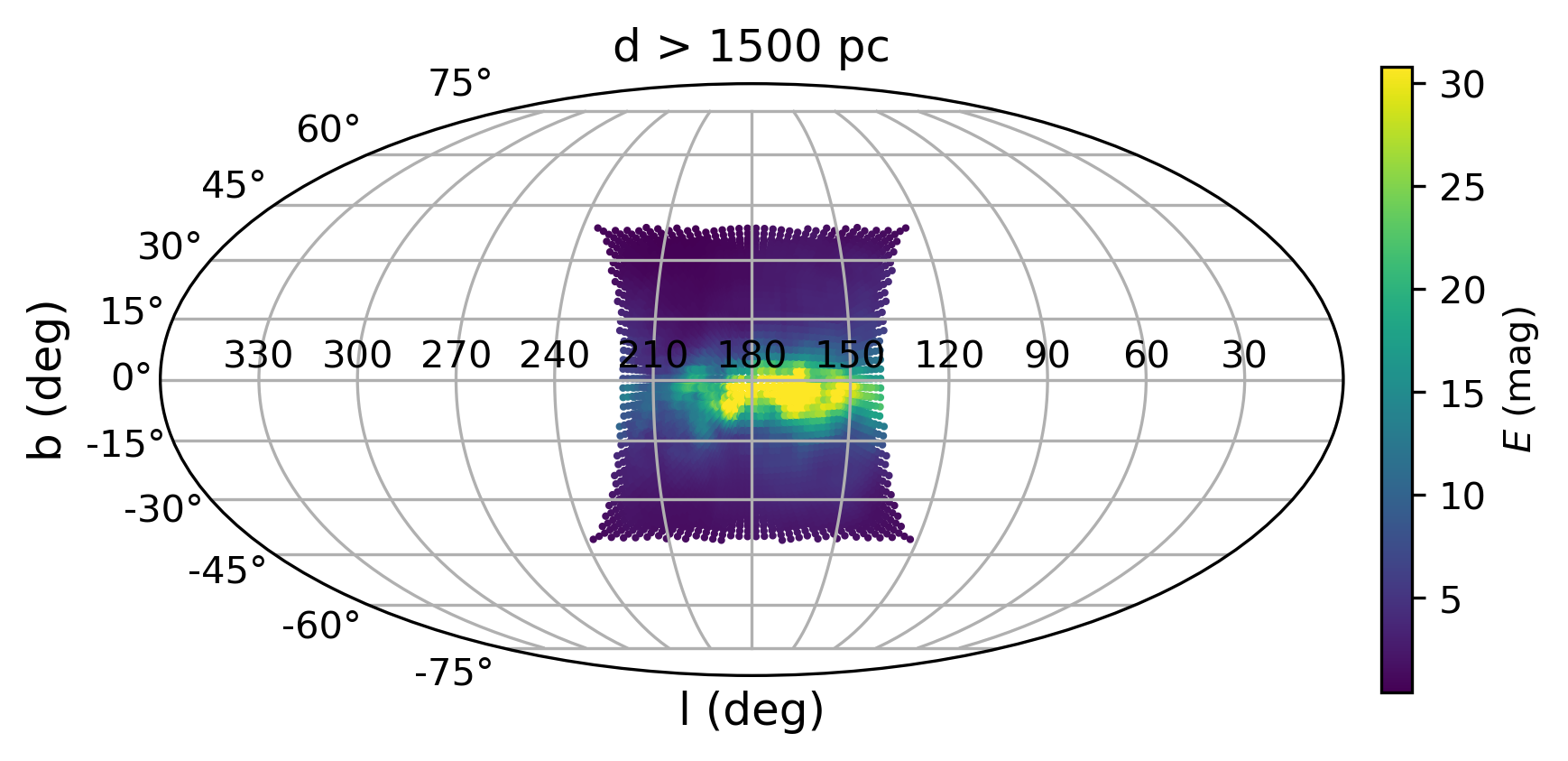}
\caption{Extinction distributions in the Galactic coordinates for the example simulation of the LR region. The upper and bottom panels show grids with distances smaller than 800\,pc and those with distances larger than 1500\,pc, respectively.  }
\label{extlbLR}
\end{figure}

\section{The mock catalogues}

This section presents the simulations utilized to train and evaluate the V-net. In the present study, the term ``extinction" is used to denote the cumulative value along the LOS, which may pertain to either extinction or reddening in different passbands (denoted as $E$). The dust density (represented by $\rho$) is expressed in mag\,pc$^{-1}$, a unit of column density that facilitates integration along the LOS.
  
Our study employs the Heliocentric Cartesian coordinates to depict spatial position. We take the Sun as the origin, with the $X$-axis pointing towards the Galactic centre, the $Y$-axis perpendicular to the $X$-axis in the Galactic plane, and the $Z$-axis perpendicular to the Galactic plane pointing towards the North Galactic Pole. We have chosen two test regions in the direction of the Galactic anti-centre for our study: a small region (referred to as `SR') defined by $X$ ranging from $-$192 to 0\,pc, $Y$ ranging from $-$96 to 96\,pc, and $Z$ ranging from $-$96 to 96\,pc, and a large region (referred to as `LR') defined by $X$ ranging from $-$1920 to 0\,pc, $Y$ ranging from $-$960 to 960\,pc, and $Z$ ranging from $-$960 to 960\,pc. The SR region has a resolution of 6\,pc, while the LR region has a resolution of 60\,pc. The 3D LOS extinction maps and dust density maps of both regions can be represented by 32 $\times$ 32 $\times$ 32 matrices, where each element in the matrices corresponds to the LOS extinction or dust density value at that specific spatial position.

\subsection{Simulate the dust density and extinction maps}

A substantial number of training samples is essential to train the V-net model. In this study, we will generate these samples by utilizing simulated dust density distributions and integrating them to derive LOS extinction distributions. Given the absence of physical models for dust distribution in the Milky Way, we employ simple mathematical models to simulate the distribution of dust within the scope of this paper. To simulate the dust distribution in the two test regions, we have adopted two different methods.

For the SR region, we generate the dust density distribution using the ``l\'{e}vy flight" method, which is similar to the technique described by \citet{Puspitarini2014}. The l\'{e}vy flight is a type of random walk that involves long-distance moves interspersed with short-distance steps, where the length and direction of each step are determined by a probability distribution known as the L\'{e}vy distribution. In this study, we employ a simulation methodology whereby ten starting points are initially selected. For each of these starting points, we conduct 10 l\'{e}vy flights, each consisting of 20,000 jumps. This process yields a cumulative total of 2 million particles. Subsequently, we constrain these particles within the specific 3D volume under investigation and proceed to calculate the particle density within small cubic volume bins of 6\,pc $\times$ 6\,pc $\times$ 6\,pc. In order to obtain the dust density map of the region, we multiply the particle density values by a constant factor carefully selected to ensure that most of the resulting matrix values do not exceed a random value between 0.01 and 0.05 mag\,pc$^{-1}$. The resulting 32 $\times$ 32 $\times$ 32 matrix is then adopted as the simulated dust density map of the region. 
An example is shown in Fig.~\ref{rhosim}. In the top panels of Fig.~\ref{rhosim} we show the dust density maps expressed in mag\,pc$^{-1}$. 

Using the simulated dust density map we obtained above, we proceed to create an LOS direction from the Sun to every position in the map. By integrating the dust column density over these LOS directions, we generate the LOS extinction map, which is also a 32 $\times$ 32 $\times$ 32 matrix.  The bottom panel of Fig.~\ref{rhosim} displays the resulting LOS extinction map of the example dust density simulation. Fig.~\ref{extlb} shows two-dimensional (2D) maps of extinction in Galactic coordinates for different grids located within the selected distance bins. The top panel of the Figure displays blue crosshatched patterns that signify areas with notably lower extinction values than adjacent pixels. These patterns correspond to pixels that represent shorter distances compared to their neighbours. The adoption of a Cartesian coordinate system results in a broad spread in the grid point distances within a confined area of the sky when mapped to Galactic coordinates. As a result, pixels representing closer distances show striking reddening contrasts when compared to those at greater distances, which gives rise to the prominent patterns observed.

For the LR region, we simulate the dust distribution by assuming a logarithmic density profile that follows a Gaussian Process as a function of position. Since the LR region covers a relatively large range in $Z$, we first determine the value of $Z$ that corresponds to the highest dust density and manually shift this $Z$ value to $Z$ = 0\,pc. Next, we convolve the generated dust density distribution using an exponential function with a scale height of 120\,pc. Finally, we scale the extinction density values by a constant factor to ensure that the majority of matrix values do not exceed a random value between 0.01 and 0.5 mag pc$^{-1}$. To obtain the LOS extinction maps, we integrate the derived extinction density maps in a manner similar to that of the SR region. An example is presented in Fig.\ref{rhoLR}, where the top panels depict the dust density maps and the bottom panels illustrate the resulting LOS extinction map. Fig.\ref{extlbLR} displays 2D maps of extinction in Galactic coordinates for various grids within the selected distance bins.

It should be noted that the dust density simulations utilized in the current work are purely mathematical constructs, devoid of any physical correspondence to the actual dust distribution of the Milky Way. Despite this limitation, the integral LOS extinction map is derived by integrating the aforementioned dust density map. Our objective, therefore, is to train the V-net to predict the dust density distribution from the integral LOS extinction map, thereby enabling it to learn this inverse mapping process. The model trained using simulated data should be able to accurately deduce the synthetic extinction density distribution from the measured integral LOS extinction map. Further elaboration on this topic is presented in Sect.~\ref{app}.

\subsection{Training and testing data sets}
 
We have employed the above-mentioned methods to generate 10,000 simulated maps of dust density ($\rho$) simultaneously for the SR and LR regions. Subsequently, each of these maps is integrated to derive LOS extinction ($E$) maps. For the SR samples, the values of $E$ span a range of 0 to about 1\,mag, while the range of $\rho$ extends only from 0 to about 0.05\,mag\,pc$^{-1}$. Conversely, for the LR samples, $E$ varies from 0 to about 50\,mag, and $\rho$ ranges from 0 to 0.5\,mag\,pc$^{-1}$. To align these values and ensure they are comparable, we renormalize $E$ using
\begin{equation}
\hat{E}=\ln(1+E),
\end{equation} 
and $\rho$ using
\begin{equation}
   \hat{\rho}=10\ln(1+\rho).
\end{equation} 
This process results in extinction values that are of a similar order of magnitude as the density values. 

Out of the 10,000 simulated maps, a subset of 1,000 datasets, or 10\,per\,cent of the total, are randomly selected as the testing dataset. The remaining 9,000 datasets are split into two subsets: 80\,per\,cent, or 7,200 datasets, are used for training the V-net, while the remaining 20\,per\,cent, or 1,800 datasets, are employed for validating the V-net.

\begin{figure*}
\centering
 \includegraphics[width=182mm]{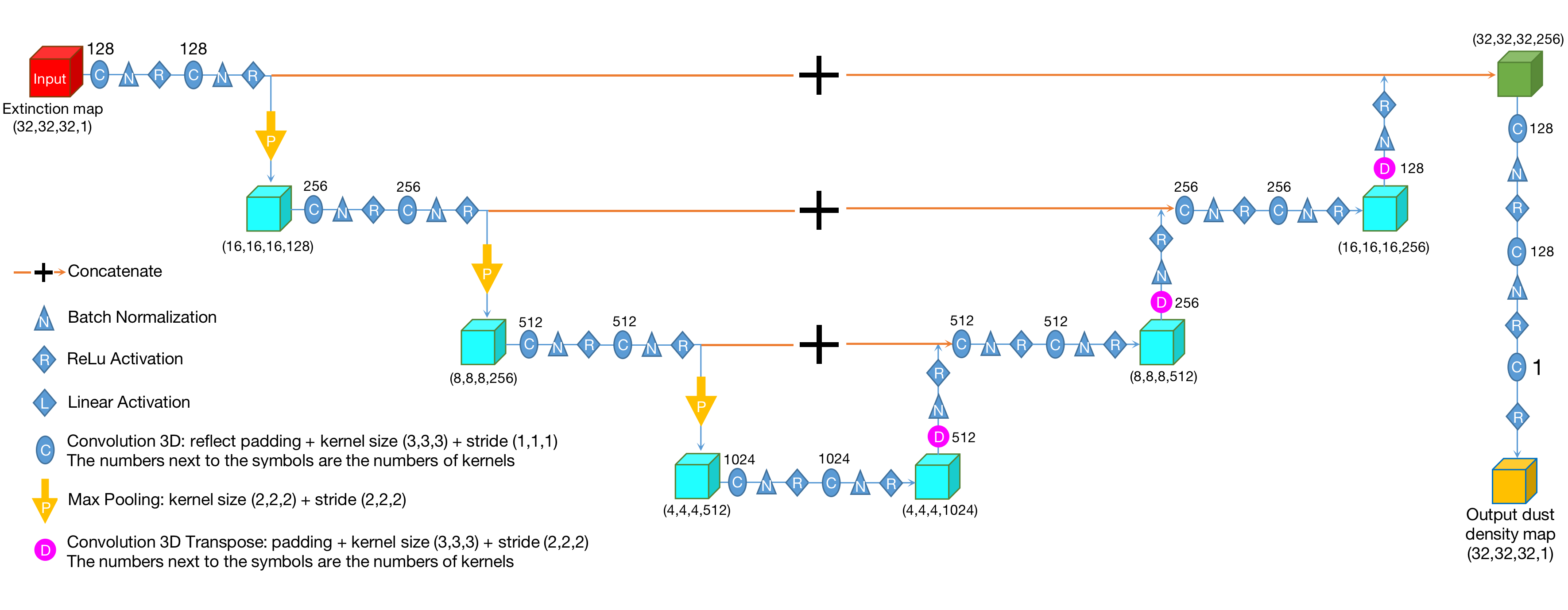}
 \caption{The V-net architecture used in this work.} 
 \label{fig1unet}
\end{figure*}

\begin{figure}
\centering
\includegraphics[width=0.45\textwidth]{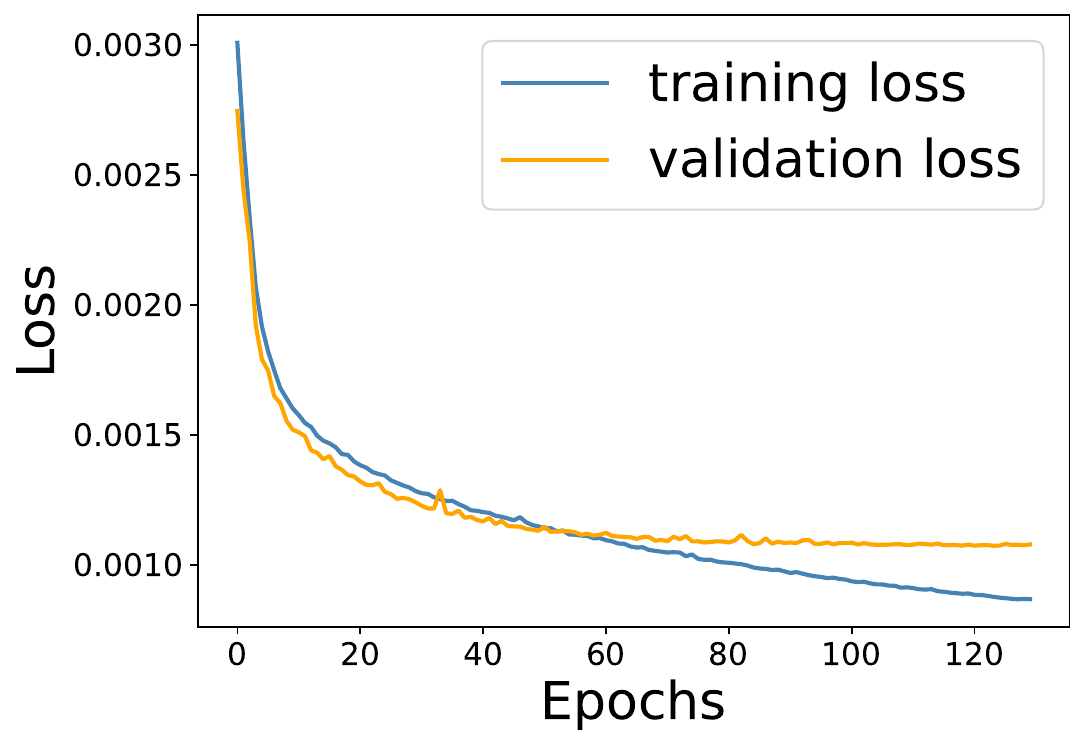}
\includegraphics[width=0.45\textwidth]{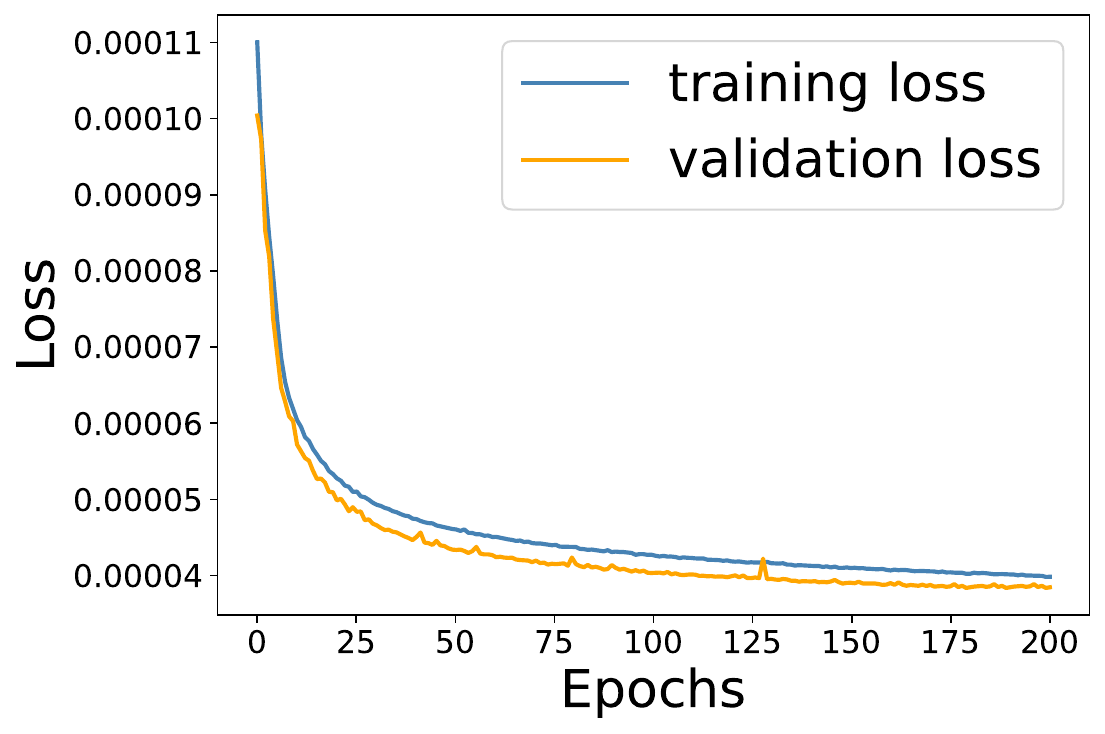}
\caption{The loss function $\mathcal{L}(\hat{\rho})$ as a function of the training epoch for the SR (upper panel) and LR (bottom panel) samples, respectively. The training loss and validation loss are represented by the blue and orange curves, respectively.}  
\label{lossfun}
\end{figure}

\begin{figure}
\centering
\includegraphics[width=0.45\textwidth]{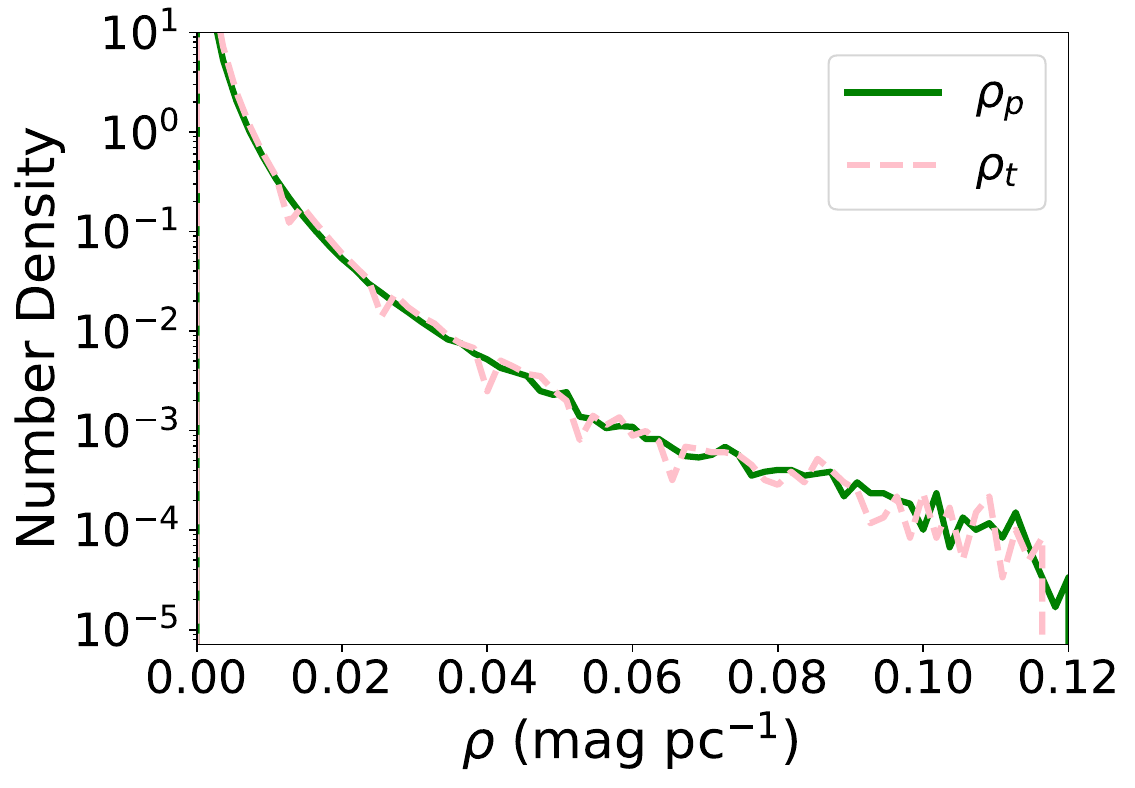}
\includegraphics[width=0.45\textwidth]{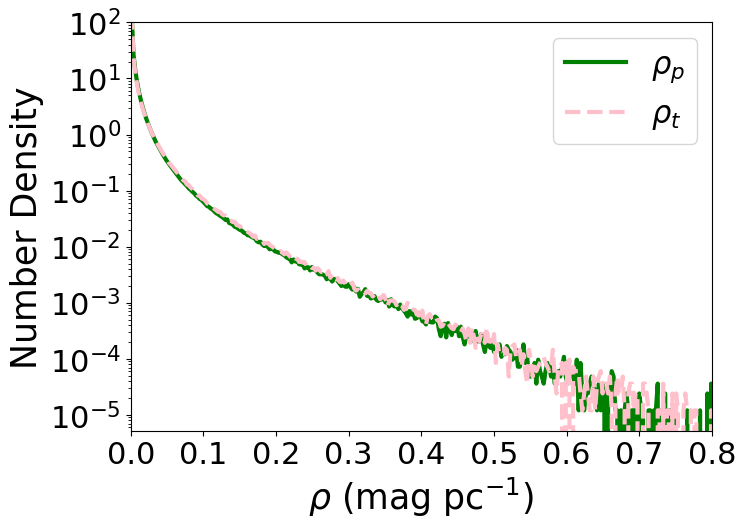}
\caption{The probability distributions of the predicted dust density $\rho_p$ (green curve) and the true density $\rho_t$ (pink curve) for the 1000 testing data sets in the SR (upper panel) and LR (bottom panel) samples, respectively.}  
\label{1dcomp}
\end{figure} 

\begin{figure}
\centering
\includegraphics[width=0.45\textwidth]{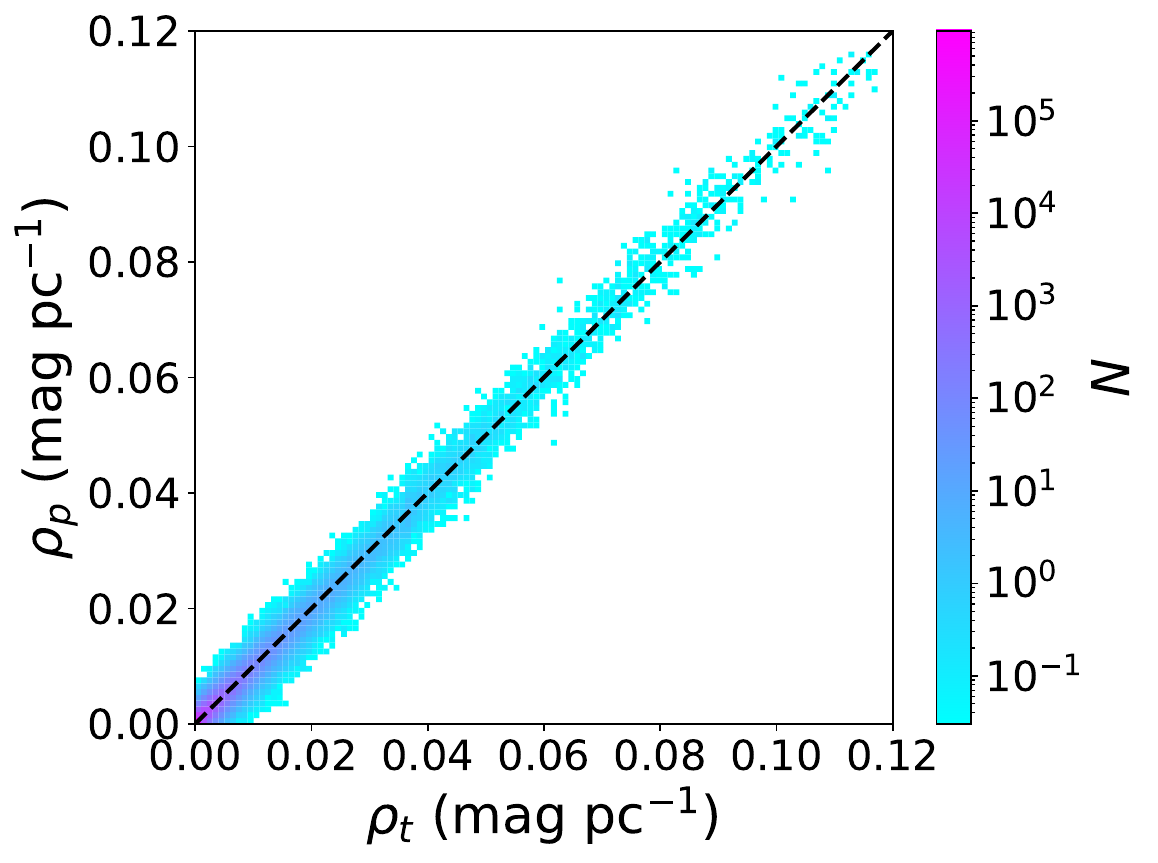}
\includegraphics[width=0.45\textwidth]{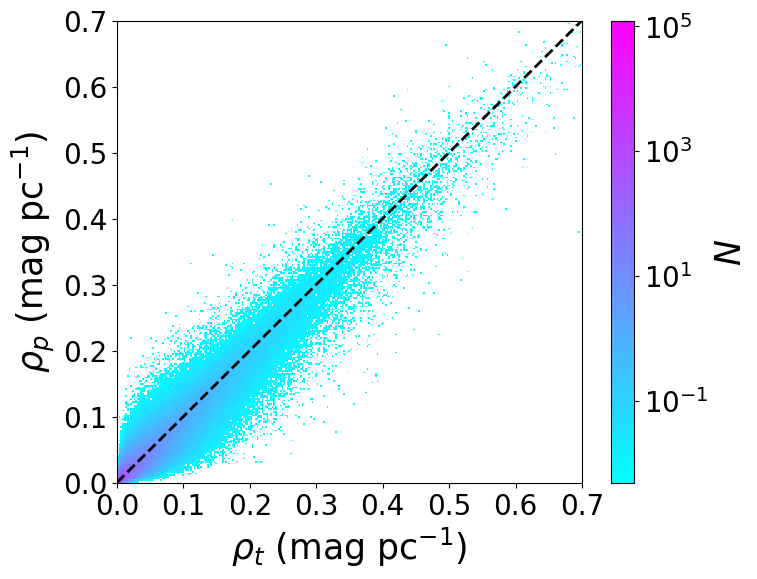}
\caption{Comparison between the predicted values of dust density $\rho_p$ and the true values $\rho_t$ for the 1000 testing data sets in the SR (upper panel) and LR (bottom panel) samples, respectively. The black dashed line is plotted to denote complete equality.}  
\label{2dcomp}
\end{figure} 

\begin{figure}
\centering
\includegraphics[width=0.45\textwidth]{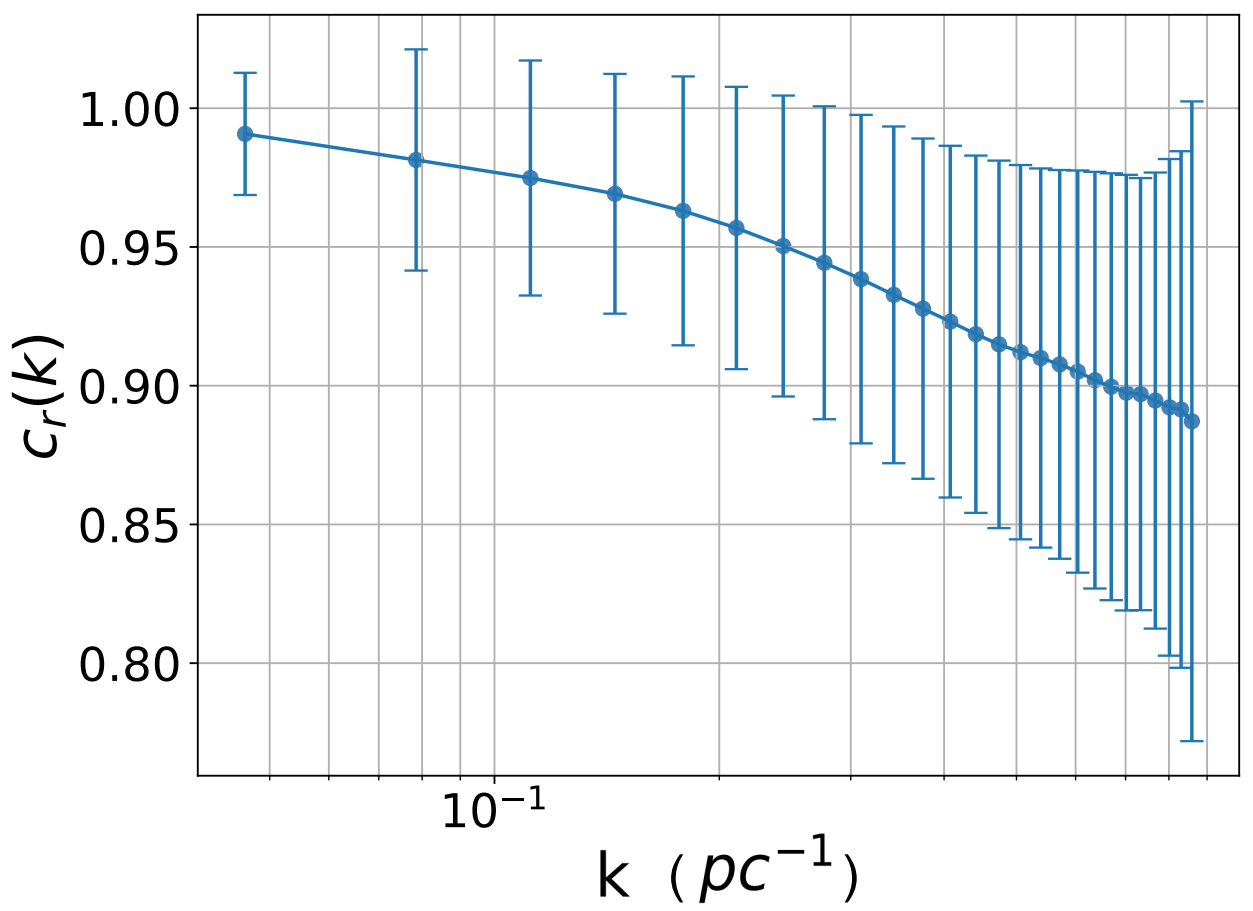}
\includegraphics[width=0.45\textwidth]{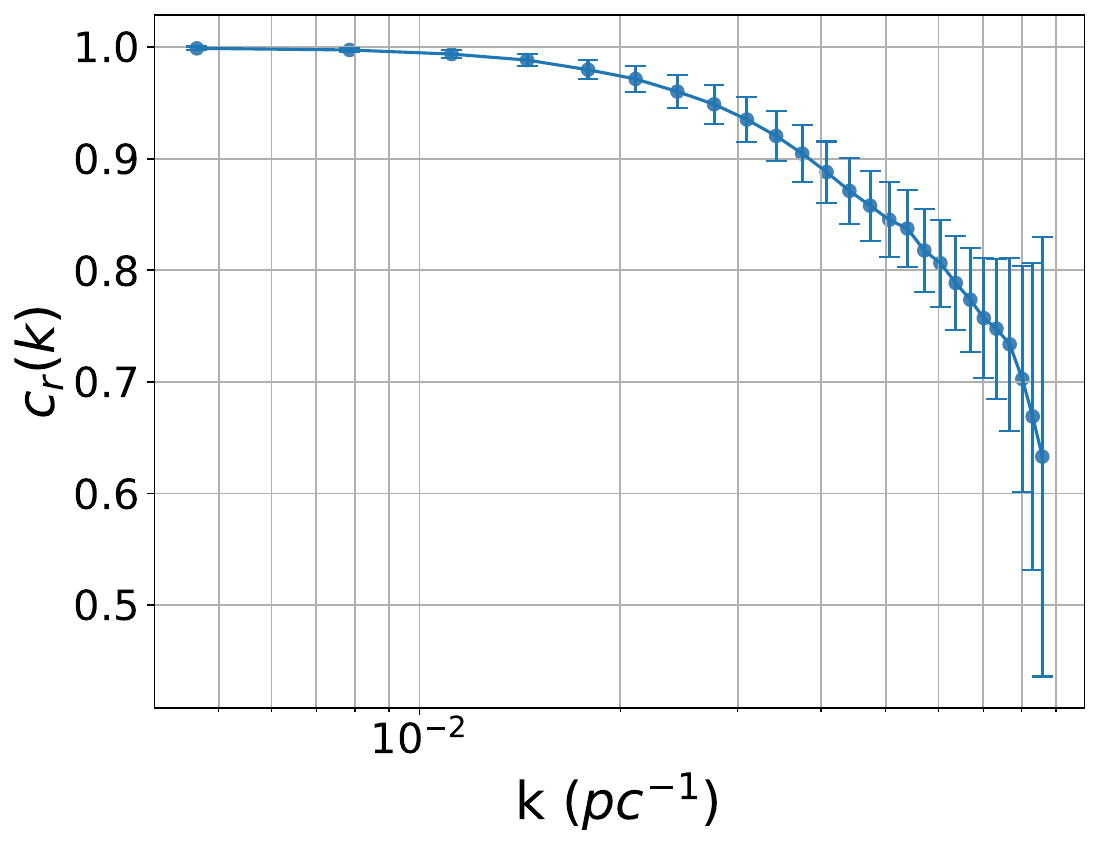}
\caption{The cross-correlation coefficient between the predicted dust density maps and the true density maps for the 1000 testing data sets in the SR (upper panel) and LR (bottom panel) samples, respectively.  }  
\label{crcomp}
\end{figure}

\begin{figure}
\centering
\includegraphics[width=0.45\textwidth]{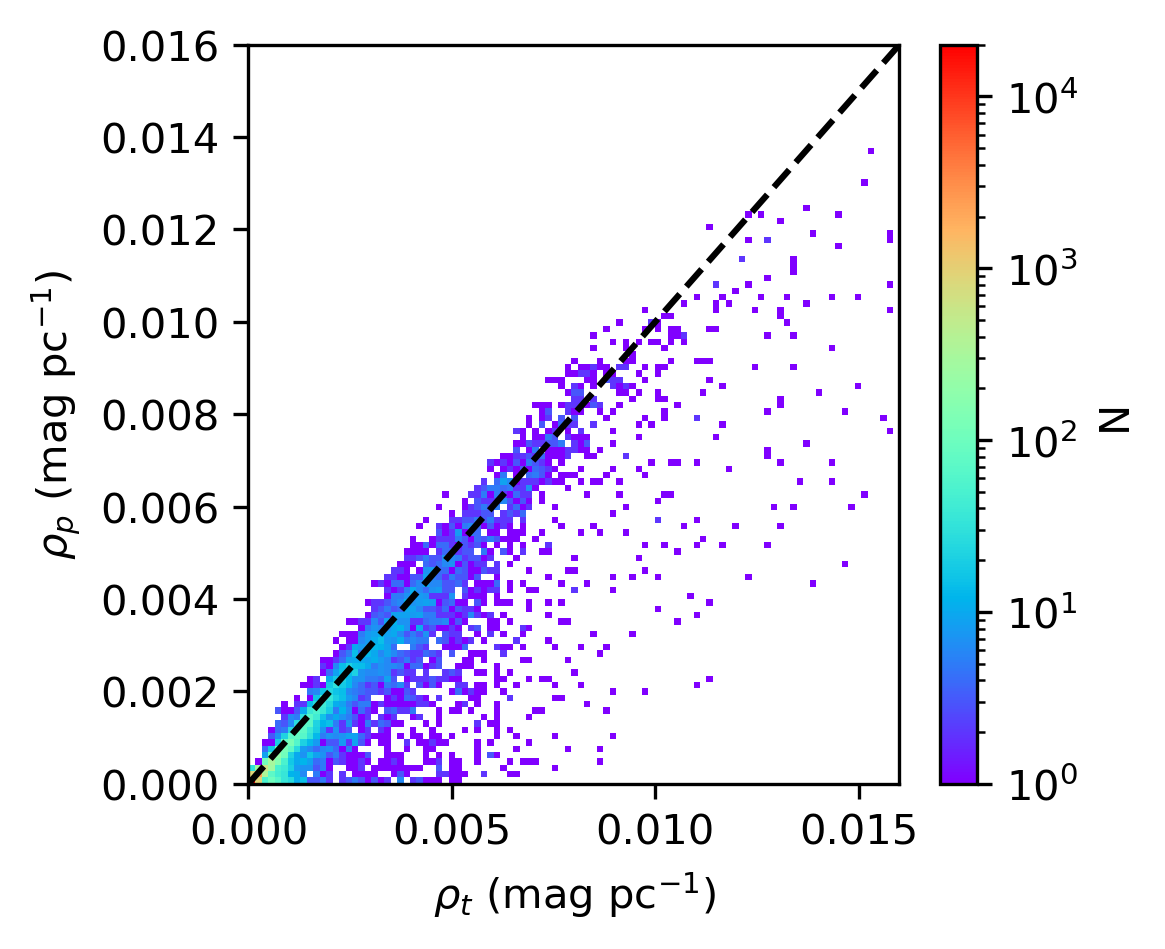}
\includegraphics[width=0.45\textwidth]{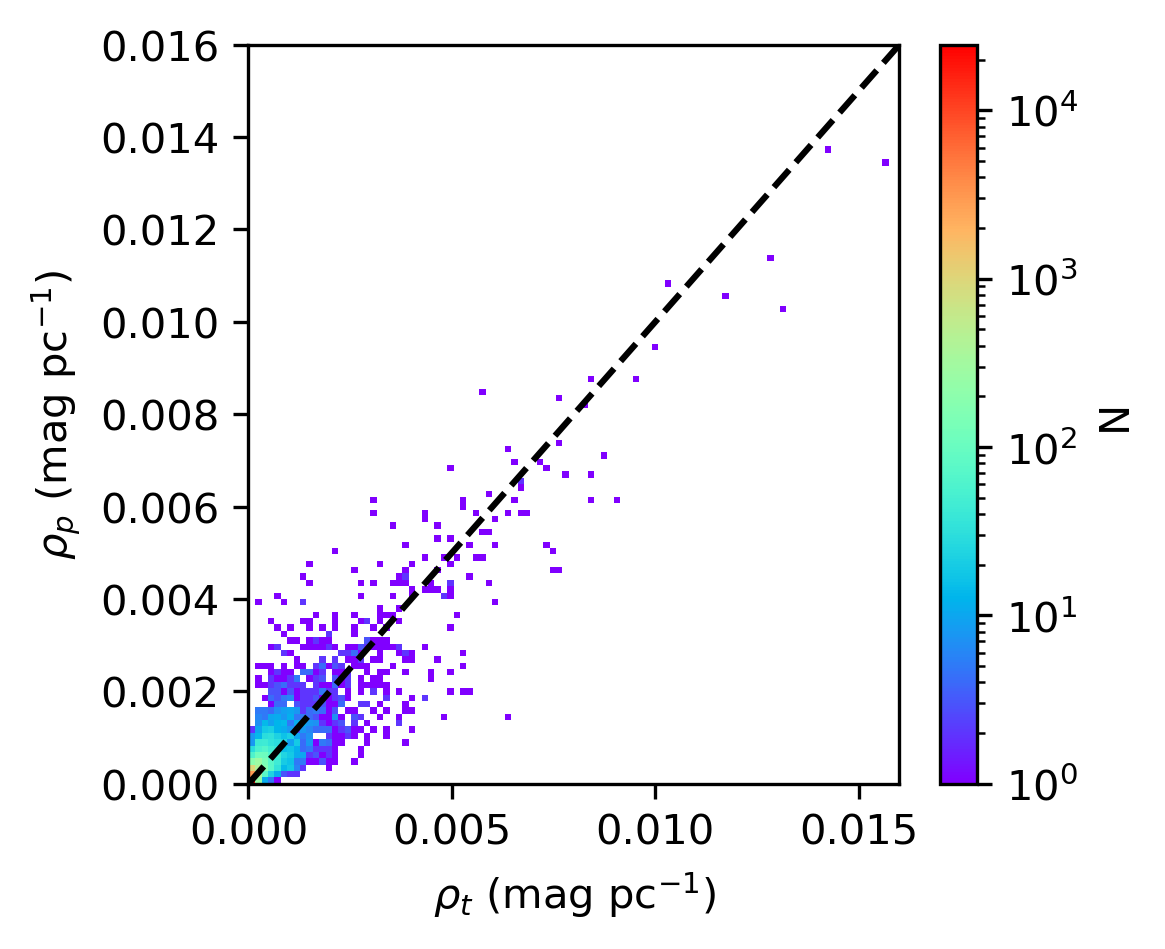}
\caption{Comparison between the predicted values of dust density $\rho_p$ and the true values $\rho_t$ for the \citet{Vergely2022} data set. The blue dashed line is plotted to denote complete equality. The upper and bottom panels demonstrate the comparisons of the SR and LR regions, respectively. }
\label{compv22}
\end{figure}

\begin{figure}
\centering
\includegraphics[width=0.45\textwidth]{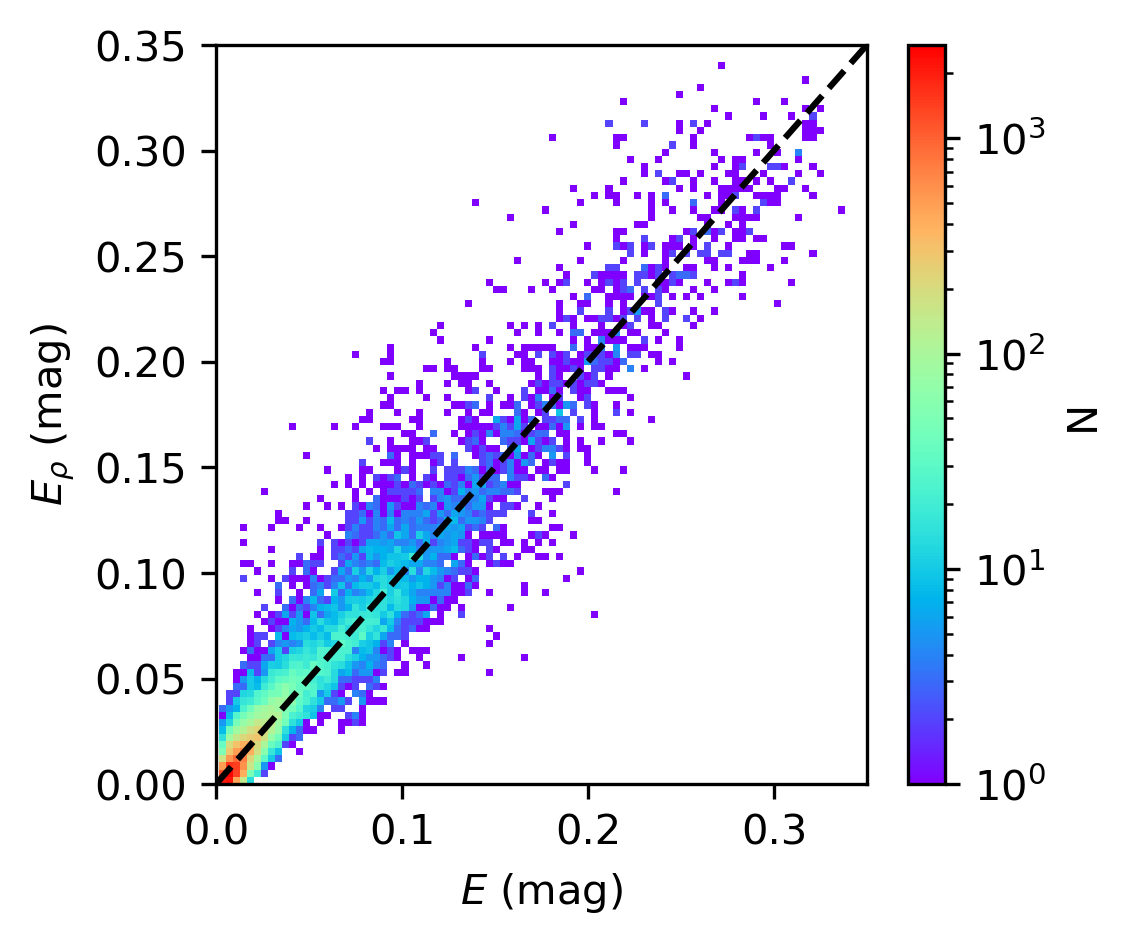}
\includegraphics[width=0.45\textwidth]{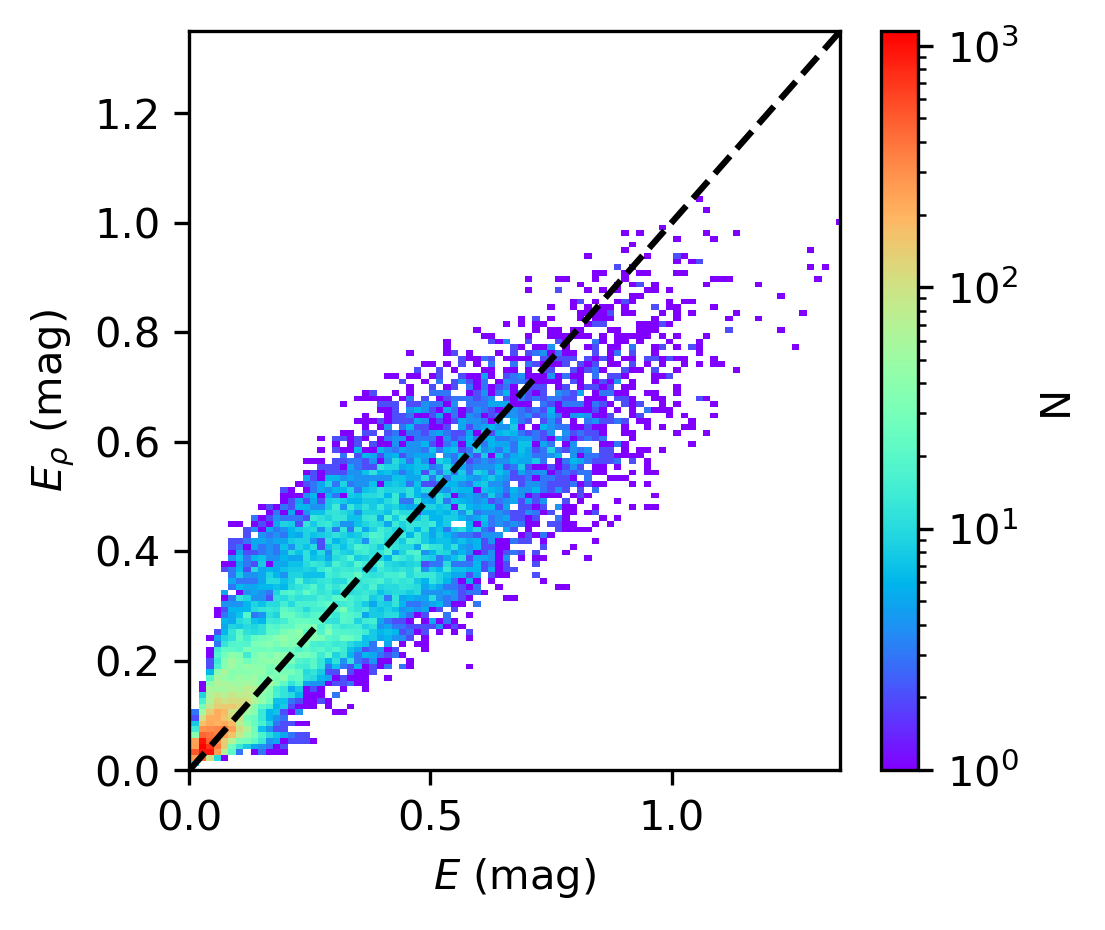}
\caption{Comparison between the extinction values $E$ obtained from the stellar sample of \citet{Zhang2023} and the LOS extinction values $E_{\rho}$ which are recovered from the generated 3D dust density maps. The blue dashed line is plotted to denote complete equality. The upper and bottom panels demonstrate the comparisons of the SR and LR regions, respectively.}  
\label{compv20002}
\end{figure}

\section{Methodology} 

\subsection{The architecture of V-net}

V-Net is a deep-learning architecture that is commonly employed for image segmentation \citep{Milletari2016}. V-net is specifically designed for 3D volumes, which is an extension of U-Net. Initially, U-Net was developed by \citet{Unet2015} for processing 2D maps. The U-Net architecture consists of two stages: an encoding stage and a decoding stage. During the encoding stage, the input map is processed by increasing the channel number and decreasing the map size of each channel, allowing the machine to extract relevant features. During the decoding stage, the machine reconstructs a new map by decreasing the channel number and increasing the map size of each channel. To prevent the loss of small-scale features, the outputs from the decoding stage are concatenated with the inputs from the encoding stage in U-Net. 

Several revised versions of V-Net have been proposed for reconstructing the cosmological density field, including those put forth by \cite{Hong2021}, \cite{Wu2021}, and \cite{Qin2023}. In this study, we employ the V-Net architecture presented by \citet{Qin2023} to construct a dust density map from an extinction map. The V-Net architecture is illustrated in Fig.~\ref{fig1unet}. The input to the V-Net consists of LOS extinction maps, each possessing a spatial dimension of $32^3$ and a single channel representing a scalar field, resulting in a shape of $(32,~32,~32,~1)$. The output of the V-Net is the dust density map, which also has a shape of $(32,~32,~32,~1)$.

A detailed description of the V-net architecture used in our study is listed below:
\begin{enumerate}
\item The input extinction maps are first passed through a convolution layer, followed by batch normalization \citep{Ioffe2015} and a rectified linear unit (ReLU, \cite{Glorot2011}) activation function. The convolution layer comprises 128 kernels of $3^3$ shape, with a scanning stride of 1. The output is then passed through the same set of operations once again, resulting in feature maps with a shape of $(32,~32,~32,~128)$. \\
\item The feature maps are then subjected to a max pooling layer to reduce their dimension. The shape of the max pooling layer is $2^3$, and the stride is 2. The output of this step has a shape of $(16,~16,~16,~128)$. \\   
\item The outputs of step~(ii) are then passed through the same set of operations as steps~(i) and (ii), three more times to further extract features. However, the convolution layer comprises 256, 512, and 1024 kernels in each operation, respectively. The last operation is not followed by any max pooling layer, as shown in Fig.~\ref{fig1unet}. The output of this step has a shape of $(4,~4,~4,~1024)$.   \\
\item The outputs of step~(iii) are then fed into a transpose convolution layer, followed by batch normalization and a ReLU activation function. The transpose convolution layer comprises 512 kernels of $3^3$ shape, with a scanning stride of 2. The output of this step has a shape of $(8,~8,~8,~512)$. \\
\item The outputs of step~(iv) are concatenated with the feature fields of the equivalent encoding stage (with a size of $(8,~8,~8,~512)$). The concatenated outputs are then passed through two convolution layers, which are the same as step~(i). The number of convolution kernels is 512.\\
\item The outputs of step~(v) are then subjected to the same set of operations as steps~(iv) and (v), two more times. However, the number of kernels for the transpose convolution used in each operation is changed to 256 and 128, respectively. No convolution layer is applied in the last operation. The output of this step has a shape of $(32,~32,~32,~256)$. \\
\item The outputs of step~(vi) are fed into two convolution layers, each with 128 kernels, using the same configuration as in step~(i). The resulting outputs are then passed through a final convolution layer with one kernel and shape $3^3$, using a stride of 1, and followed by a ReLU activation function \citep{Qin2023,Mao2021,Pan2020,Wu2021,Hong2021,Ganeshaiah2022}. In our simulations, the density and LOS extinction values are inherently greater than zero. Consequently, the V-net should only be activated when the predicted values exceed zero. To achieve this, we employ the ReLU activation function, which allows for smooth handling of density values above zero while remaining inactive for values below zero.
\end{enumerate}
Finally, we obtain the 3D dust density maps constructed from the 3D extinction maps. To minimize the loss of information around the boundaries and keep the outputs the same dimension as the inputs, we apply reflect padding in the convolution layers. The optimization method used for the training of V-net is the Adam algorithm \citep{Kingma2014}. We set the mini-batch size to 32 during training and stop training when the validation loss tends to be flat. However, to avoid overfitting, we save the trained V-net model corresponding to the minimum validation loss instead of the last epoch.

\subsection{Training the V-net}

In this study, the mean squared error (MSE) is adopted as the chosen loss function, defined as,
\begin{equation} \label{lossden}
\mathcal{L}(\hat{\rho})=\frac{1}{N_{\rm pix}} {\sum^{N_{\rm pix}}_{i=1} \left(\hat{\rho}_{{\rm p},i}-\hat{\rho}_{{\rm t},i}  \right)^2} ,
\end{equation}
where $\hat{\rho}_{{\rm p},i}$ refers to the predicted density value obtained by the V-net model and $\hat{\rho}_{{\rm t},i}$ represents the true density value. Fig.~\ref{lossfun} illustrates the evolution of the loss function as a function of the training epoch for both the SR and LR samples. For the SR region, the validation loss (orange-coloured curve) becomes stagnant beyond the 60th epoch, leading us to terminate the training at Epoch 130. For the LR region, we stop at Epoch 200. The trained V-net model, which corresponds to the minimum validation loss rather than the last epoch, is saved as the optimal model. This model is then applied to the 1000 testing extinction data sets to generate the predicted dust density maps. Using the Cloud Tensor Processing Units (TPUs) with around 4.5 GB of memory on Google Colaboratory, the V-net training process usually takes around 3.5 hours to complete\footnote{In our current investigation, we are using a small cubic (32$^3$) for testing. However, as we progress, our intention is to employ larger cubes, such as 64$^3$ and even 128$^3$. Consequently, the estimated training duration may increase by a few hours, depending on the size of the convolutional kernel and the number of generations we train.}. Once we achieve the optimal model, we can generate the resulting extinction density maps within seconds.

\subsection{Testing the V-net}

Fig.~\ref{1dcomp} presents the probability distributions of the predicted dust density $\rho_p$ and the true density $\rho_t$ of the 1000 testing data sets in both the SR and LR samples. The distributions of $\rho_p$ and $\rho_t$ exhibit close agreement, indicating that the V-net predicts the dust density values with high accuracy.      
Fig.~\ref{2dcomp} displays a comparison between the predicted dust density values $\rho_p$ and their corresponding true values $\rho_t$. Overall, the predicted density values are in good agreement with the true values. There is no systematic difference between the values predicted by our trained V-net model $\rho_p$ and the true values of the simulated data $\rho_t$, and the difference between them exhibits a small dispersion. In the SR region, the median value of the difference between the predicted and true values is about 0, with a dispersion of only 0.0003. Similarly, in the LR region, the median value is about 0, and the dispersion is merely 0.0004.

The cross-correlation coefficient between predicted and true density values can be determined by analyzing their power spectrum. This can be expressed as,
\begin{equation}
c_r(k)\equiv\frac{P_{\rm tp}(k)}{\sqrt{P_{\rm tt}(k)P_{\rm pp}(k)}}, 
\end{equation}
where $P_{\rm pp}(k)$ and $P_{\rm tt}(k)$ represent the auto power spectrum \citep{Howlett2019,Qin2019} of predicted and true dust density maps, respectively. $P_{\rm tp}(k)$ denotes the cross-power spectrum of predicted and true dust density maps, $k$ is the wave number defined as $2\pi/r$, and $r$ is the pair separations of the points in the maps measured in units of pc. Our results, presented in Fig.~\ref{crcomp}, demonstrate a strong correlation between the predicted and true dust density maps. In the case of the SR samples, there is a minimum correlation of 90\% across the range of $k\in [0.046,~0.893]$\,pc$^{-1}$ or $r\in [7,~136.6]$\,pc. As for the LR samples, all correlation values are larger than 60\%. The correlation values are larger than 90\% for $r > 167.6$\,pc. These findings suggest that our V-net model accurately predicts the dust density maps and successfully reconstructs the structural characteristics of the dust distributions.

\section{Application of the model}
\label{app}

In this study, we present a V-net model that has been trained using mathematically constructed dummy data. In this section, we evaluate the performance of the trained model in constructing 3D maps of dust density by applying it to the empirical data from the literature.  
 
\subsection{Data from previous 3D dust maps}

We first examine a realistic data set by directly utilizing the 3D dust maps from \citet{Vergely2022} as the dust density distribution in the regions of interest. \citet{Vergely2022} employed the hierarchical inversion technique to a merged catalogue and obtained 3D extinction density maps of the Milky Way. The resolution of their maps for the specific region we are analyzing is 10\,pc. For the SR region, we interpolate the extinction density maps onto our position grids using the 3D nearest method. For the LR region, we smooth the data to a resolution of 60\,pc and then interpolate the smoothed maps onto our position grids, also using the 3D nearest method. The resulting 3D dust density maps from \citet{Vergely2022} for both the SR and LR regions are shown in Appendix~A (see upper panels of Figures~\ref{v22} and \ref{v22large}). We then construct the LOS direction for each position and integrate the extinction density along the LOS to produce the 3D extinction maps of the SR and LR regions.

We employ the trained V-net model to derive dust density maps from the LOS extinction maps. The resulting dust density maps for the SR and LR regions are presented in the middle panels of Fig.~\ref{v22} and \ref{v22large}, respectively. The comparisons between the values of the derived dust density maps and those of the actual ones are depicted in Fig.~\ref{compv22}. The density values predicted by the V-net model exhibit considerable consistency with the true values for both the SR and LR regions. In the SR region, the dispersion of the difference values between the predicted density values and the actual values is merely 0.0005. About 2\% of the pixels in this region exhibit predicted values that are 3$\sigma$ smaller than the true values. In the LR region, the dispersion of the differences is about 0.0002, and around 2\% of the pixels have predicted values that fall outside the 3$\sigma$ range. Notably, there is a greater density of points located below the blue dashed line. This bias suggests that the V-net tends to underestimate the true values. The underlying causes of this phenomenon are yet to be elucidated and may warrant further refinement of our model. Nonetheless, it is noteworthy that the proportion of points exhibiting substantial deviations is minimal relative to the entirety of the dataset, affirming the satisfactory overall performance of our model. In the bottom panels of Fig.~\ref{v22} and \ref{v22large}, we present the residuals obtained by comparing our results for the SR and LR regions with the literature maps. Regarding the SR region, the resulting extinction density maps and residual maps exhibit the presence of straight grid lines. It is worth noting that this phenomenon is a common challenge encountered in AI reconstruction methods, especially near the cube boundaries and in denser regions, as observed in the V-net predictions. In our future work, we plan to address this by generating larger cubes with dimensions of, like, 128$^3$. This will allow us to directly eliminate pixels that are in close proximity to the boundaries.

\subsection{Data from the previous stellar catalogue}

In this section, our trained model is employed to generate a 3D extinction density map of the target region by utilizing a sample of stars with distance and extinction measurements. To do so, we have used the extinction and distance estimates of individual stars from \citet{Zhang2023}. They derived stellar atmospheric parameters (\teff, \logg, and \feh), distances, and extinction values for 220 million stars from the Gaia XP spectra. We have selected stars located within our region of interest to construct a 3D extinction map. To each grid in both the SR and LR regions, we select the three (for the SR region) and 30 (for the LR region) nearest stars and compute their weighted average extinction value to represent the extinction at that particular position. The weights assigned to the individual stars are determined by a combination of their distances from the grid and the associated extinction errors, given by the equation 
\begin{equation}
w=(1 - 0.75(d/d_{\rm max})^2)/(\sigma_E^2 + 10^{-4}),
\end{equation}
where $d$ is the distance of a given star to the grid, $d_{\rm max}$ the maximum value of $d$ among the three stars, and $\sigma_E$ the extinction uncertainty. The resulted LOS extinction maps are illustrated in the upper panels of Fig.~\ref{g22} and \ref{g22large}.

We utilize the trained V-net model to derive dust density maps from the resultant LOS extinction map. The generated dust density maps are presented in the bottom panels of Fig.~\ref{g22} and \ref{g22large}. Notably, our maps exhibit more detailed structures compared to those obtained by \citet{Vergely2022}, likely due to the higher resolution of our maps for the SR region and the use of Gaussian kernel priors in their analysis. Our approach does not rely on any prior, enabling us to obtain a more realistic representation of the dust structure. To validate the accuracy of our results, we compare the observed extinction values ($E$) with the values integrated from the generated dust density maps ($E_{\rho}$) in Fig.~\ref{compv20002}. The excellent agreement between $E$ and $E_{\rho}$ demonstrates the robustness of our 3D dust density maps. 

We have conducted a comparative analysis of the dust density results obtained in our study with those reported by \citet{Zhang2023}. Specifically, \citet{Zhang2023} present the distribution of their preliminary 3D dust density maps for the Galactic plane within $|Z| < 400$\,pc in their Fig.~24, which we have adjusted to match the spatial extent of our LR region. By summing the dust densities within pixels of our 3D dust density maps for the same volume and plotting the results in the $X$-$Y$ plane, we facilitate a direct comparison with \citet{Zhang2023}'s findings, as depicted in Fig.~\ref{g22descomp}. Our analysis reveals a dust structure that is largely consistent with that of \citet{Zhang2023}, albeit with a notable difference: their maps exhibit a pronounced elongation along the line of sight, an effect not observed in our maps. 

The V-net model does not provide uncertainty outputs. In order to determine the errors associated with the derived extinction density maps, we have utilized a Monte Carlo method. We generate 300 extinction maps of the LR region from the \citet{Zhang2023} catalogue, taking into account their extinction errors. Subsequently, we apply the trained V-net model to these 300 maps to obtain extinction density maps. The dispersion values of the resulting 300 extinction density maps are then adopted as the errors. Figure~\ref{g22errorlarge} displays the resulting density error distributions.

\section{Conclusions}

In this study, we propose a novel machine-learning technique, known as the V-net, for creating 3D maps of dust density from LOS extinction maps. To train and test the V-net, we simulate dust density and extinction maps within two regions of interest. Furthermore, we apply the V-net model to data from the literature. Our results demonstrate that the V-net is capable of producing 3D dust extinction maps with high precision.

With the continuing progress in conducting large-scale photometric, spectroscopic, astrometric, and time-domain surveys of the Milky Way, it has become possible to acquire accurate extinction and distance values of large samples of stars. Consequently, this has enabled the construction of cutting-edge 3D maps that depict the distribution of dust density in the Milky Way. Moving forward, we plan to apply our method to deliver a comprehensive view of both the large- and small-scale (fine) structures of the Galactic dust. In our current study, we have opted for a modestly sized test cube (32$^3$), which unfortunately restricts us from simultaneously achieving an extensive map scale of over 2\,kpc and a fine spatial resolution under 5\,pc. Looking ahead, we aim to generate 3D extinction maps that offer both a large scale and high resolution. To accomplish this, we will utilize larger data cubes that will necessitate greater computational memory. On our usual personal computers, we are able to process 128$^3$ cubes with 24\,GB of memory, enabling us to reach a spatial resolution greater than 4\,pc for an extinction map that spans 500\,pc. In forthcoming studies, we plan to divide our target area of the sky into sectors measuring 500$^3$\,pc$^3$ each and train the V-net models to these subdivisions. This approach will allow us to accurately map the 3D dust density distribution across the Milky Way on both macro and micro scales. This will enable us to gain insights into the evolution of molecular clouds, the processes involved in star formation, and other complex physical phenomena that have contributed to the evolution of our Galaxy.

\section*{Acknowledgements}
We want to thank the referee for his/her fruitful comments. We extend our appreciation to Mr. Xiangyu Zhang and his collaborators for granting permission and assisting with the creation of Fig.~A5. This work is partially supported by the National Key R\&D Program of China No. 2019YFA0405500, National Natural Science Foundation of China 12173034, 11833006 and 12322304, and Yunnan University grant No.~C619300A034. We acknowledge the science research grants from the China Manned Space Project with NO.\,CMS-CSST-2021-A09, CMS-CSST-2021-A08 and CMS-CSST-2021-B03. 

\section*{Data availability}

The data underlying this article is available in the manuscript.

\bibliographystyle{mnras}
\bibliography{extin3d}

\begin{thebibliography}{}
\makeatletter
\relax
\def\mn@urlcharsother{\let\do\@makeother \do\$\do\&\do\#\do\^\do\_\do\%\do\~}
\def\mn@doi{\begingroup\mn@urlcharsother \@ifnextchar [ {\mn@doi@}
  {\mn@doi@[]}}
\def\mn@doi@[#1]#2{\def\@tempa{#1}\ifx\@tempa\@empty \href
  {http://dx.doi.org/#2} {doi:#2}\else \href {http://dx.doi.org/#2} {#1}\fi
  \endgroup}
\def\mn@eprint#1#2{\mn@eprint@#1:#2::\@nil}
\def\mn@eprint@arXiv#1{\href {http://arxiv.org/abs/#1} {{\tt arXiv:#1}}}
\def\mn@eprint@dblp#1{\href {http://dblp.uni-trier.de/rec/bibtex/#1.xml}
  {dblp:#1}}
\def\mn@eprint@#1:#2:#3:#4\@nil{\def\@tempa {#1}\def\@tempb {#2}\def\@tempc
  {#3}\ifx \@tempc \@empty \let \@tempc \@tempb \let \@tempb \@tempa \fi \ifx
  \@tempb \@empty \def\@tempb {arXiv}\fi \@ifundefined
  {mn@eprint@\@tempb}{\@tempb:\@tempc}{\expandafter \expandafter \csname
  mn@eprint@\@tempb\endcsname \expandafter{\@tempc}}}

\bibitem[\protect\citeauthoryear{{Chen}, {Schultheis}, {Jiang}, {Gonzalez},
  {Robin}, {Rejkuba}  \& {Minniti}}{{Chen} et~al.}{2013}]{Chen2013}
{Chen} B.~Q.,  {Schultheis} M.,  {Jiang} B.~W.,  {Gonzalez} O.~A.,  {Robin}
  A.~C.,  {Rejkuba} M.,   {Minniti} D.,  2013, \mn@doi [\aap]
  {10.1051/0004-6361/201219682}, \href
  {https://ui.adsabs.harvard.edu/abs/2013A&A...550A..42C} {550, A42}

\bibitem[\protect\citeauthoryear{Chen et~al.,}{Chen et~al.}{2014}]{Chen2014}
Chen B.-Q.,  et~al., 2014, \mn@doi [Monthly Notices of the Royal Astronomical
  Society] {10.1093/mnras/stu1192}, 443, 1192

\bibitem[\protect\citeauthoryear{{Chen}, {Liu}, {Yuan}, {Huang}  \&
  {Xiang}}{{Chen} et~al.}{2015}]{Chen2015}
{Chen} B.~Q.,  {Liu} X.~W.,  {Yuan} H.~B.,  {Huang} Y.,   {Xiang} M.~S.,  2015,
  \mn@doi [\mnras] {10.1093/mnras/stv103}, \href
  {https://ui.adsabs.harvard.edu/abs/2015MNRAS.448.2187C} {448, 2187}

\bibitem[\protect\citeauthoryear{{Chen} et~al.,}{{Chen}
  et~al.}{2019}]{Chen2019}
{Chen} B.~Q.,  et~al., 2019, \mn@doi [\mnras] {10.1093/mnras/sty3341}, \href
  {https://ui.adsabs.harvard.edu/abs/2019MNRAS.483.4277C} {483, 4277}

\bibitem[\protect\citeauthoryear{{Chen} et~al.,}{{Chen}
  et~al.}{2020}]{Chen2020}
{Chen} B.~Q.,  et~al., 2020, \mn@doi [\mnras] {10.1093/mnras/staa235}, \href
  {https://ui.adsabs.harvard.edu/abs/2020MNRAS.493..351C} {493, 351}

\bibitem[\protect\citeauthoryear{{Cornu}, {Montillaud}, {Marshall}, {Robin}  \&
  {Cambr{\'e}sy}}{{Cornu} et~al.}{2022}]{Cornu2022}
{Cornu} D.,  {Montillaud} J.,  {Marshall} D.~J.,  {Robin} A.~C.,
  {Cambr{\'e}sy} L.,  2022, \mn@doi [arXiv e-prints]
  {10.48550/arXiv.2201.05571}, \href
  {https://ui.adsabs.harvard.edu/abs/2022arXiv220105571C} {p. arXiv:2201.05571}

\bibitem[\protect\citeauthoryear{{Edenhofer}, {Zucker}, {Frank}, {Saydjari},
  {Speagle}, {Finkbeiner}  \& {En{\ss}lin}}{{Edenhofer}
  et~al.}{2023}]{Edenhofer2023}
{Edenhofer} G.,  {Zucker} C.,  {Frank} P.,  {Saydjari} A.~K.,  {Speagle} J.~S.,
   {Finkbeiner} D.,   {En{\ss}lin} T.,  2023, \mn@doi [arXiv e-prints]
  {10.48550/arXiv.2308.01295}, \href
  {https://ui.adsabs.harvard.edu/abs/2023arXiv230801295E} {p. arXiv:2308.01295}

\bibitem[\protect\citeauthoryear{{Ganeshaiah Veena}, {Lilow}  \&
  {Nusser}}{{Ganeshaiah Veena} et~al.}{2022}]{Ganeshaiah2022}
{Ganeshaiah Veena} P.,  {Lilow} R.,   {Nusser} A.,  2022, arXiv e-prints, \href
  {https://ui.adsabs.harvard.edu/abs/2022arXiv221206439G} {p. arXiv:2212.06439}

\bibitem[\protect\citeauthoryear{{Ganeshaiah Veena}, {Lilow}  \&
  {Nusser}}{{Ganeshaiah Veena} et~al.}{2023}]{Veena2023}
{Ganeshaiah Veena} P.,  {Lilow} R.,   {Nusser} A.,  2023, \mn@doi [\mnras]
  {10.1093/mnras/stad1222}, \href
  {https://ui.adsabs.harvard.edu/abs/2023MNRAS.522.5291G} {522, 5291}

\bibitem[\protect\citeauthoryear{Glorot, Bordes  \& Bengio}{Glorot
  et~al.}{2011}]{Glorot2011}
Glorot X.,  Bordes A.,   Bengio Y.,  2011, in ICML. pp 513--520, \url
  {https://icml.cc/2011/papers/342_icmlpaper.pdf}

\bibitem[\protect\citeauthoryear{{Green} et~al.,}{{Green}
  et~al.}{2015}]{Green2015}
{Green} G.~M.,  et~al., 2015, \mn@doi [\apj] {10.1088/0004-637X/810/1/25},
  \href {https://ui.adsabs.harvard.edu/abs/2015ApJ...810...25G} {810, 25}

\bibitem[\protect\citeauthoryear{{Green}, {Schlafly}, {Zucker}, {Speagle}  \&
  {Finkbeiner}}{{Green} et~al.}{2019}]{Green2019}
{Green} G.~M.,  {Schlafly} E.,  {Zucker} C.,  {Speagle} J.~S.,   {Finkbeiner}
  D.,  2019, \mn@doi [\apj] {10.3847/1538-4357/ab5362}, \href
  {https://ui.adsabs.harvard.edu/abs/2019ApJ...887...93G} {887, 93}

\bibitem[\protect\citeauthoryear{{Guo} et~al.,}{{Guo} et~al.}{2021}]{Guo2021}
{Guo} H.~L.,  et~al., 2021, \mn@doi [\apj] {10.3847/1538-4357/abc68a}, \href
  {https://ui.adsabs.harvard.edu/abs/2021ApJ...906...47G} {906, 47}

\bibitem[\protect\citeauthoryear{{Hanson} et~al.,}{{Hanson}
  et~al.}{2016}]{Hanson2016}
{Hanson} R.~J.,  et~al., 2016, \mn@doi [\mnras] {10.1093/mnras/stw2240}, \href
  {https://ui.adsabs.harvard.edu/abs/2016MNRAS.463.3604H} {463, 3604}

\bibitem[\protect\citeauthoryear{{He}, {Li}, {Feng}, {Ho}, {Ravanbakhsh},
  {Chen}  \& {P{\'o}czos}}{{He} et~al.}{2019}]{He2019}
{He} S.,  {Li} Y.,  {Feng} Y.,  {Ho} S.,  {Ravanbakhsh} S.,  {Chen} W.,
  {P{\'o}czos} B.,  2019, \mn@doi [Proceedings of the National Academy of
  Science] {10.1073/pnas.1821458116}, \href
  {https://ui.adsabs.harvard.edu/abs/2019PNAS..11613825H} {116, 13825}

\bibitem[\protect\citeauthoryear{{Hong}, {Jeong}, {Hwang}  \& {Kim}}{{Hong}
  et~al.}{2021}]{Hong2021}
{Hong} S.~E.,  {Jeong} D.,  {Hwang} H.~S.,   {Kim} J.,  2021, \mn@doi [\apj]
  {10.3847/1538-4357/abf040}, \href
  {https://ui.adsabs.harvard.edu/abs/2021ApJ...913...76H} {913, 76}

\bibitem[\protect\citeauthoryear{{Howlett}}{{Howlett}}{2019}]{Howlett2019}
{Howlett} C.,  2019, \mn@doi [\mnras] {10.1093/mnras/stz1403}, \href
  {https://ui.adsabs.harvard.edu/abs/2019MNRAS.487.5209H} {487, 5209}

\bibitem[\protect\citeauthoryear{Ioffe \& Szegedy}{Ioffe \&
  Szegedy}{2015}]{Ioffe2015}
Ioffe S.,  Szegedy C.,  2015, in Bach F.,  Blei D.,  eds,  Proceedings of
  Machine Learning Research Vol. 37, Proceedings of the 32nd International
  Conference on Machine Learning. PMLR, Lille, France, pp 448--456, \url
  {https://proceedings.mlr.press/v37/ioffe15.html}

\bibitem[\protect\citeauthoryear{{Kingma} \& {Ba}}{{Kingma} \&
  {Ba}}{2014}]{Kingma2014}
{Kingma} D.~P.,  {Ba} J.,  2014, arXiv e-prints, \href
  {https://ui.adsabs.harvard.edu/abs/2014arXiv1412.6980K} {p. arXiv:1412.6980}

\bibitem[\protect\citeauthoryear{{Lallement}, {Vergely}, {Valette},
  {Puspitarini}, {Eyer}  \& {Casagrande}}{{Lallement}
  et~al.}{2014}]{Lallement2014}
{Lallement} R.,  {Vergely} J.~L.,  {Valette} B.,  {Puspitarini} L.,  {Eyer} L.,
    {Casagrande} L.,  2014, \mn@doi [\aap] {10.1051/0004-6361/201322032}, \href
  {https://ui.adsabs.harvard.edu/abs/2014A&A...561A..91L} {561, A91}

\bibitem[\protect\citeauthoryear{{Lallement}, {Babusiaux}, {Vergely}, {Katz},
  {Arenou}, {Valette}, {Hottier}  \& {Capitanio}}{{Lallement}
  et~al.}{2019}]{Lallement2019}
{Lallement} R.,  {Babusiaux} C.,  {Vergely} J.~L.,  {Katz} D.,  {Arenou} F.,
  {Valette} B.,  {Hottier} C.,   {Capitanio} L.,  2019, \mn@doi [Astronomy \&
  Astrophysics] {10.1051/0004-6361/201834695}, \href
  {https://ui.adsabs.harvard.edu/abs/2019A&A...625A.135L} {625, A135}

\bibitem[\protect\citeauthoryear{{Leike} \& {En{\ss}lin}}{{Leike} \&
  {En{\ss}lin}}{2019}]{Leike2019}
{Leike} R.~H.,  {En{\ss}lin} T.~A.,  2019, \mn@doi [\aap]
  {10.1051/0004-6361/201935093}, \href
  {https://ui.adsabs.harvard.edu/abs/2019A&A...631A..32L} {631, A32}

\bibitem[\protect\citeauthoryear{{Mao}, {Wang}, {Li}, {Cai}, {Falck},
  {Neyrinck}  \& {Szalay}}{{Mao} et~al.}{2021}]{Mao2021}
{Mao} T.-X.,  {Wang} J.,  {Li} B.,  {Cai} Y.-C.,  {Falck} B.,  {Neyrinck} M.,
  {Szalay} A.,  2021, \mn@doi [\mnras] {10.1093/mnras/staa3741}, \href
  {https://ui.adsabs.harvard.edu/abs/2021MNRAS.501.1499M} {501, 1499}

\bibitem[\protect\citeauthoryear{{Marshall}, {Robin}, {Reyl{\'e}}, {Schultheis}
   \& {Picaud}}{{Marshall} et~al.}{2006}]{Marshall2006}
{Marshall} D.~J.,  {Robin} A.~C.,  {Reyl{\'e}} C.,  {Schultheis} M.,   {Picaud}
  S.,  2006, \mn@doi [\aap] {10.1051/0004-6361:20053842}, \href
  {https://ui.adsabs.harvard.edu/abs/2006A&A...453..635M} {453, 635}

\bibitem[\protect\citeauthoryear{{Milletari}, {Navab}  \& {Ahmadi}}{{Milletari}
  et~al.}{2016}]{Milletari2016}
{Milletari} F.,  {Navab} N.,   {Ahmadi} S.-A.,  2016, arXiv e-prints, \href
  {https://ui.adsabs.harvard.edu/abs/2016arXiv160604797M} {p. arXiv:1606.04797}

\bibitem[\protect\citeauthoryear{{Pan}, {Liu}, {Forero-Romero}, {Sabiu}, {Li},
  {Miao}  \& {Li}}{{Pan} et~al.}{2020}]{Pan2020}
{Pan} S.,  {Liu} M.,  {Forero-Romero} J.,  {Sabiu} C.~G.,  {Li} Z.,  {Miao} H.,
    {Li} X.-D.,  2020, \mn@doi [Science China Physics, Mechanics, and
  Astronomy] {10.1007/s11433-020-1586-3}, \href
  {https://ui.adsabs.harvard.edu/abs/2020SCPMA..6310412P} {63, 110412}

\bibitem[\protect\citeauthoryear{{Puspitarini}}{{Puspitarini}}{2014}]{Puspitarini2014}
{Puspitarini} L.,  2014, PhD thesis, Observatoire De Paris

\bibitem[\protect\citeauthoryear{{Qin}, {Howlett}  \& {Staveley-Smith}}{{Qin}
  et~al.}{2019}]{Qin2019}
{Qin} F.,  {Howlett} C.,   {Staveley-Smith} L.,  2019, \mn@doi [\mnras]
  {10.1093/mnras/stz1576}, \href
  {https://ui.adsabs.harvard.edu/abs/2019MNRAS.487.5235Q} {487, 5235}

\bibitem[\protect\citeauthoryear{{Qin}, {Parkinson}, {Hong}  \& {Sabiu}}{{Qin}
  et~al.}{2023}]{Qin2023}
{Qin} F.,  {Parkinson} D.,  {Hong} S.~E.,   {Sabiu} C.~G.,  2023, \mn@doi
  [\jcap] {10.1088/1475-7516/2023/06/062}, \href
  {https://ui.adsabs.harvard.edu/abs/2023JCAP...06..062Q} {2023, 062}

\bibitem[\protect\citeauthoryear{{Rezaei Kh.}, {Bailer-Jones}, {Hanson}  \&
  {Fouesneau}}{{Rezaei Kh.} et~al.}{2017}]{Kh2017}
{Rezaei Kh.} S.,  {Bailer-Jones} C.~A.~L.,  {Hanson} R.~J.,   {Fouesneau} M.,
  2017, \mn@doi [\aap] {10.1051/0004-6361/201628885}, \href
  {https://ui.adsabs.harvard.edu/abs/2017A&A...598A.125R} {598, A125}

\bibitem[\protect\citeauthoryear{{Ronneberger}, {Fischer}  \&
  {Brox}}{{Ronneberger} et~al.}{2015}]{Unet2015}
{Ronneberger} O.,  {Fischer} P.,   {Brox} T.,  2015, arXiv e-prints, \href
  {https://ui.adsabs.harvard.edu/abs/2015arXiv150504597R} {p. arXiv:1505.04597}

\bibitem[\protect\citeauthoryear{{Sale} \& {Magorrian}}{{Sale} \&
  {Magorrian}}{2014}]{SaleM2014}
{Sale} S.~E.,  {Magorrian} J.,  2014, \mn@doi [\mnras] {10.1093/mnras/stu1728},
  \href {https://ui.adsabs.harvard.edu/abs/2014MNRAS.445..256S} {445, 256}

\bibitem[\protect\citeauthoryear{{Sale} et~al.,}{{Sale}
  et~al.}{2014}]{Sale2014}
{Sale} S.~E.,  et~al., 2014, \mn@doi [\mnras] {10.1093/mnras/stu1090}, \href
  {https://ui.adsabs.harvard.edu/abs/2014MNRAS.443.2907S} {443, 2907}

\bibitem[\protect\citeauthoryear{{Vergely}, {Lallement}  \& {Cox}}{{Vergely}
  et~al.}{2022}]{Vergely2022}
{Vergely} J.~L.,  {Lallement} R.,   {Cox} N.~L.~J.,  2022, \mn@doi [\aap]
  {10.1051/0004-6361/202243319}, \href
  {https://ui.adsabs.harvard.edu/abs/2022A&A...664A.174V} {664, A174}

\bibitem[\protect\citeauthoryear{{Wu} et~al.,}{{Wu} et~al.}{2021}]{Wu2021}
{Wu} Z.,  et~al., 2021, \mn@doi [\apj] {10.3847/1538-4357/abf3bb}, \href
  {https://ui.adsabs.harvard.edu/abs/2021ApJ...913....2W} {913, 2}

\bibitem[\protect\citeauthoryear{{Zhang}, {Green}  \& {Rix}}{{Zhang}
  et~al.}{2023}]{Zhang2023}
{Zhang} X.,  {Green} G.~M.,   {Rix} H.-W.,  2023, \mn@doi [\mnras]
  {10.1093/mnras/stad1941}, \href
  {https://ui.adsabs.harvard.edu/abs/2023MNRAS.tmp.1869Z} {}

\bibitem[\protect\citeauthoryear{{Zucker} et~al.,}{{Zucker}
  et~al.}{2022}]{Zucker2022}
{Zucker} C.,  et~al., 2022, \mn@doi [\nat] {10.1038/s41586-021-04286-5}, \href
  {https://ui.adsabs.harvard.edu/abs/2022Natur.601..334Z} {601, 334}

\makeatother
\end{thebibliography}

\appendix
\section{}
\begin{figure*}
\centering
\includegraphics[width=0.45\textwidth]{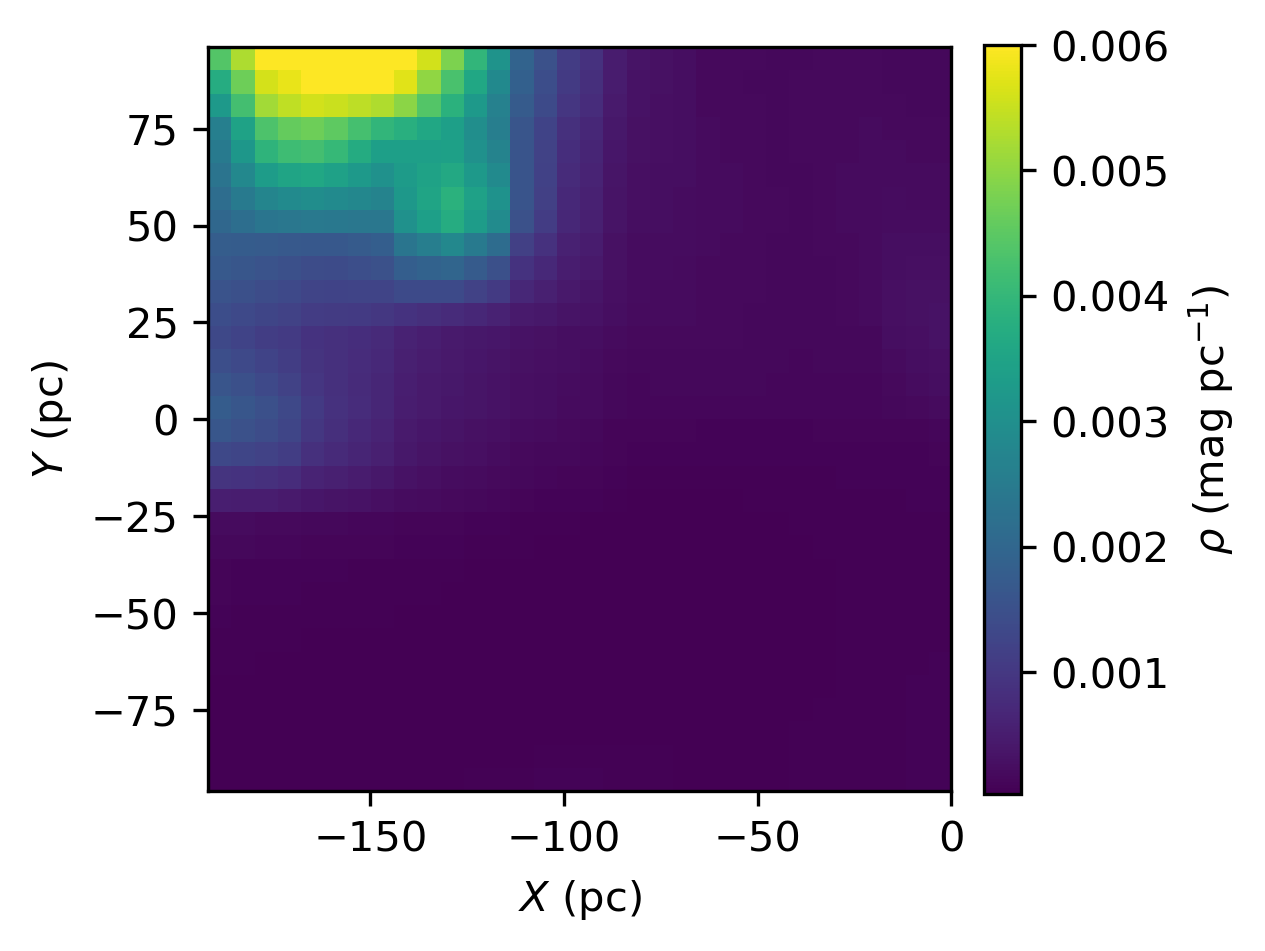}
\includegraphics[width=0.45\textwidth]{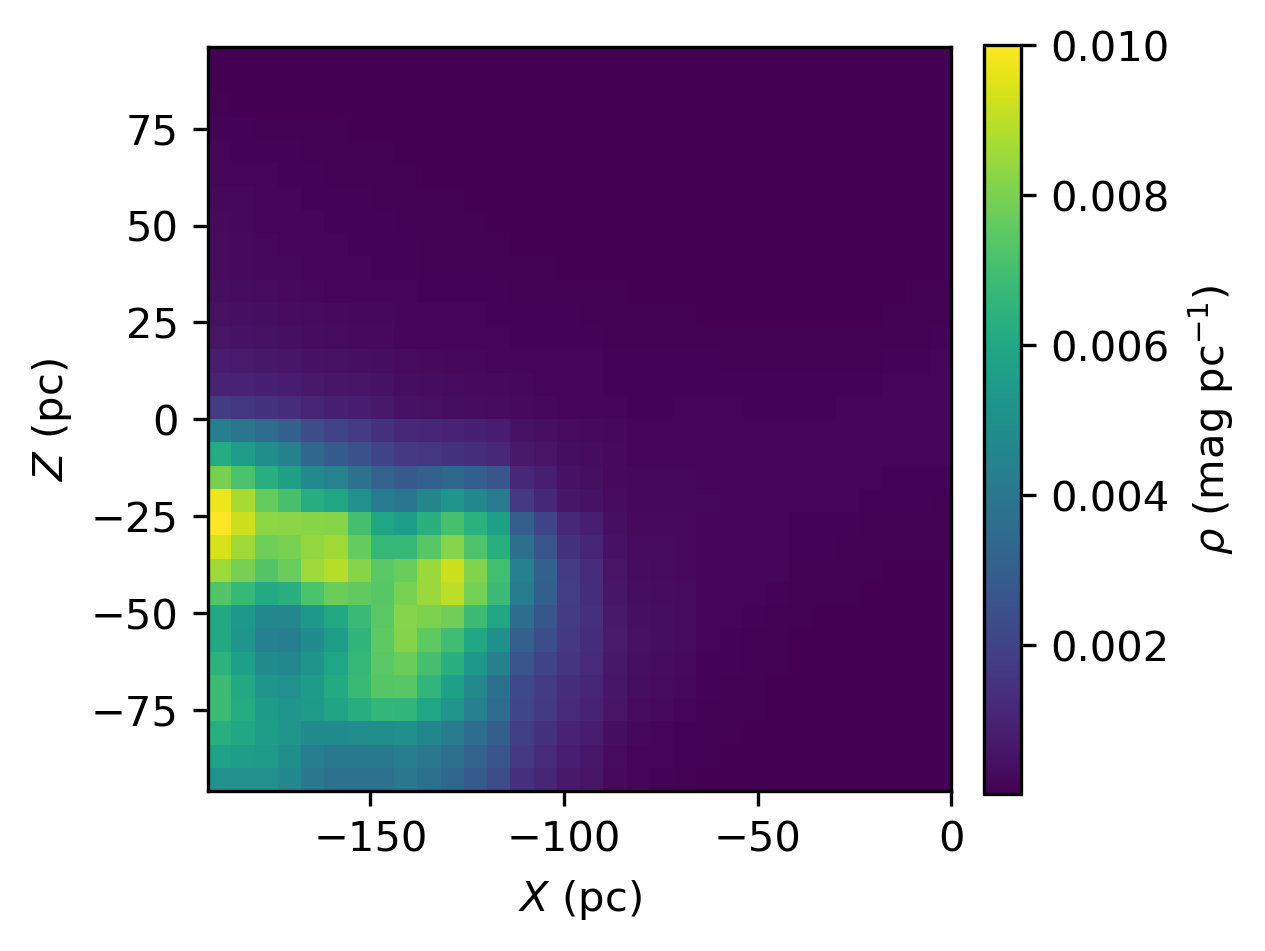}
\includegraphics[width=0.45\textwidth]{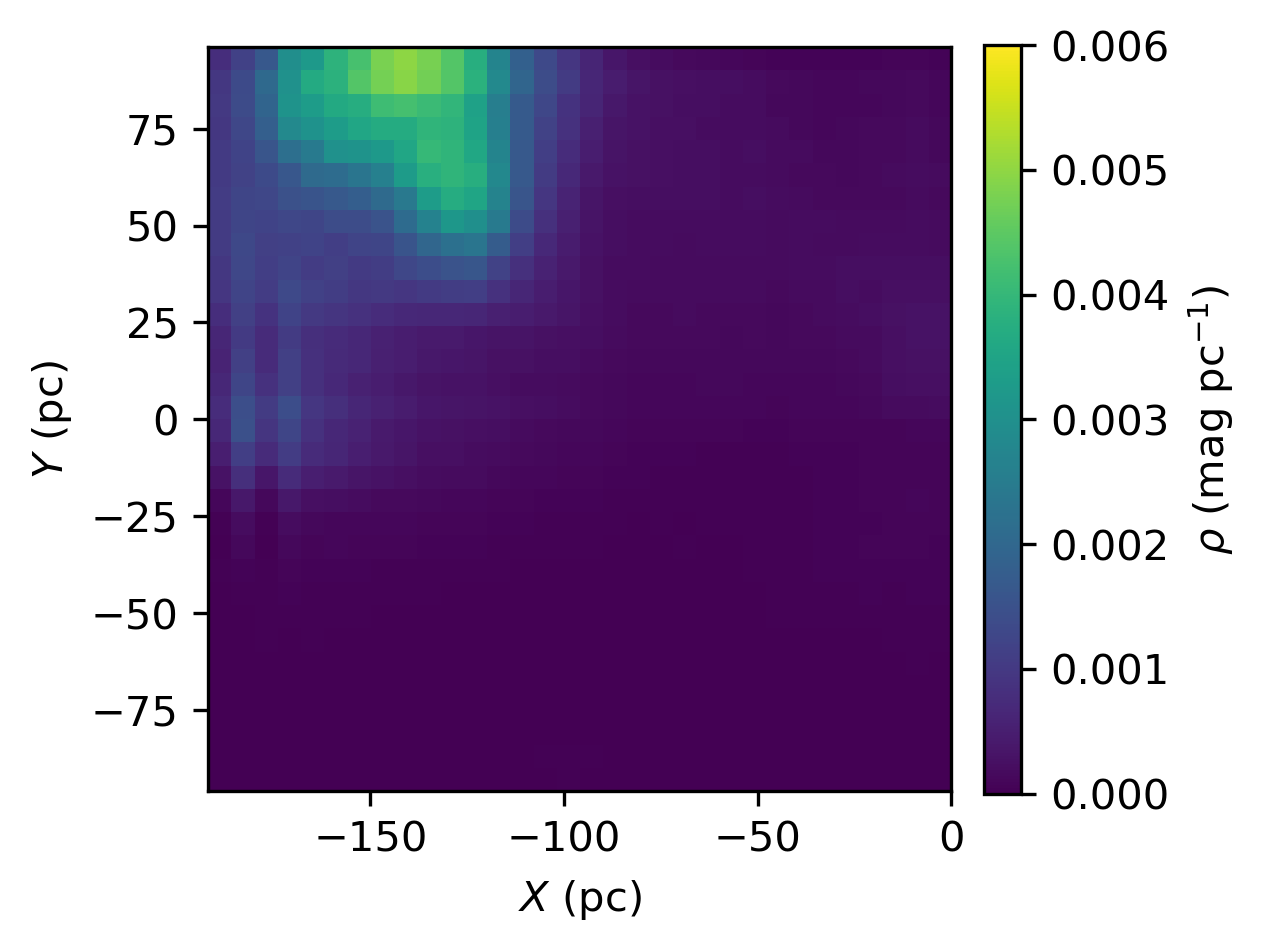}
\includegraphics[width=0.45\textwidth]{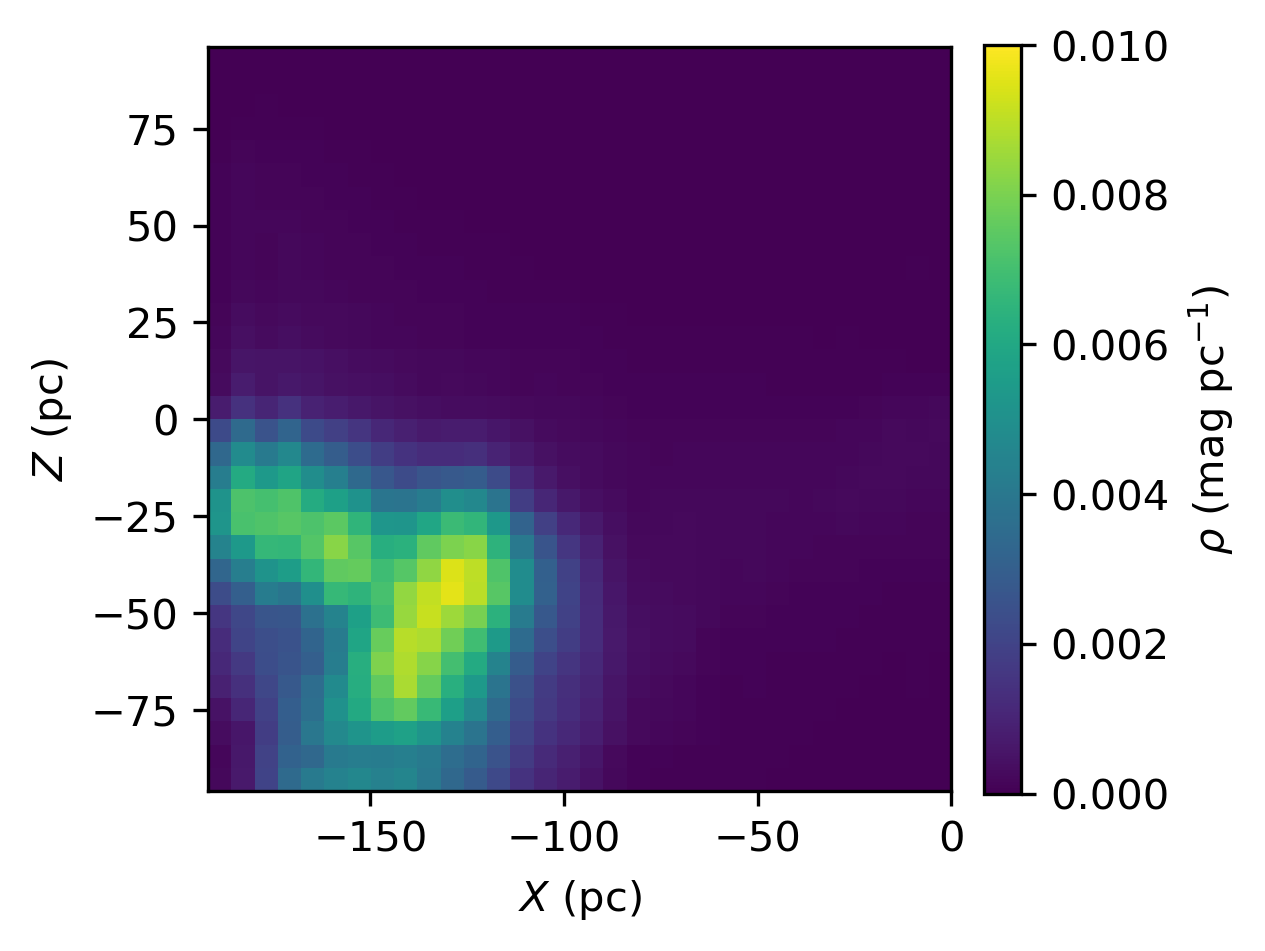}
\includegraphics[width=0.45\textwidth]{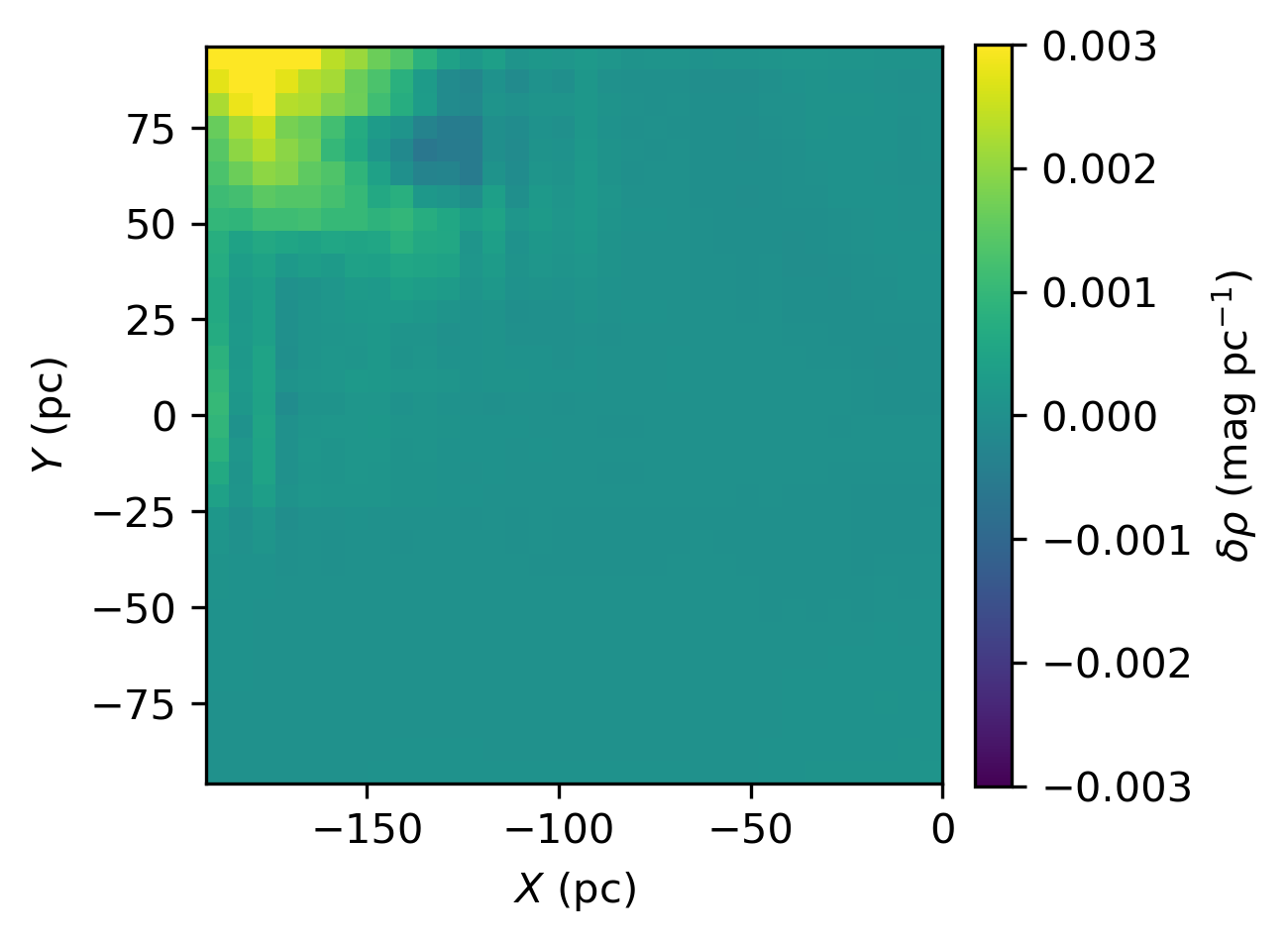}
\includegraphics[width=0.45\textwidth]{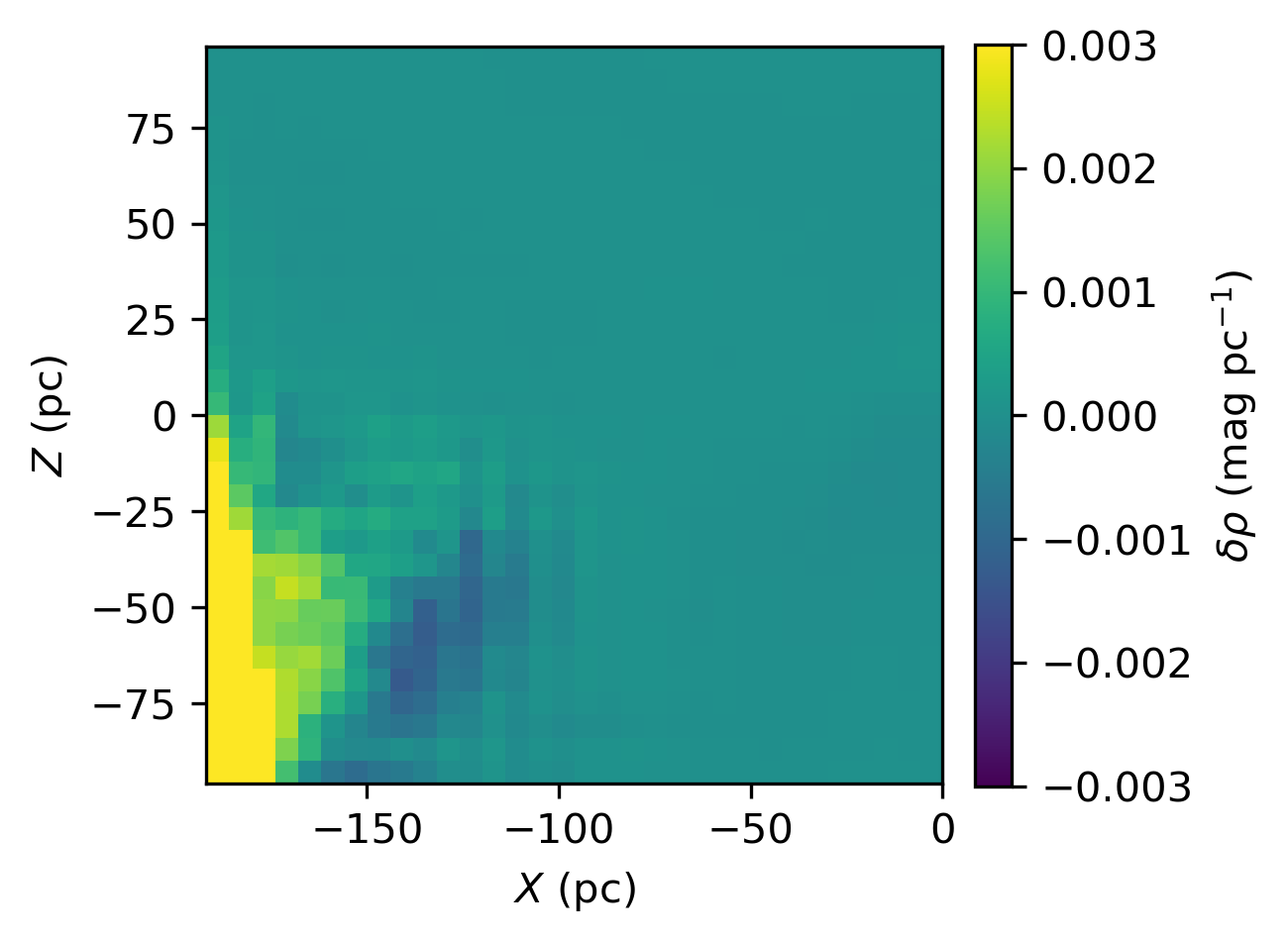}
\caption{Upper panels show the 3D extinction density maps interpolated from \citet{Vergely2022}, middle panels display the maps derived from our trained V-net model and bottom panels depict the residuals between the Vergely et al. maps and our results. The left panels illustrate the dust density maps and residue maps in the $XY$-plane, with $Z$ values ranging from 0 to 6\,pc. Conversely, the right panels depict the same quantities in the $XZ$-plane, with $Y$ values ranging from 0 to 6\,pc. }  
\label{v22}
\end{figure*}

\begin{figure*}
\centering
\includegraphics[width=0.45\textwidth]{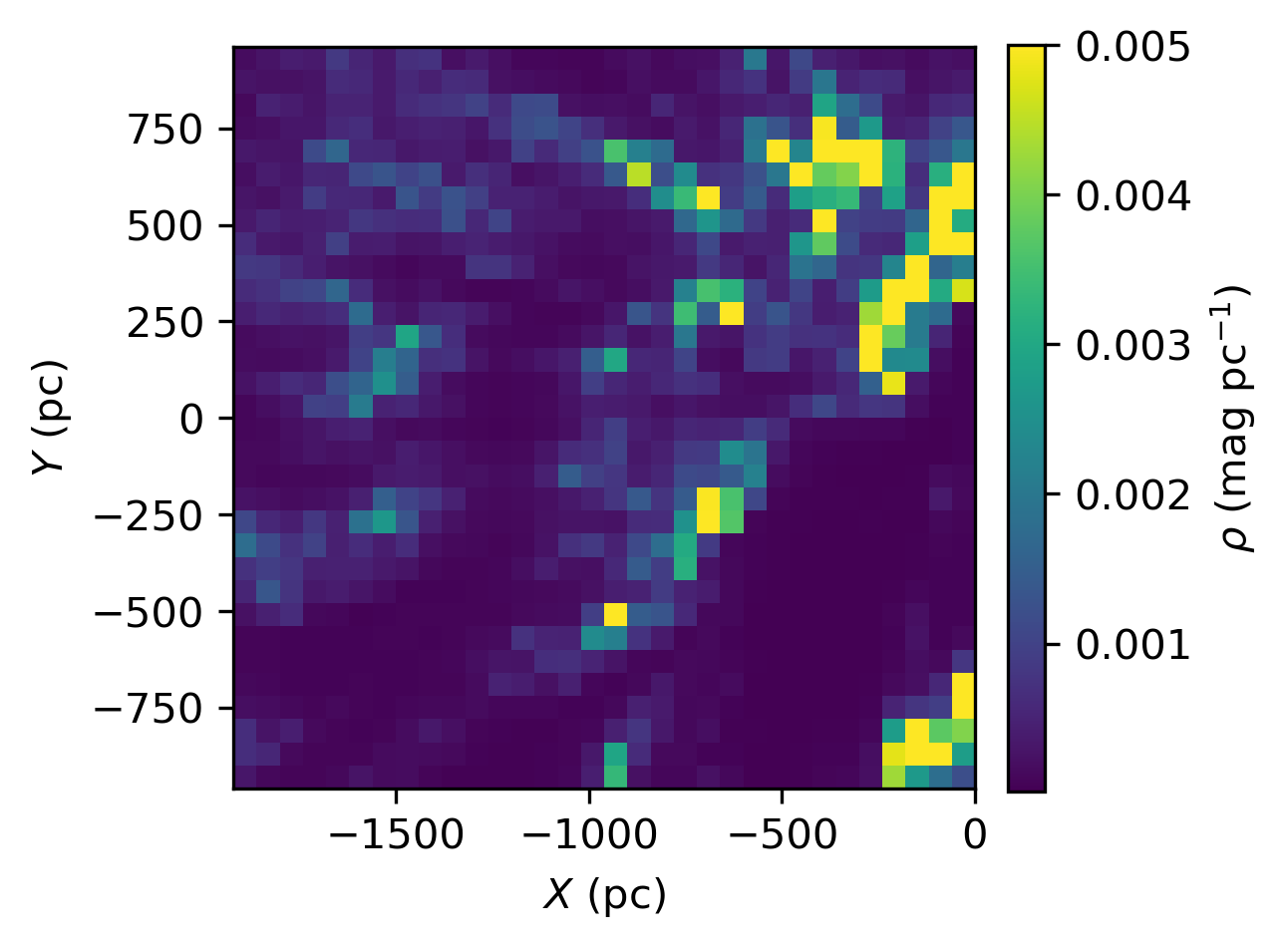}
\includegraphics[width=0.45\textwidth]{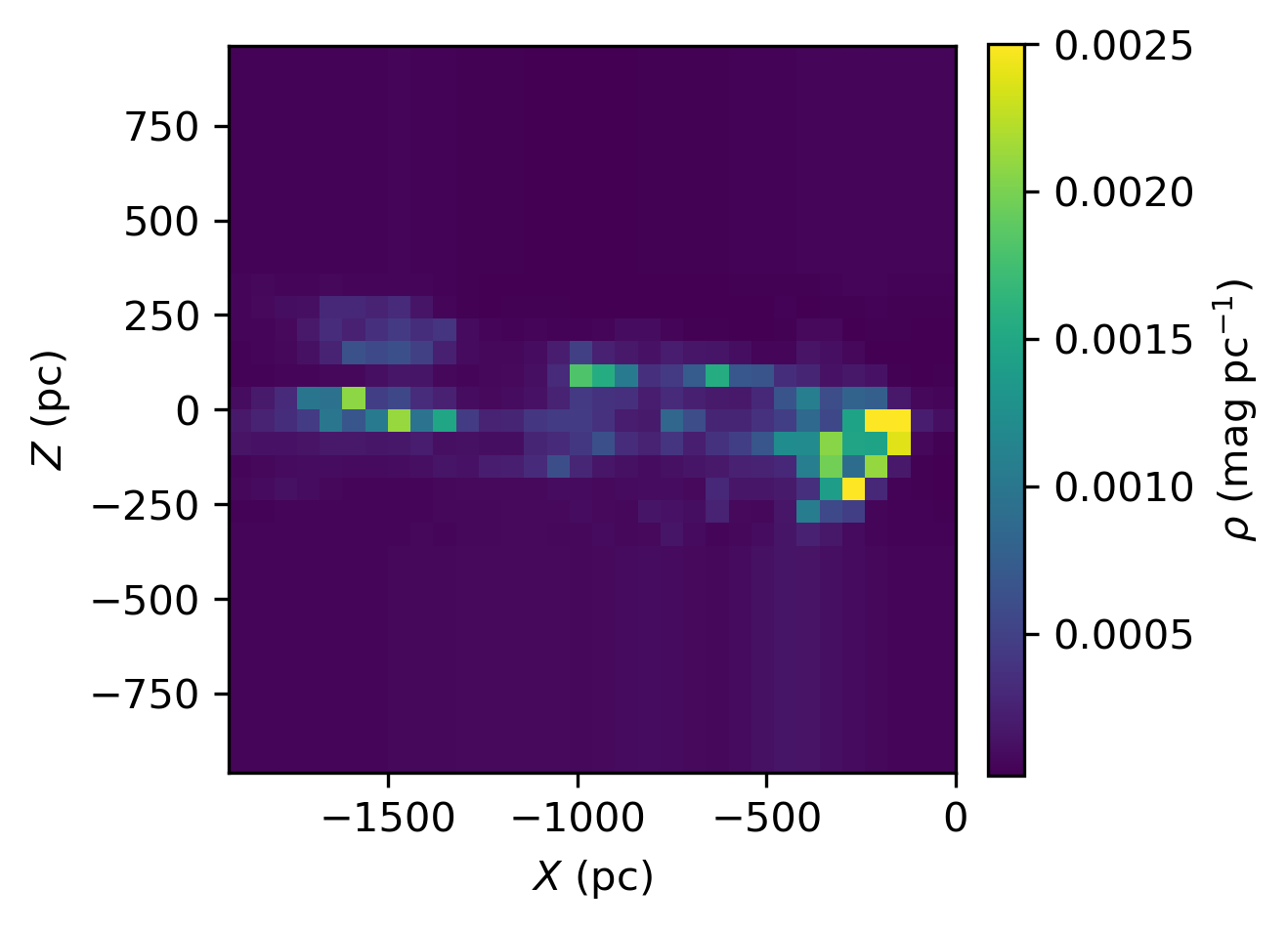}
\includegraphics[width=0.45\textwidth]{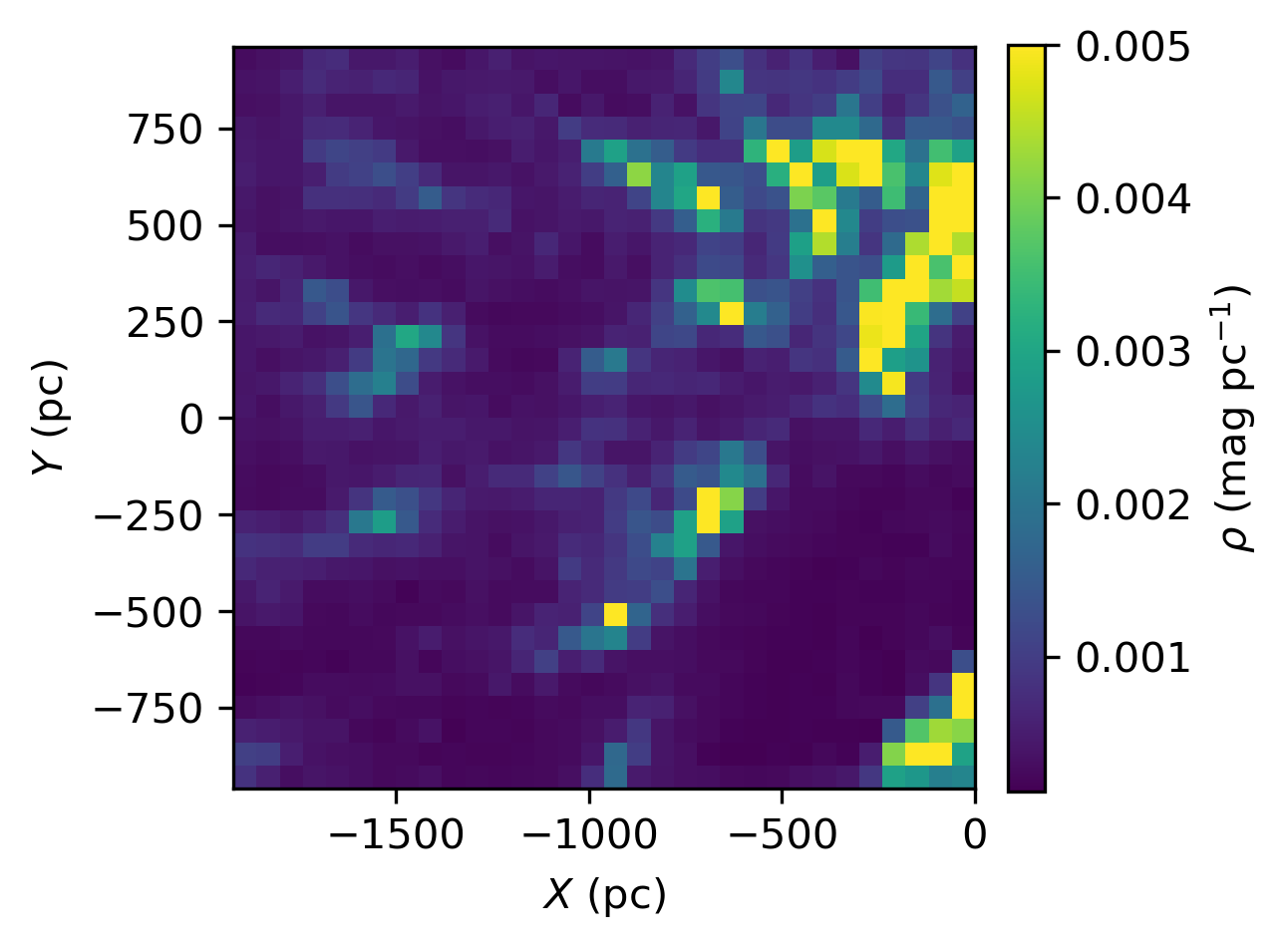}
\includegraphics[width=0.45\textwidth]{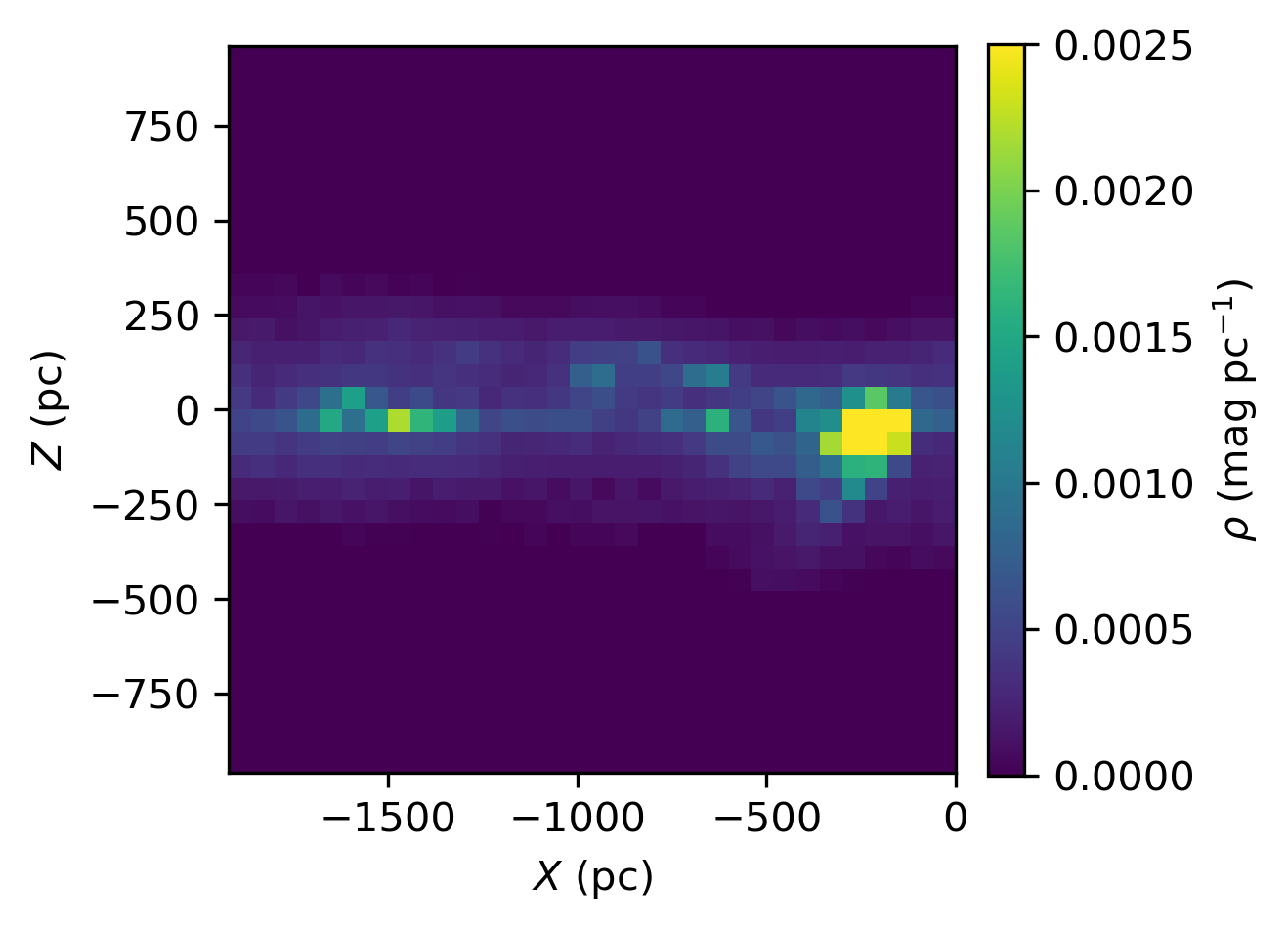}
\includegraphics[width=0.45\textwidth]{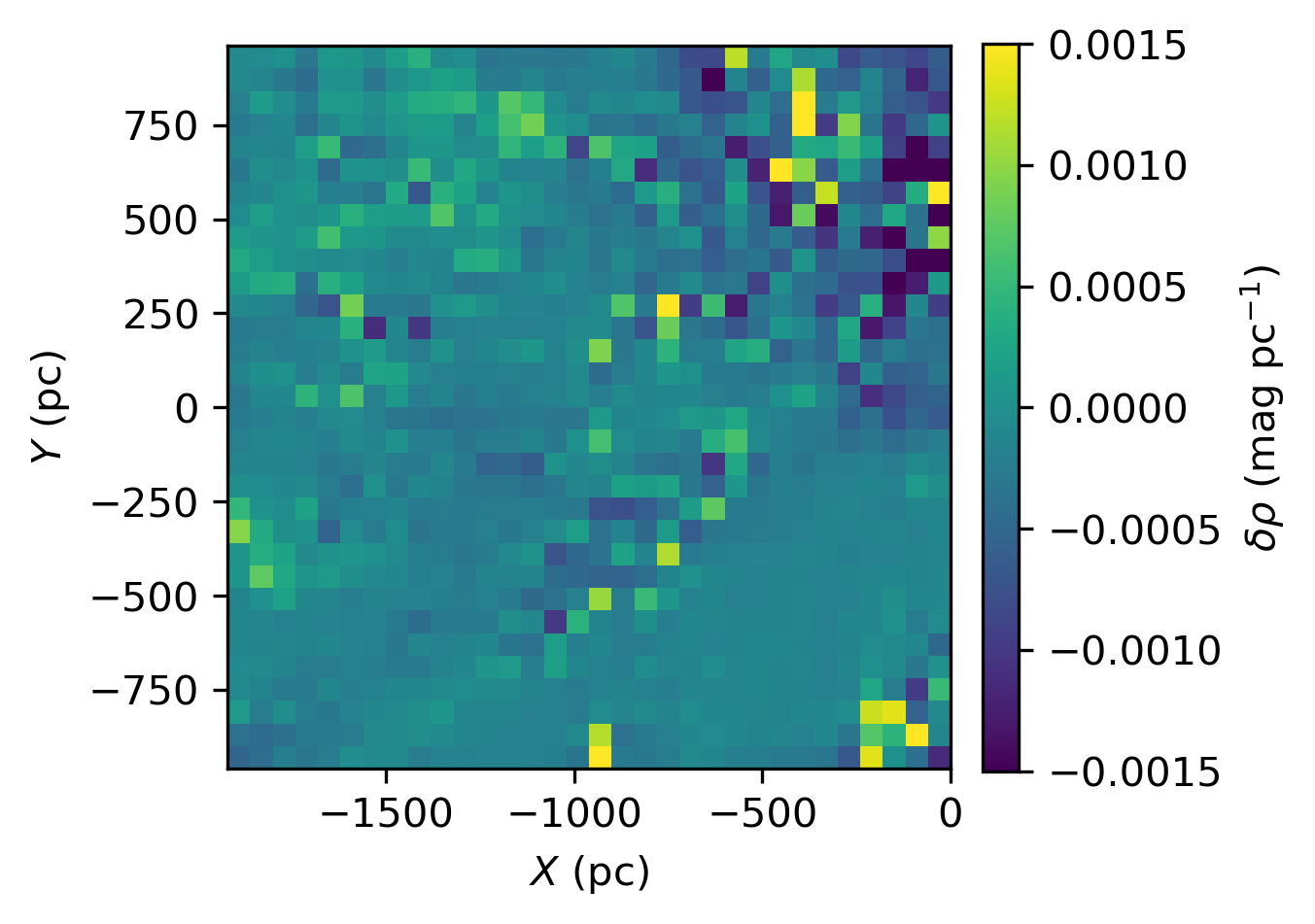}
\includegraphics[width=0.45\textwidth]{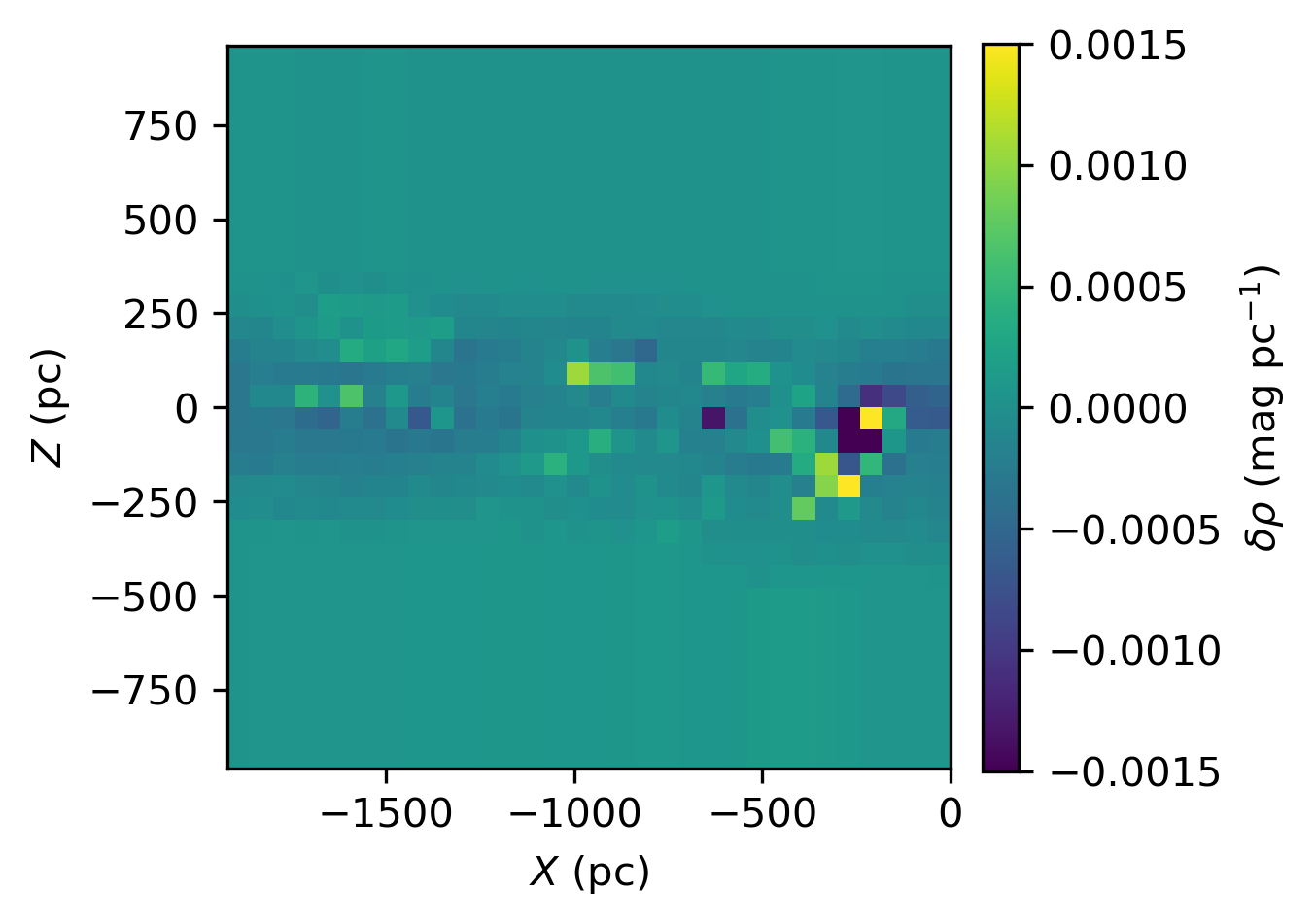}
\caption{Similar as Fig.~\ref{v22} but for the LR region. The left panels illustrate the dust density maps and residue maps in the $XY$-plane, with $Z$ values ranging from 0 to 60\,pc. Conversely, the right panels depict the same quantities in the $XZ$-plane, with $Y$ values ranging from 0 to 60\,pc. }
\label{v22large}
\end{figure*}

\begin{figure*}
\centering
\includegraphics[width=0.45\textwidth]{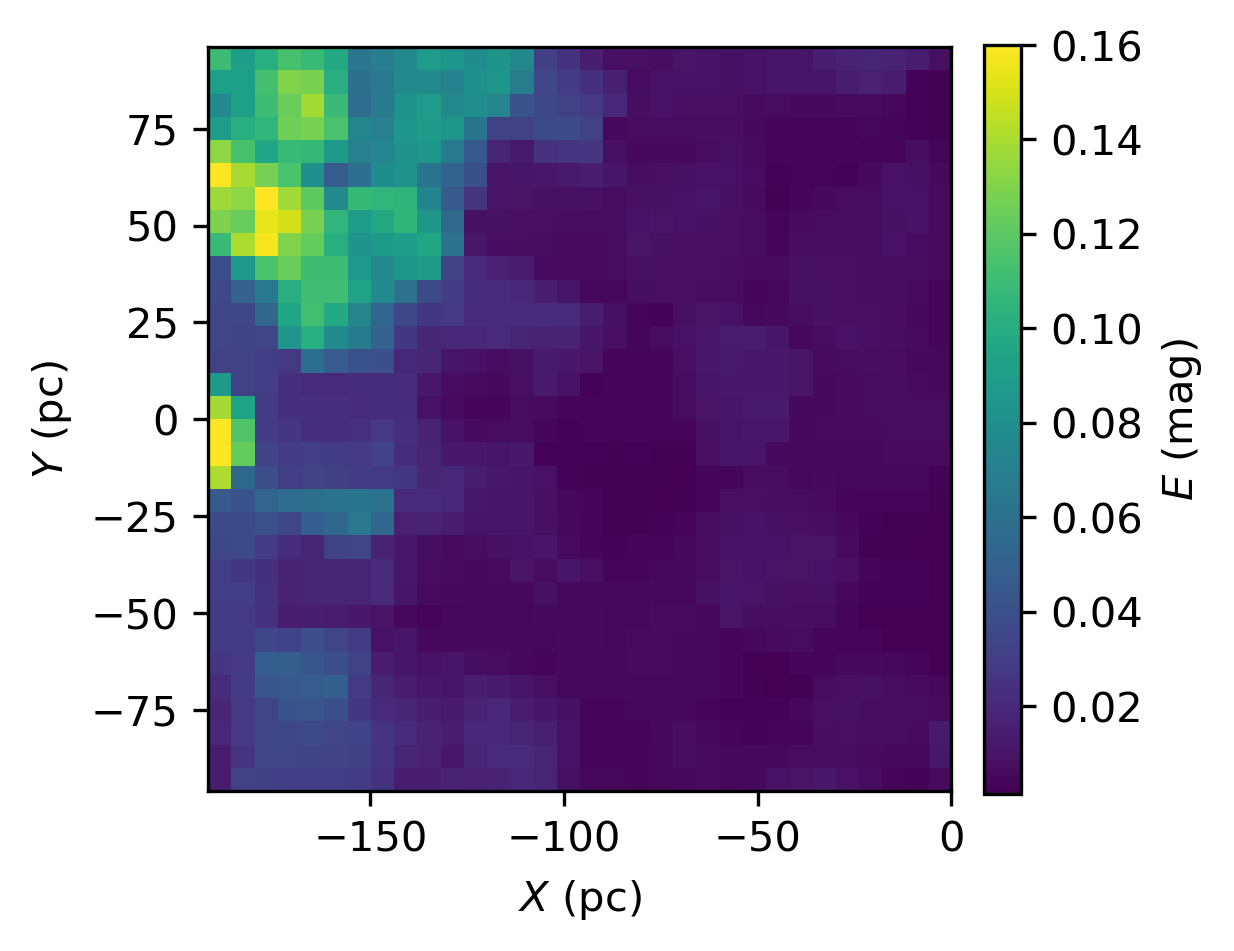}
\includegraphics[width=0.45\textwidth]{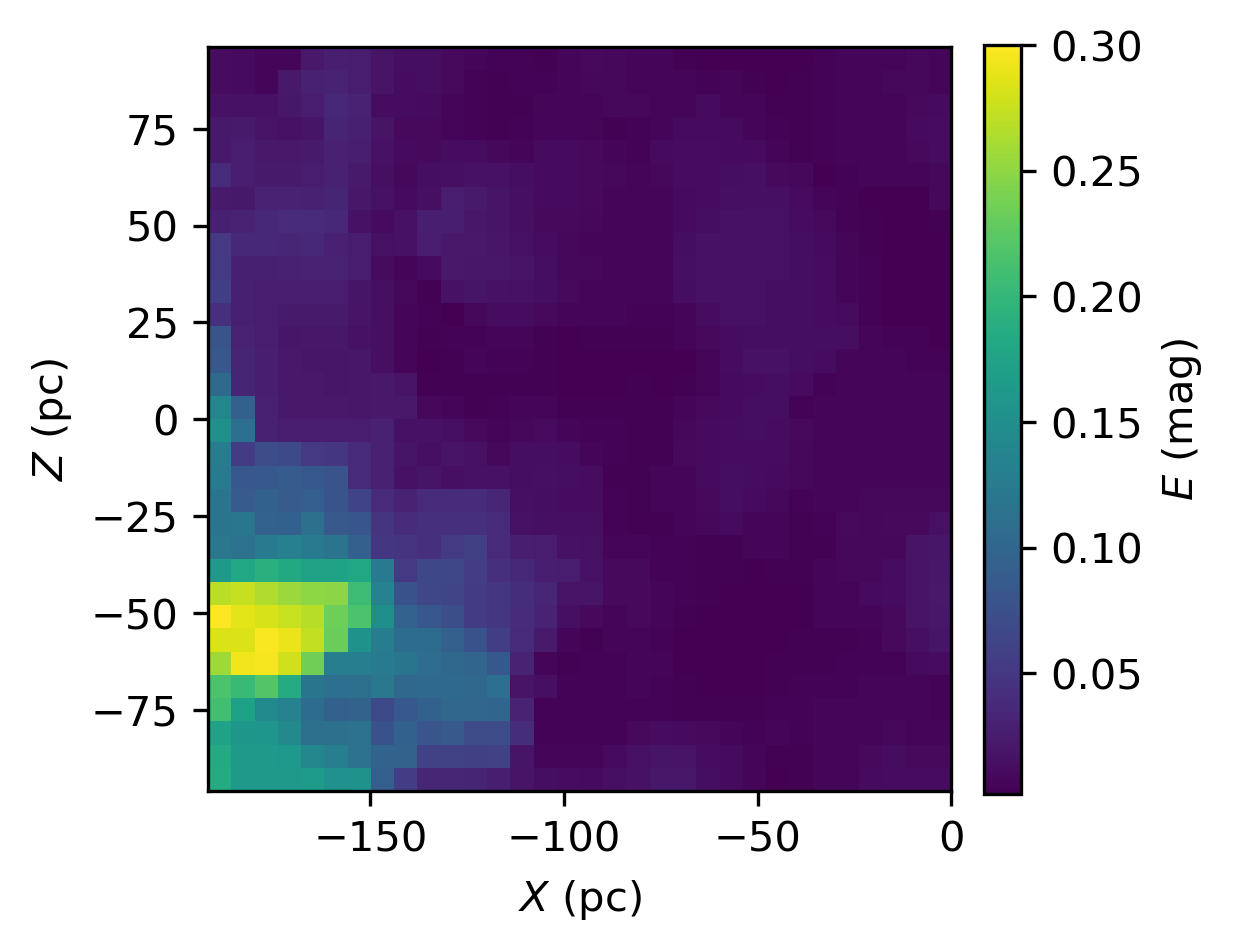}
\includegraphics[width=0.45\textwidth]{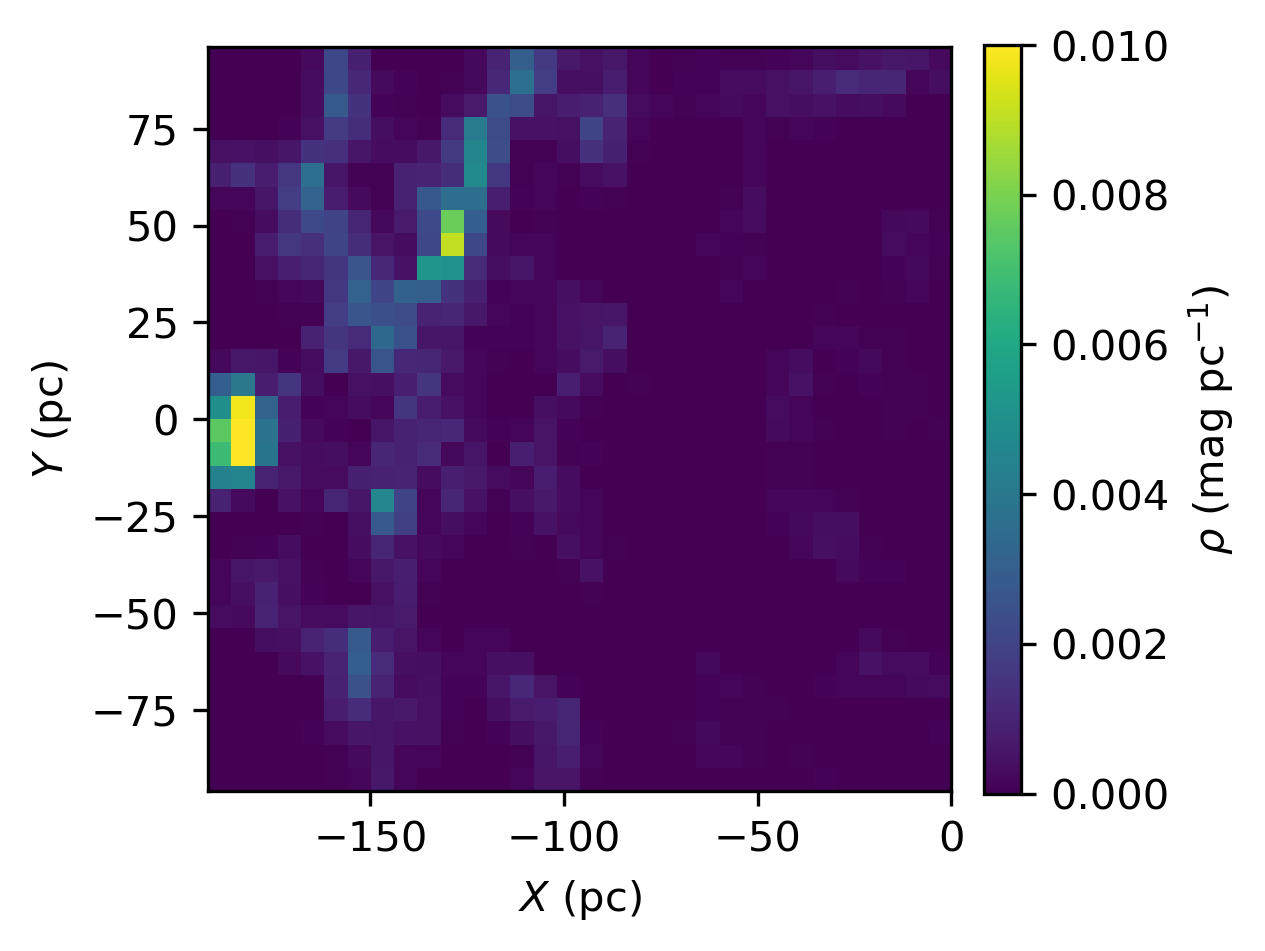}
\includegraphics[width=0.45\textwidth]{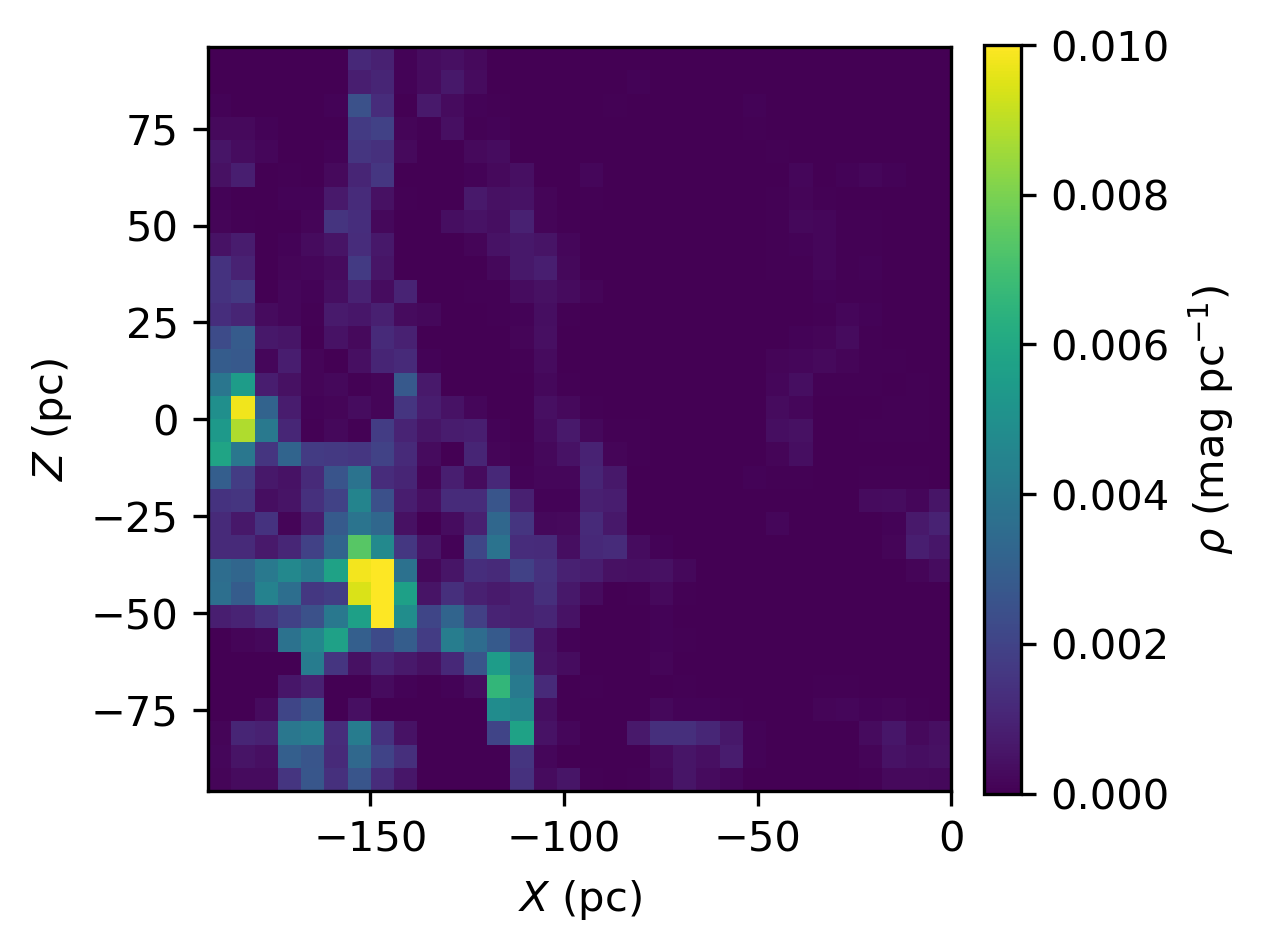}
\caption{The LOS extinction distributions calculated from the \citet{Zhang2023} catalogue (upper panels) and the predicted 3D dust density maps from the V-net (bottom panels) for the SR region. The left panels illustrate the dust density and extinction distributions in the $XY$-plane, with $Z$ values ranging from 0 to 6\,pc. Conversely, the right panels depict the same quantities in the $XZ$-plane, with $Y$ values ranging from 0 to 6\,pc.  }
\label{g22}
\end{figure*}

\begin{figure*}
\centering
\includegraphics[width=0.45\textwidth]{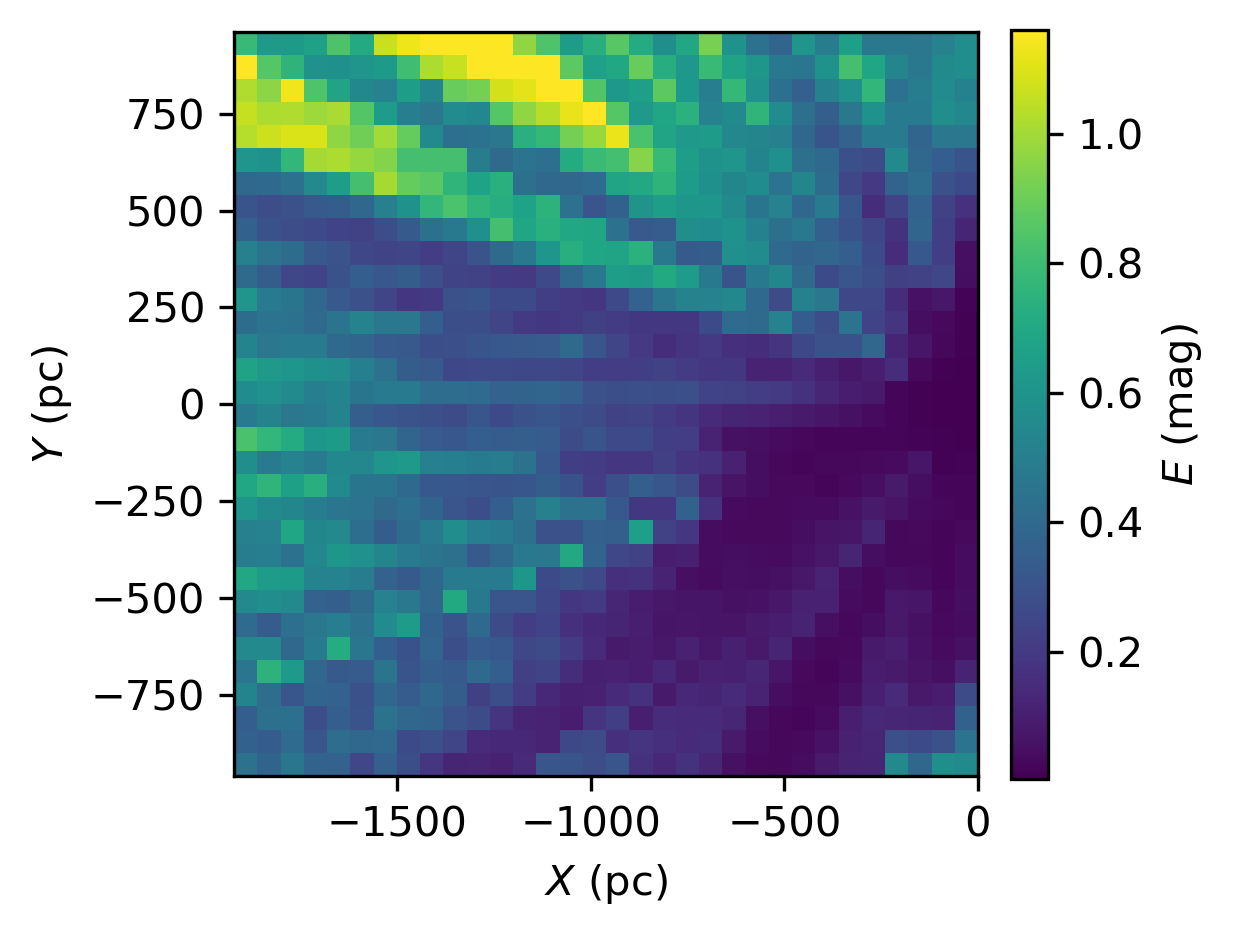}
\includegraphics[width=0.45\textwidth]{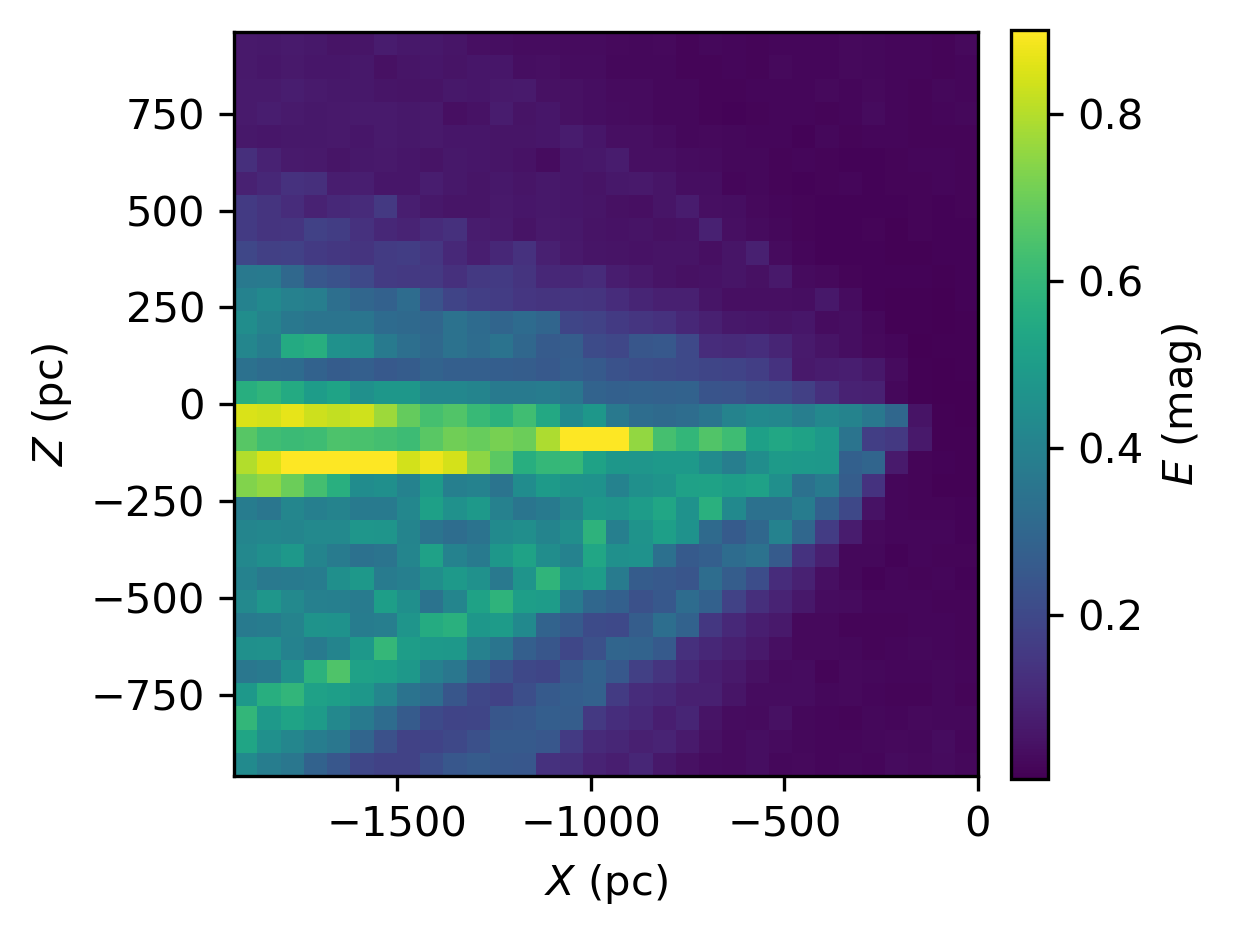}
\includegraphics[width=0.45\textwidth]{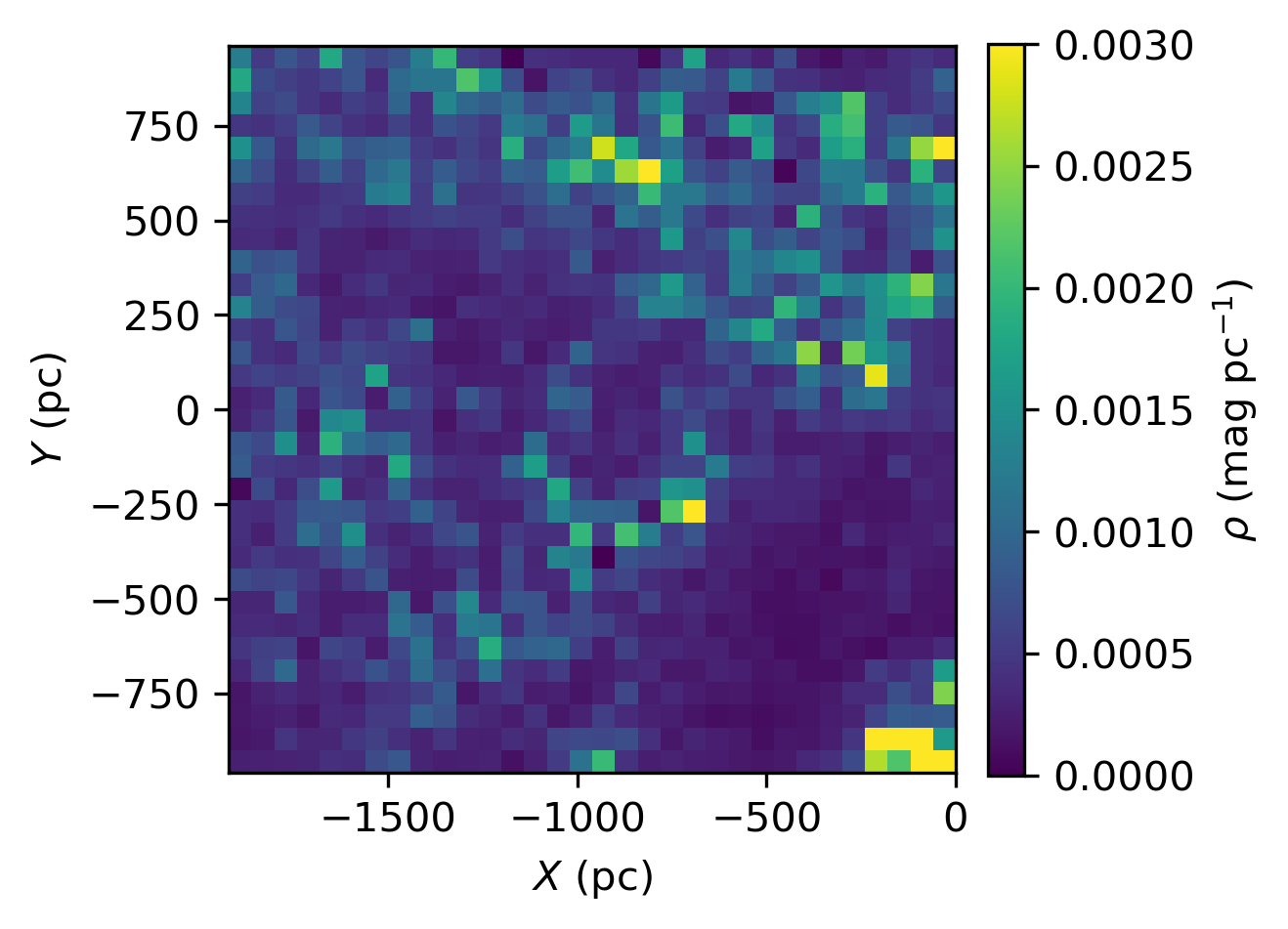}
\includegraphics[width=0.45\textwidth]{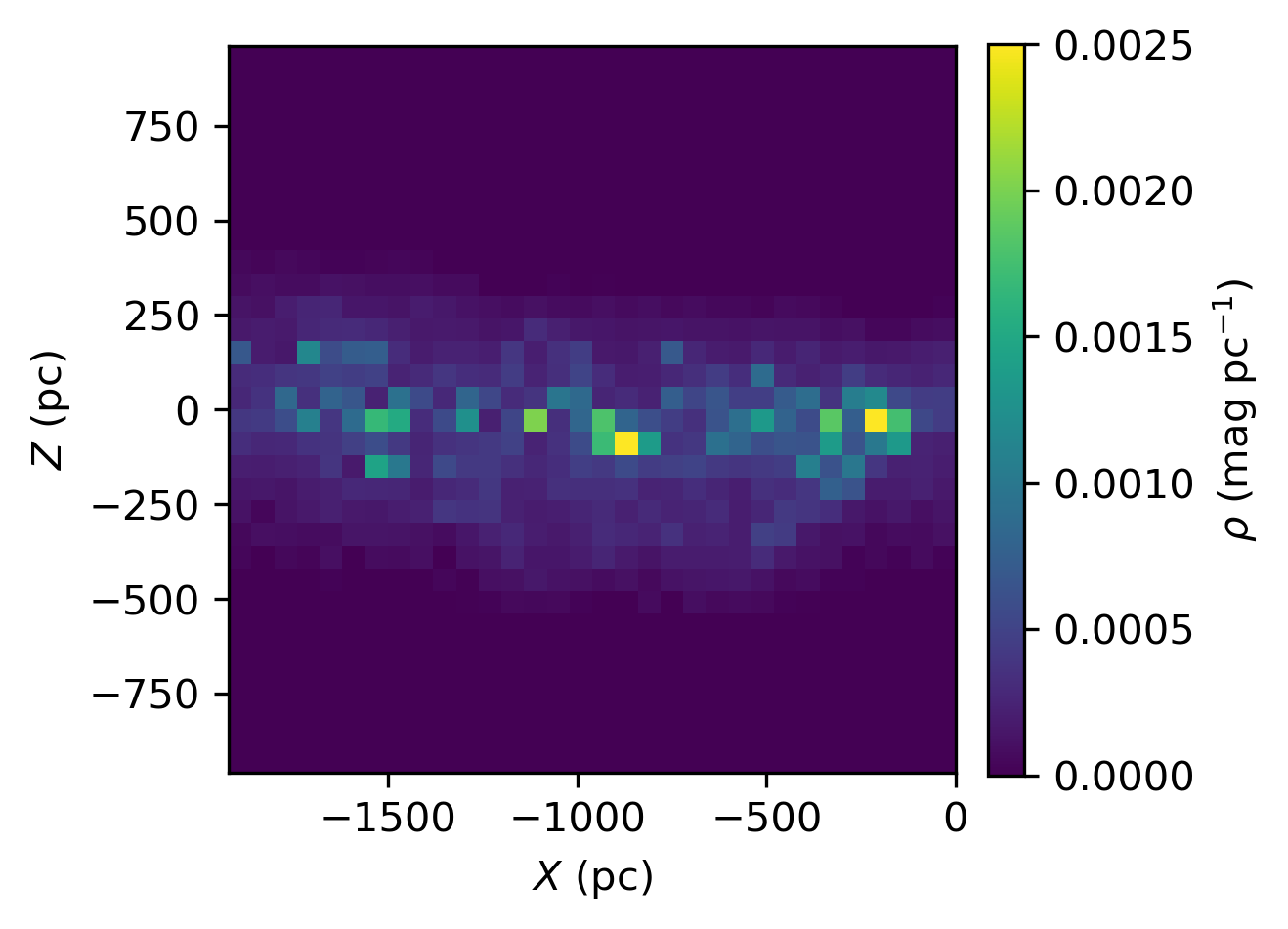}
\caption{Similar as Fig.~\ref{g22} but for the LR region. The left panels illustrate the dust density and extinction distributions in the $XY$-plane, with $Z$ values ranging from 0 to 60\,pc. Conversely, the right panels depict the same quantities in the $XZ$-plane, with $Y$ values ranging from 0 to 60\,pc. } 
\label{g22large}
\end{figure*}

\begin{figure*}
\centering
 \hspace{2.2cm}
  \begin{minipage}{0.272\textwidth}
    \centering
    \includegraphics[width=\textwidth]{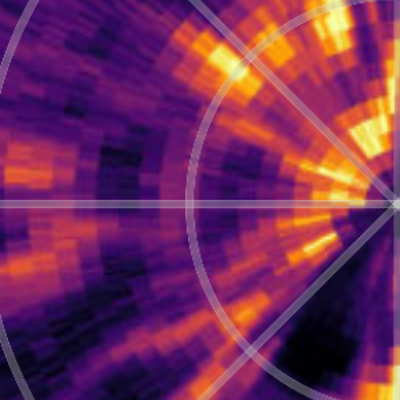} 
    \vspace{0.5cm}
  \end{minipage}
  \hfill
  \begin{minipage}{0.36\textwidth}
    \centering
    \includegraphics[width=\textwidth]{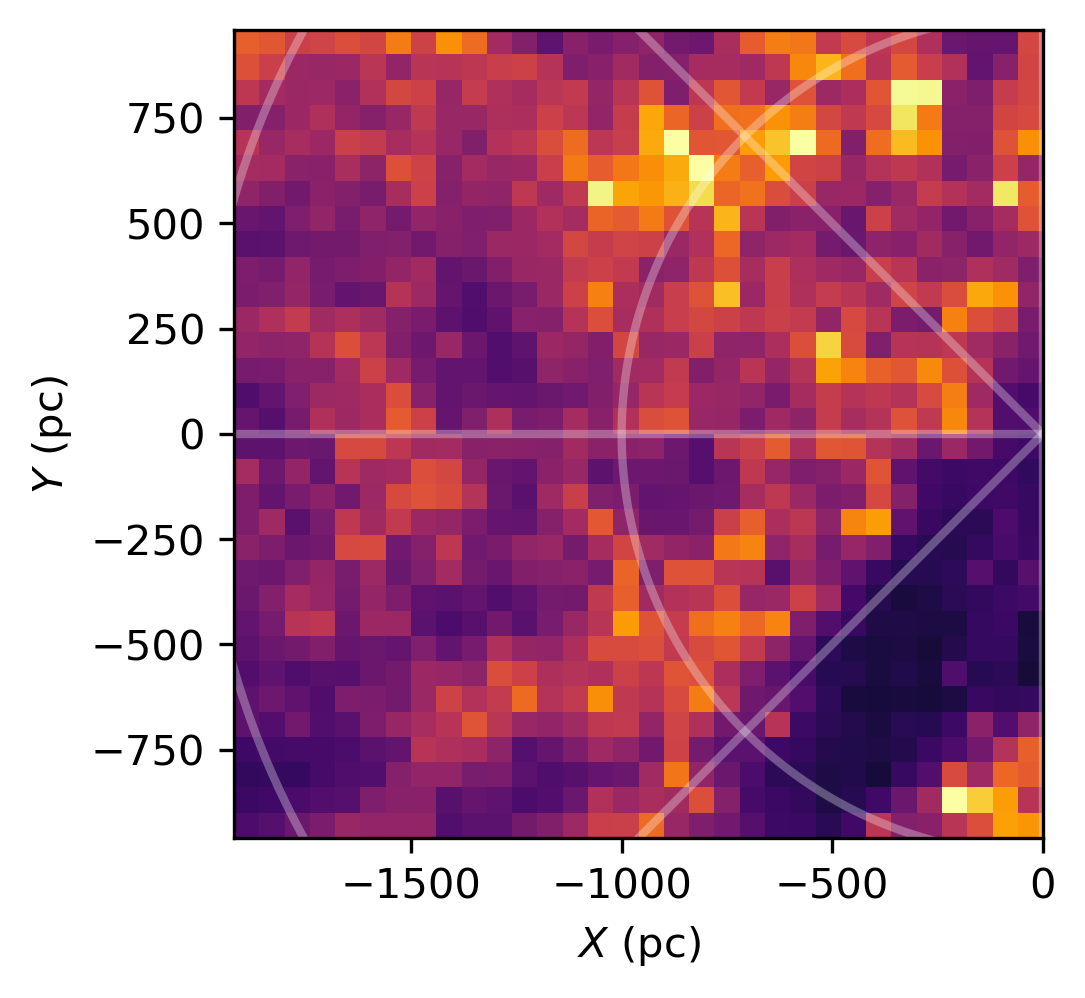}
  \end{minipage}
   \hspace{2.2cm}
\caption{Side-by-side comparison of 3D dust density results: our work (right) versus \citet{Zhang2023} (left, adapted from their Fig.~24). Both images represent the average dust density integrated along the $Z$-axis for $|Z|<400$\,pc, viewed from above the Galactic plane. The three white lines denote Galactic longitudes of $l$ = 135\degr, 180\degr, and 225\degr, while the concentric white circles indicate distances of 1 and 2\,kpc from the Sun.}
\label{g22descomp}
\end{figure*}

\begin{figure*}
\centering
\includegraphics[width=0.45\textwidth]{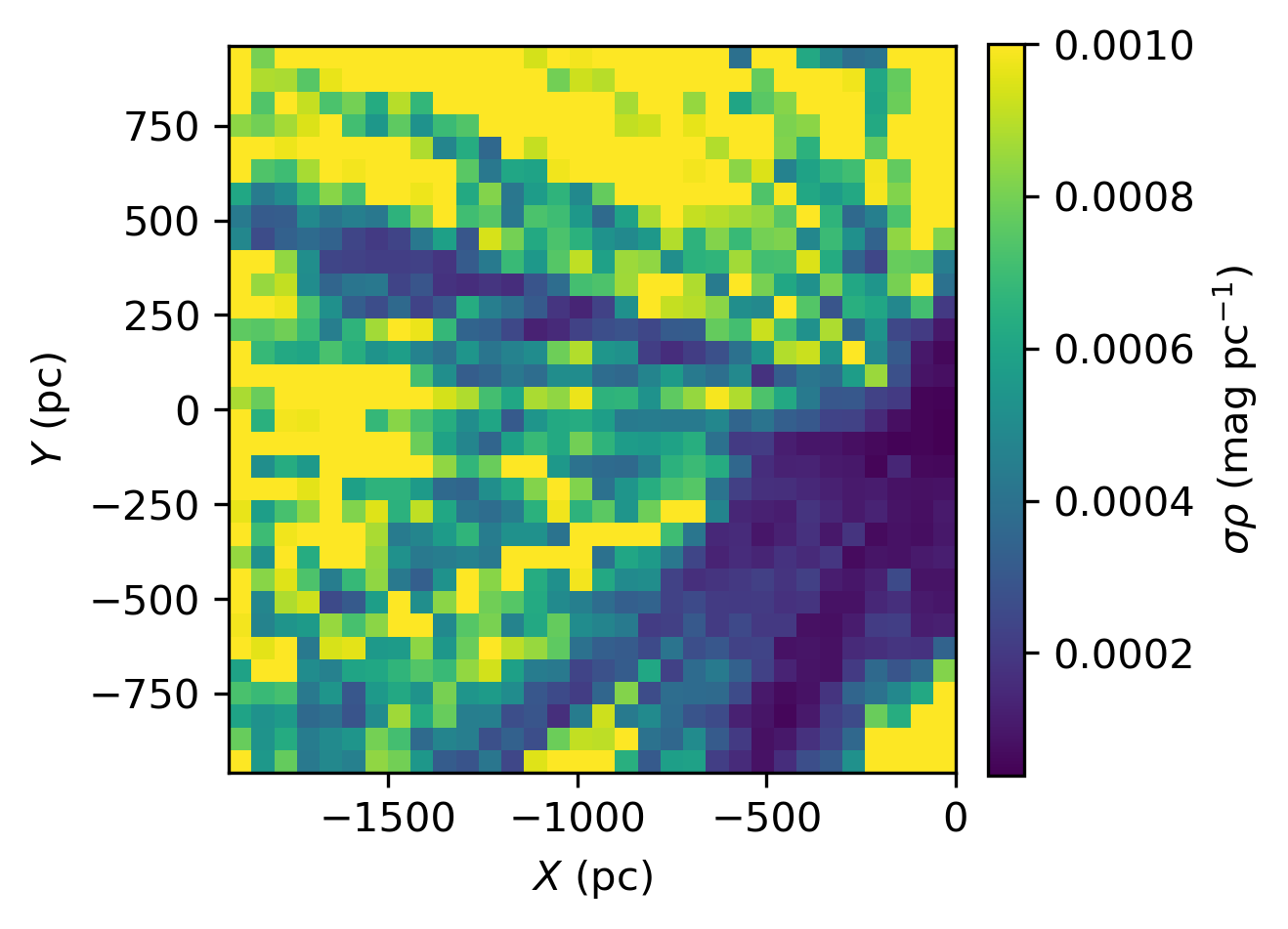}
\includegraphics[width=0.45\textwidth]{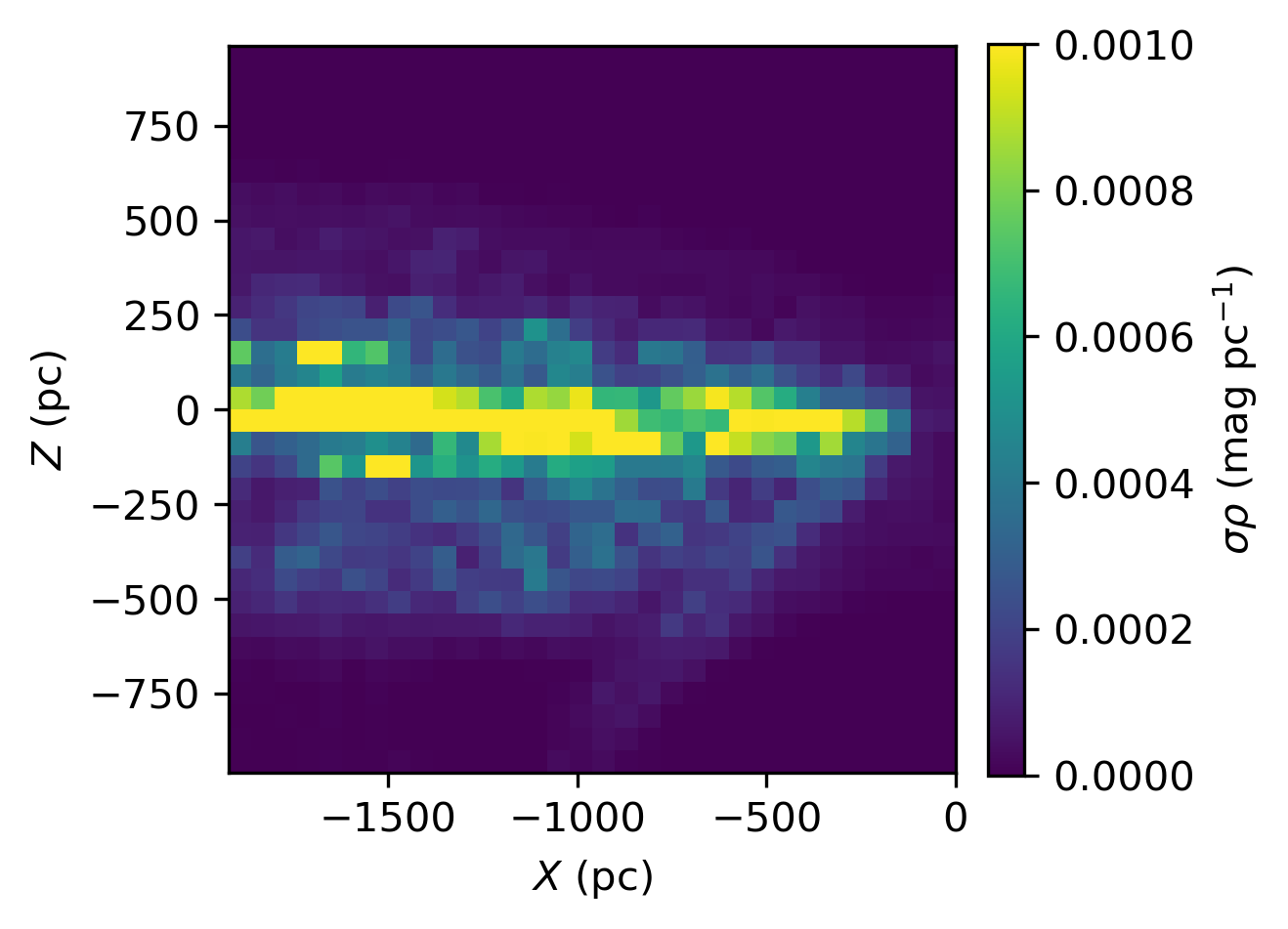}
\caption{Unertainties of the derived 3D dust density maps for the LR region. The left panels illustrate the dust density errors in the $XY$-plane, with $Z$ values ranging from 0 to 60\,pc. Conversely, the right panels depict the same quantities in the $XZ$-plane, with $Y$ values ranging from 0 to 60\,pc. } 
\label{g22errorlarge}
\end{figure*}

\begin{figure*}
\centering
\includegraphics[width=0.45\textwidth]{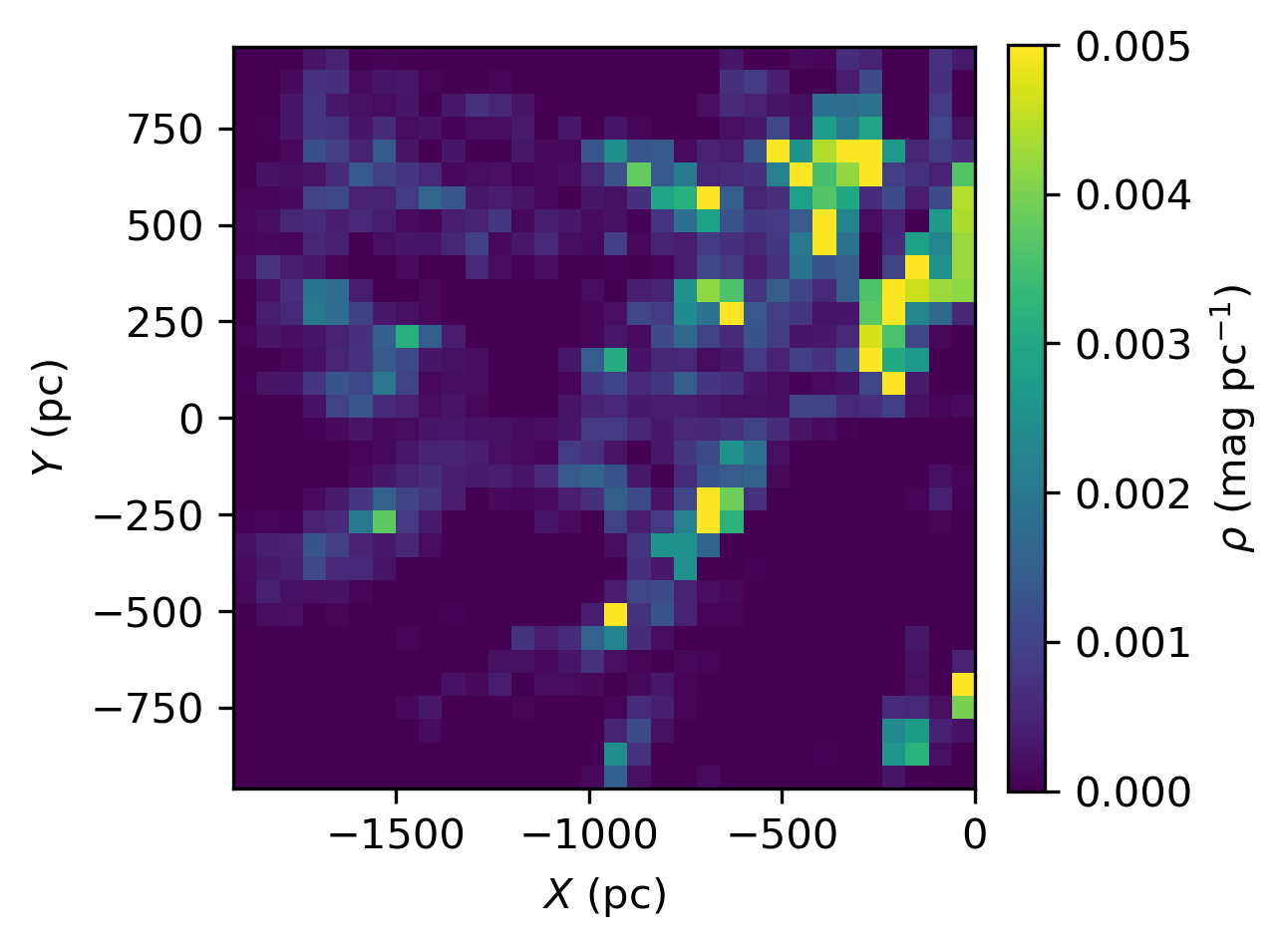}
\includegraphics[width=0.45\textwidth]{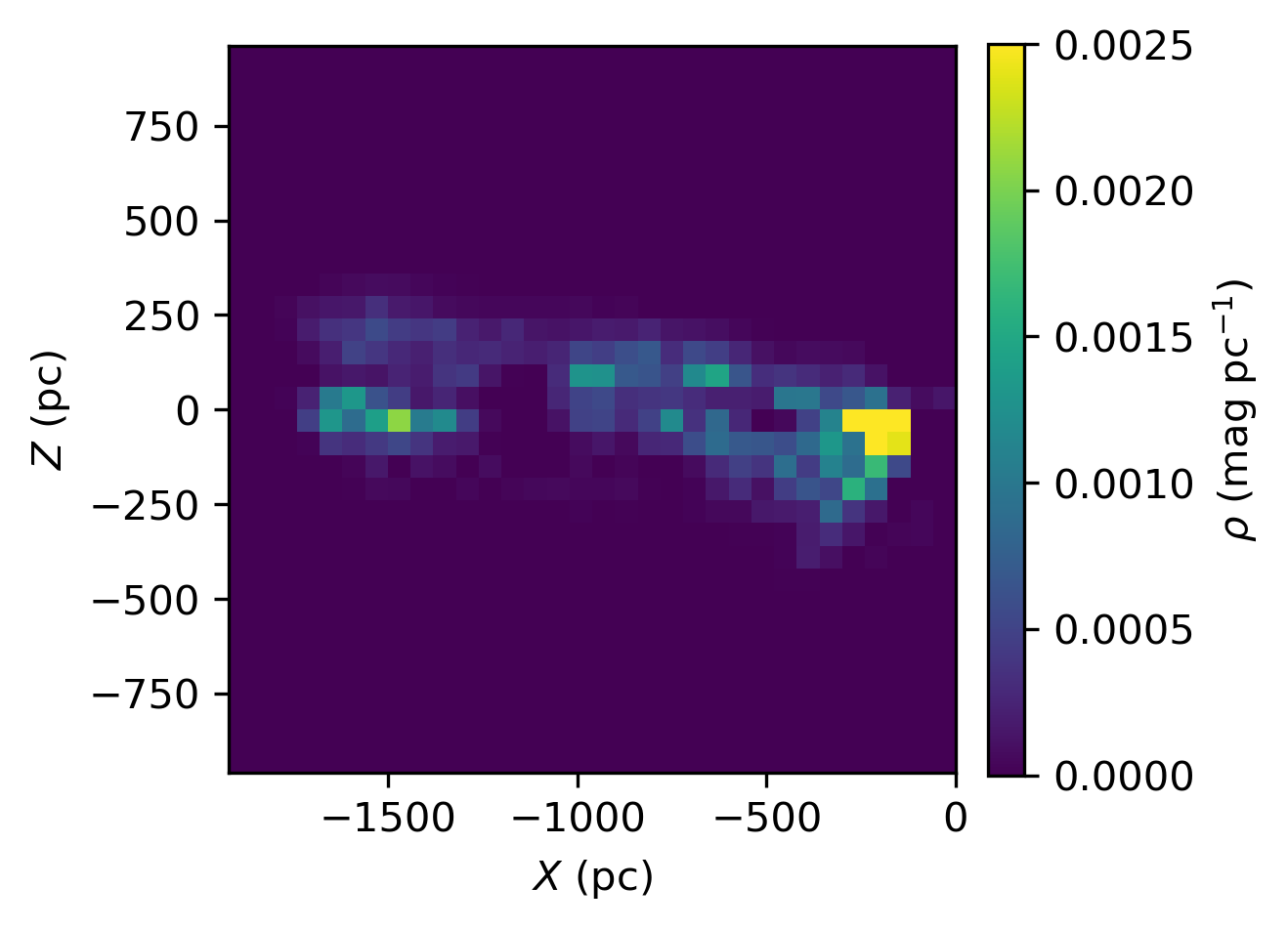}
\includegraphics[width=0.45\textwidth]{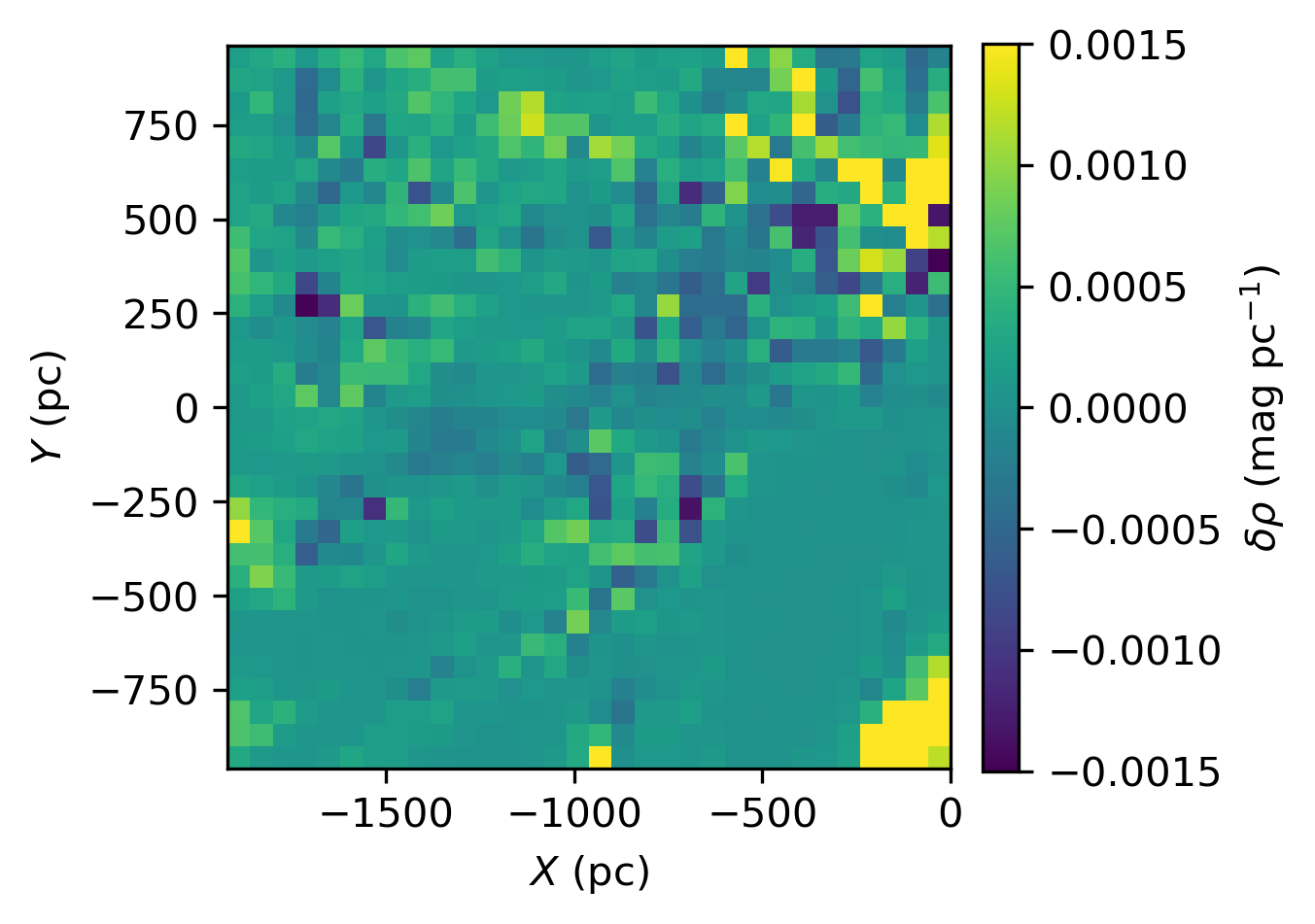}
\includegraphics[width=0.45\textwidth]{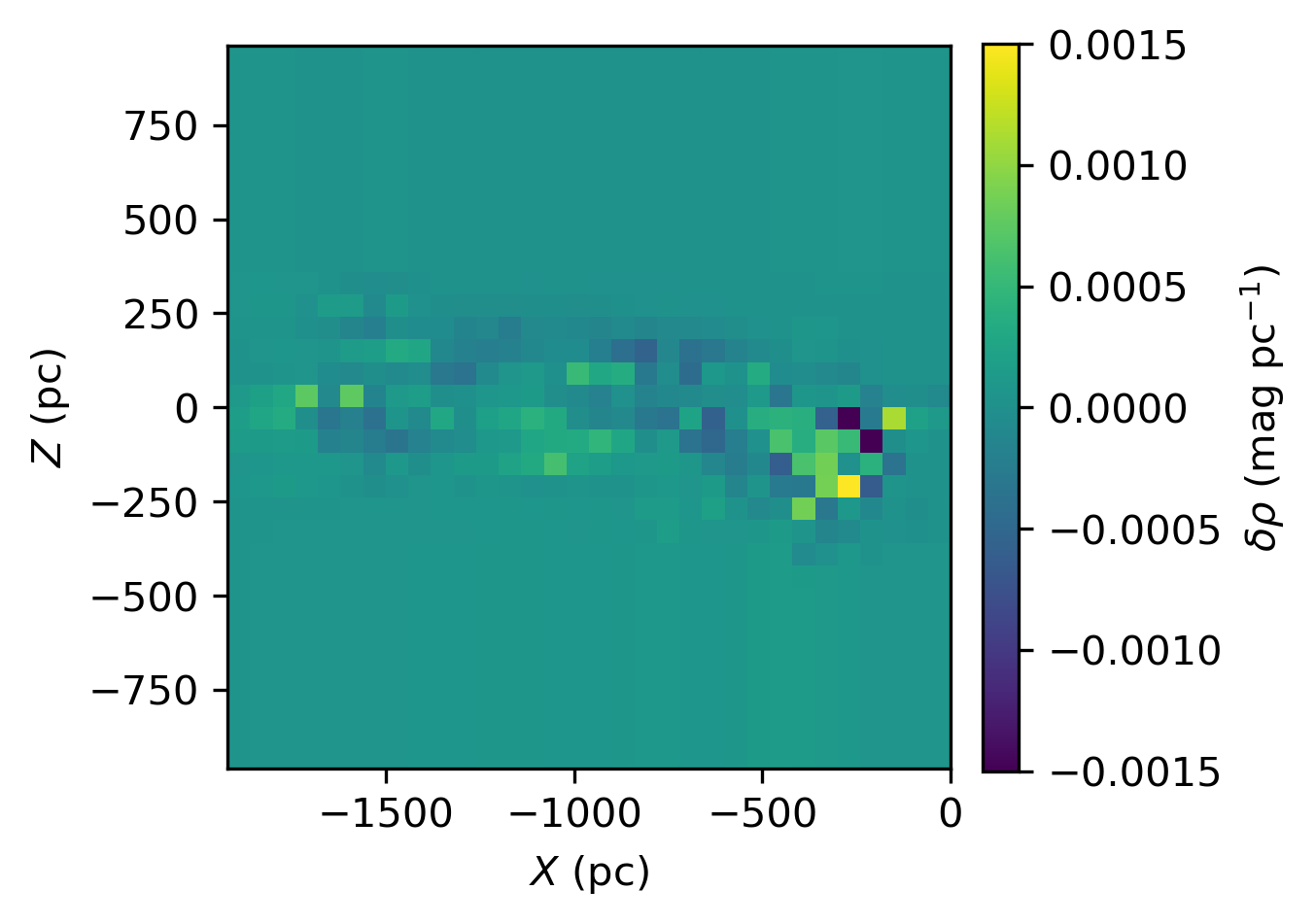}
\caption{Testing results by adopting different methods for simulating dust density distributions. The upper panels show the dust density maps derived from new V-net model using the l\'{e}vy flight-based training samples, while the lower panels show the residual maps comparing these results with those from \citet{Vergely2022}. The left panels represent the $XY$-plane dust density and residual maps for $Z$ values from 0 to 60\,pc, and the right panels for the $XZ$-plane with $Y$ values in the same range.}
\label{v22largeRWTest}
\end{figure*}

\begin{figure*}
\centering
\includegraphics[width=0.45\textwidth]{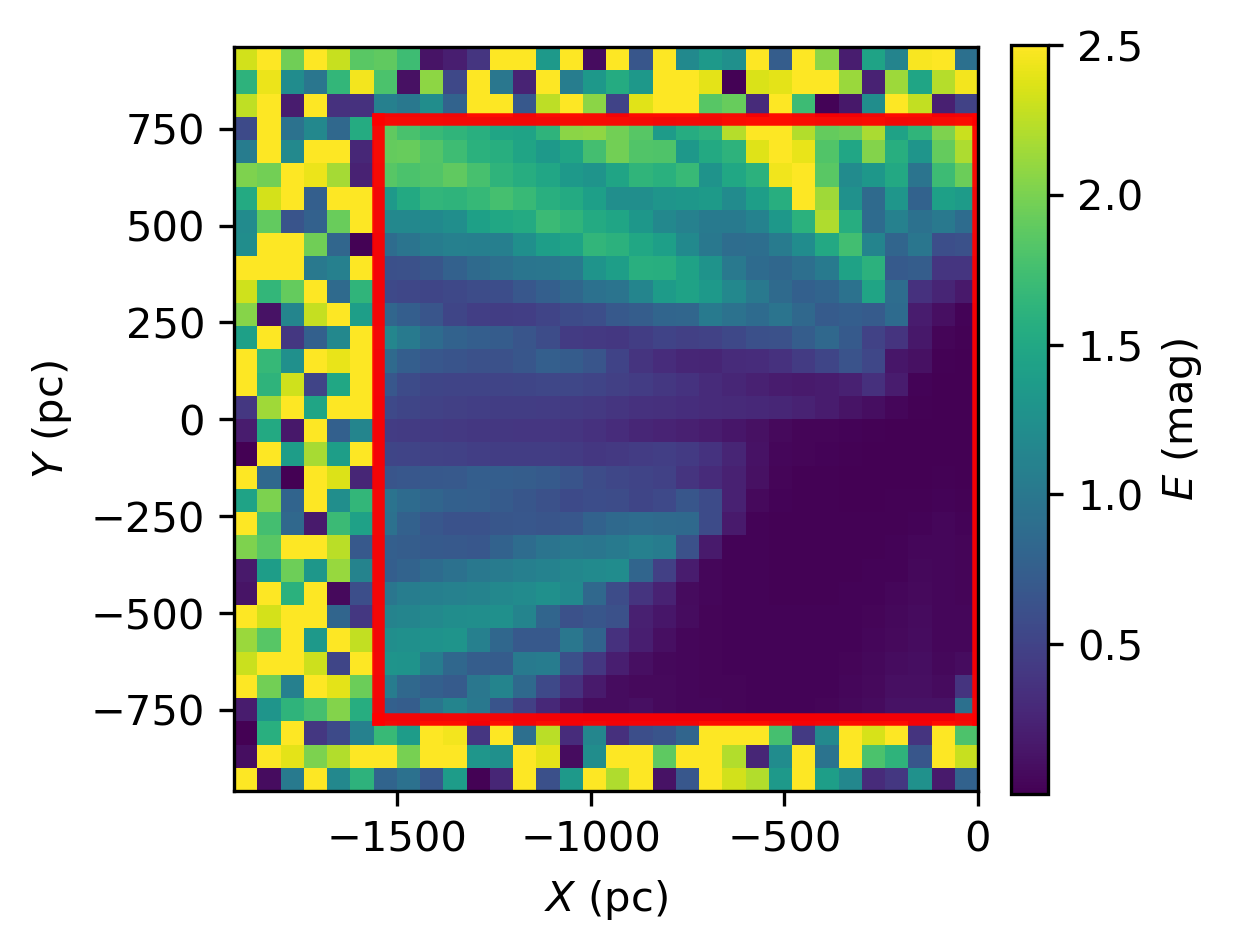}
\includegraphics[width=0.45\textwidth]{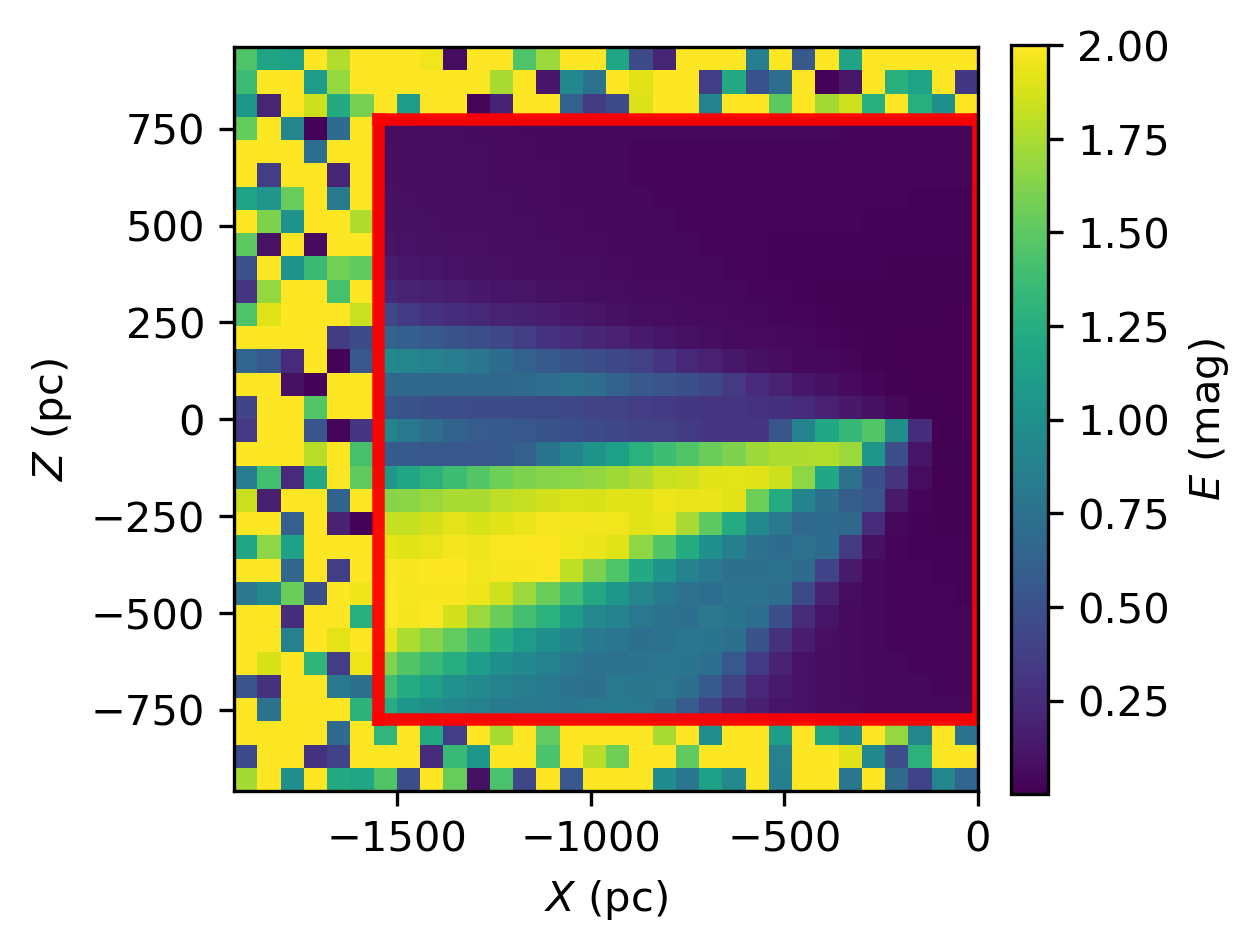}
\includegraphics[width=0.45\textwidth]{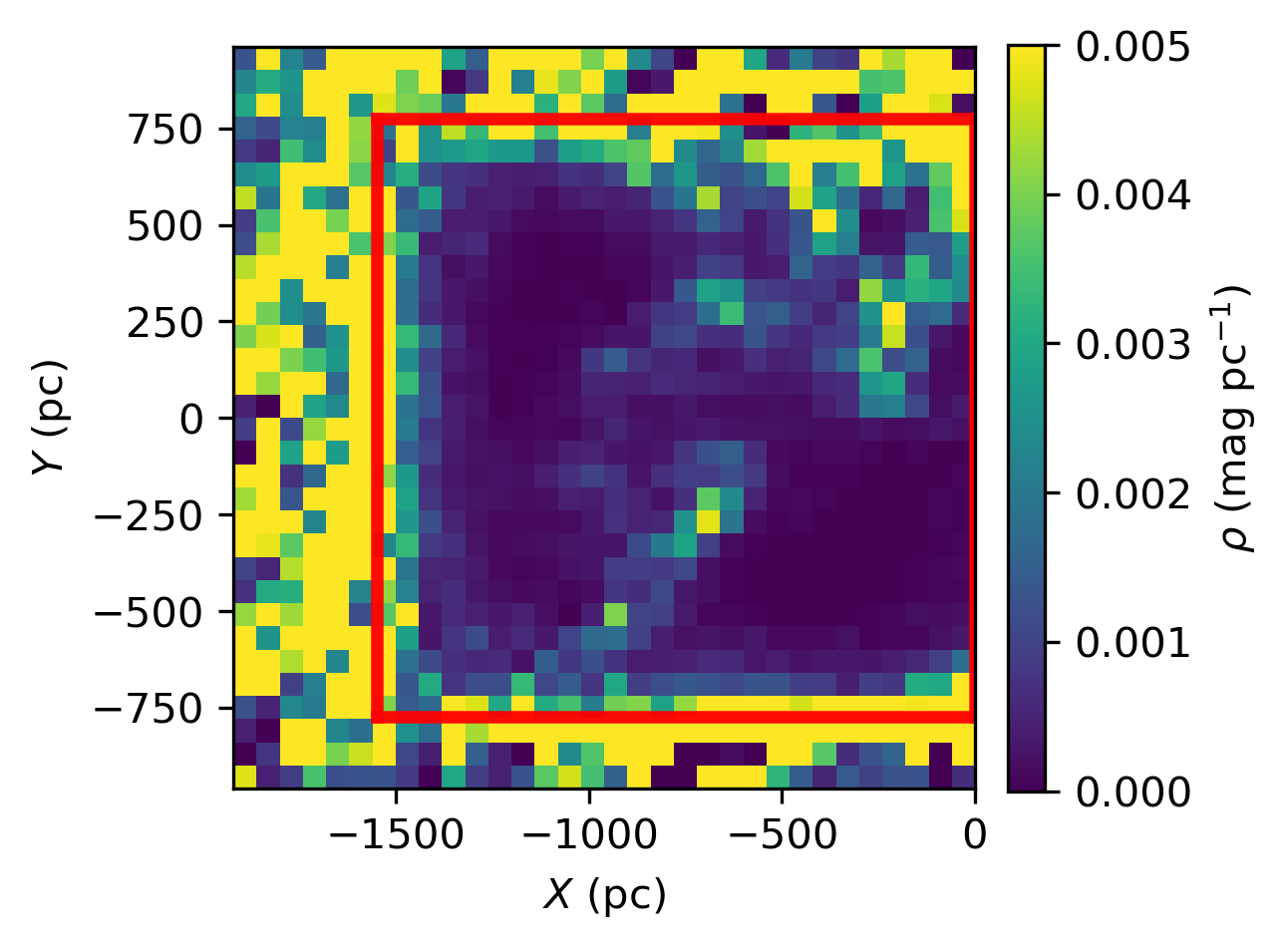}
\includegraphics[width=0.45\textwidth]{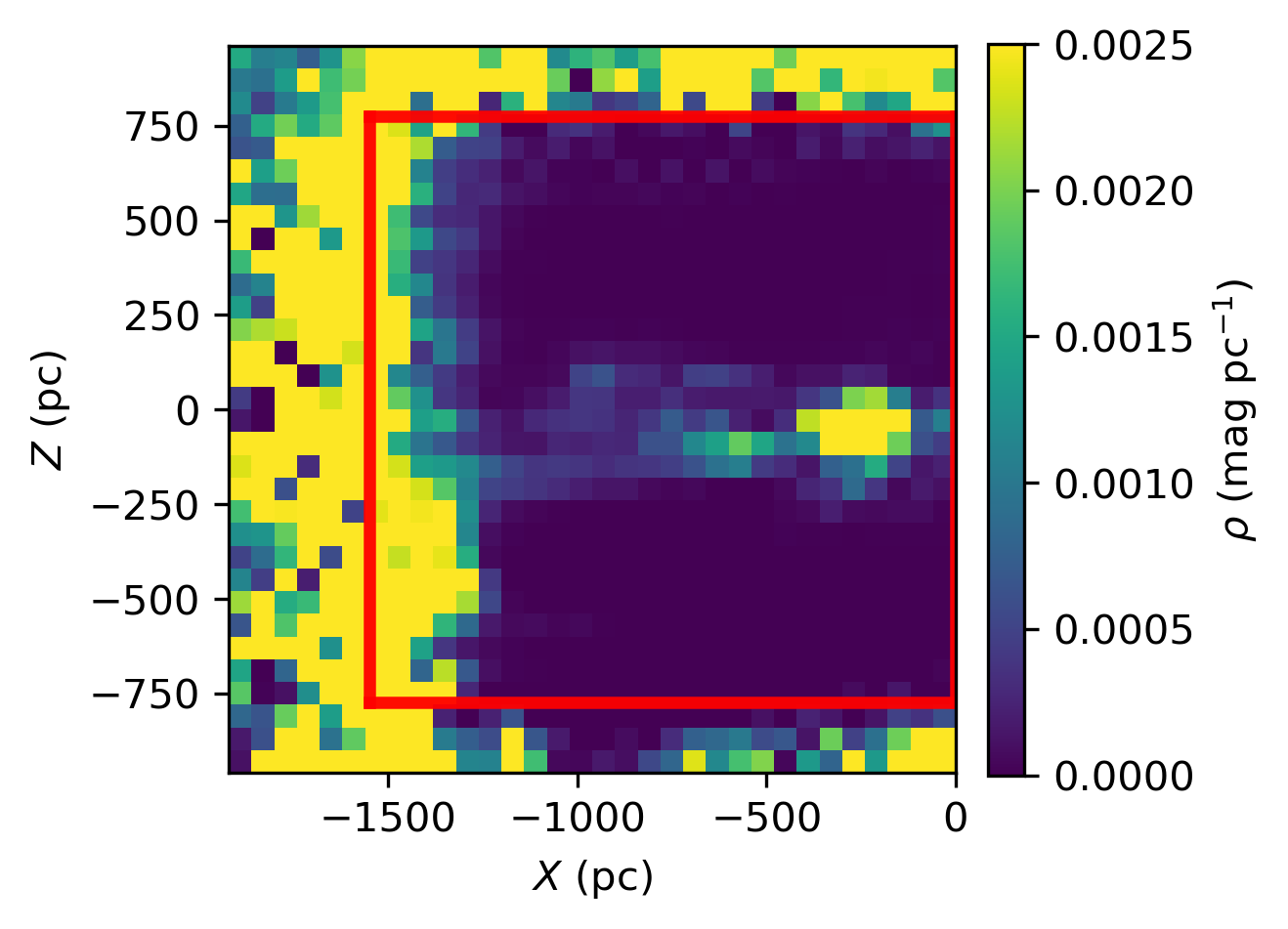}
\caption{Demonstration of derived 3D dust density maps for localized regions from the V-net model trained on larger fields. The upper panels display the LOS extinction distributions calculated from the \citet{Vergely2022} maps for the LR region, while the lower panels show the corresponding predicted 3D dust density maps from the V-net. The left panels represent the distributions in the $XY$-plane for $Z$ values between 0 and 60\,pc, whereas the right panels present them in the $XZ$-plane for $Y$ values in the same range. The red squares outline the boundaries of the tested subfield.}
\label{v22subfield}
\end{figure*}

In order to condense the main text, we have incorporated pertinent visuals in the form of 3D LOS extinction and dust density maps for both the LR and SR regions from the empirical data in Appendix~A. These figures encompass 3D LOS extinction and dust density maps obtained from the works of \citet{Vergely2022} and \citet{Zhang2023}, as well as those generated by our trained V-net model. Furthermore, we have incorporated the comparisons and residual maps that illustrate the differences between the literature findings and our own results. We have also included the corresponding uncertainty maps and maps derived from supplementary tests.

To assess the influence of different dust density modelling approaches on our results, we have conducted a test for the LR region using a simulation method analogous to that of the SR region. We generate new 3D dust density maps for the LR region and subsequently integrate them to produce the LOS extinction maps. These outputs are adopted as the training data for a new V-net model. 

Considering the LR region's extensive coverage in $Z$-space, we have constrained the initial points of our l\'{e}vy flights within $|Z| < 200$\,pc. We select 20 random starting points and execute 10 l\'{e}vy flights from each, with every flight comprising 2000 steps. Moreover, akin to the methodology applied in the logarithmic profile dust simulations, we convolve the resulting dust density distribution with an exponential function characterized by a scale height of 120\,pc. This process yielded over 5000 simulated 3D dust maps for the LR region. 

After integrating, we have derived the corresponding LOS extinction maps. These are utilized as training samples to refine a new V-net model. We then apply this updated V-net to the LOS extinction data from \citet{Vergely2022} to infer a new 3D dust density distribution for the LR region. The results, including residual comparisons with the actual data from \citet{Vergely2022}, are presented in Fig.~\ref{v22largeRWTest}. Notably, the V-net model, trained with l\'{e}vy flight-derived samples, yields outcomes highly consistent with those obtained using samples from the logarithmic profile method (see Fig.~\ref{v22large}). This consistency suggests that the choice of dust simulation technique has negligible impact on the efficacy of our V-net algorithms. 

In the inhomogeneous expanse of the Milky Way, the extinction caused by dust varies greatly depending on the LOS from our position. A uniform 3D dust density distribution can manifest distinct integral 3D extinction maps when situated in different Milky Way locales. Consequently, it is imperative to tailor the training samples for the V-net model to each specific region, using both the dust density and integral extinction distributions. An alternative, albeit less efficient, method would be to train a singular, all-encompassing model on a vast sky area and then derive the 3D extinction distributions for more confined regions. This method, however, is suboptimal as it necessitates excessive training durations and potentially reduces the resolution of the outcomes.

We propose and evaluate our technique using the V-net model trained on extensive sky areas to extrapolate the dust density distribution for more localized sectors. Our method involves inputting an extinction map with random data values for areas outside the target sub-region into the trained V-net model. This process ensures the dust density values within the sub-region are accurately inferred. An exposition of this technique is provided here, showcasing a scenario where data is limited to a specific sub-region within the LR sky area. For this test, we designate an arbitrary sky sector defined by $X < -1600$\,pc, $-800 < Y < 800$\,pc, and $-800 < Z < 800$\,pc. We then substitute random values for the data outside the designated $XYZ$ boundaries in the 3D extinction maps by \citet{Vergely2022}, aligning the random numbers' range with the maximum and minimum values present within the known sub-region. Upon processing this fabricated 3D extinction map through our V-net model, we successfully retrieve a dust density map, which is shown in Fig.~\ref{v22subfield}. The reconstruction is particularly accurate, with the exception of the test subfield's periphery.

\end{document}